\documentclass[english,11pt,oneside]{article}

\pdfoutput=1
\usepackage[T1]{fontenc}
\usepackage[latin9]{inputenc}
\usepackage[usenames,dvipsnames]{xcolor}
\usepackage{color}
\usepackage{babel}
\usepackage{setspace}
\usepackage{amstext}
\usepackage{amssymb}
\PassOptionsToPackage{normalem}{ulem}
\usepackage{ulem}
\usepackage[unicode=true,pdfusetitle,
 bookmarks=true,bookmarksnumbered=false,bookmarksopen=false,citecolor=Turquoise,
 breaklinks=false,pdfborder={0 0 1},backref=false,colorlinks=true,pdfpagemode=FullScreen,linktoc=page]
 {hyperref}
\usepackage{graphicx}
\usepackage{mathtools}
\usepackage[small]{caption}
\usepackage{changepage}
\usepackage{slashed}
\definecolor{note_fontcolor}{rgb}{0.80078125, 0.80078125, 0.80078125}
\usepackage[top=3 cm, bottom=3 cm, left=3.1416 cm, right=3.1416 cm]{geometry}

\newcommand{\ka}[1]{\textbf{\textcolor{blue}{(#1 --KA)}}}

\def\beq{\begin{equation}}
\def\eeq{\end{equation}}
\def\bea{\begin{eqnarray}}
\def\eea{\end{eqnarray}}

\begin{document}

\baselineskip=17pt

%%%%%%%%%%
%%%%%%%%%%    Title page
%%%%%%%%%%

\thispagestyle{empty}
\vspace{20pt}
\font\cmss=cmss10 \font\cmsss=cmss10 at 7pt

\begin{flushright}
%\today \\
%
UMD-PP-017-017\\
\end{flushright}

\hfill
%\vspace{20pt}

\begin{center}
{\Large \textbf
{LHC signals for Singlet Neutrinos from a Natural Warped Seesaw (I)
}}
\end{center}

\vspace{15pt}

\begin{center}
{\large Kaustubh Agashe$\, ^{a}$, Peizhi Du$\, ^{a}$, Sungwoo Hong$\, ^{a}$
%
%Luca Vecchi$\, ^{b, c}$ 
%
\\
\vspace{15pt}
$^{a}$\textit{Maryland Center for Fundamental Physics,
     Department of Physics,
     University of Maryland,
     College Park, MD 20742, U.~S.~A.} \\
     %
%      $^{b}$\textit{Dipartimento di Fisica e Astronomia, Universita` di Padova,\\ and INFN Sezione di Padova, Via Marzolo 8, 35131 Padova, Italy} 
 %     \\ 
  %    $^{c}$\textit{SISSA, Via Bonomea 265, 34136 Trieste, Italy} 
  %    \\ 
  %
      
\vspace{0.3cm}
      
{\it email addresses}: kagashe@umd.edu; pdu@umd.edu; sungwoo83hong@gmail.com
%
%; vecchi@pd.infn.it
%
}

\end{center}

\vspace{5pt}

\begin{center}
\textbf{Abstract}
\end{center}
\vspace{5pt} {\small \noindent

%\begin{center}

%{\bf Estimate for singlet neutrino LHC signals}

%\end{center}

%%%%%%%%%%%%%%%%%%%%%%%%%%%%%%%%%%%%%%%%%%%%%%%%%%%%%%%%%%%%%%%%%%%%%%%%%%%%%%%%%%%%%%%%%%%%%%%

%\ka{Slightly shorter one is below; in particular, removing the mention of $Z^{ \prime }$.}

%%%%%%%%%%%%%%%%%%%%%%%%%%%%%%%%%%%%%%%%%%%%%%%%%%%%%%%%%%%%%%%%%%%%%%%%%%%%%%%%%%%%%%%%%%%%%%%%

Recently, it was shown in arXiv:1512.06742 that a straightforward implementation of the type I seesaw mechanism in a warped extra dimensional framework is in reality a {\em natural} realization of ``inverse'' seesaw, i.e., the Standard Model (SM) neutrino mass is dominantly generated by exchange of pseudo-Dirac {\em TeV}-mass SM singlet neutrinos. By the AdS/CFT correspondence, this scenario is {\em dual} to these singlet particles being composites of some new strong dynamics, along with the SM Higgs boson (and possibly the top quark), with the rest of the SM particles being mostly elementary. We study signals from production of these  heavy neutrinos at the Large Hadron Collider (LHC). We focus on the scenario where the strong sector has a global $SU(2)_{\rm L} \times SU(2)_{\rm R} \times U(1)_{\rm X}$ symmetry; such a left-right (LR) structure being motivated by consistency with the electroweak (EW) precision tests. The singlet neutrinos are charged under $SU(2)_{\rm R} \times U(1)_{\rm X}$ symmetry, thus can be produced from $W^{ \pm }_R$ exchange, as in four-dimensional (4D) LR symmetric models. However, the direct coupling of light quarks to $W^{ \pm }_R$ is negligible, due to $W^{ \pm }_R$ also being composite (cf.~4D LR models); nonetheless, a sizable coupling can be induced by mixings among the various types of $W^{ \pm }$ bosons. Furthermore, $W^{ \pm }_R$ decays dominantly into the singlet and {\em composite} partner of charged lepton (cf.~SM lepton itself in 4D LR model). This heavy charged lepton, in turn, decays into SM lepton, {\em plus} $Z$/Higgs, thus the latter can be used for extra identification of the signal. For a benchmark scenario with $W^{ \pm }_R$ of mass 2 TeV and singlet neutrino of mass 750 GeV, we find that, in both the di-lepton + di-jet + Higgs and tri-lepton + Higgs channels, significant evidence can be seen at the LHC14 for an integrated luminosity of 300/fb and that even discovery is possible with slightly more luminosity.

}
 
\vfill\eject
\noindent

%%%%%%%%%%
%%%%%%%%%%    Main Text
%%%%%%%%%%

\section{Introduction}
%
%Summary}
%

The seesaw mechanism \cite{original} is a very attractive and hence perhaps the most popular one for explaining the extreme smallness of the Standard Model (SM) neutrino masses relative to those of the charged fermions. The basic idea is illustrated by the following schematic formula
\bea 
\hbox{generic seesaw}: \; 
m_{ \nu }  & \sim & \frac{ m^2_D }{ M_N } 
\label{generic}
\eea
where $m_D$ denotes the Dirac mass term between the SM doublet left-handed (LH) neutrino ($\nu_L$) and a SM singlet right-handed (RH) neutrino ($N_R$ or simply $N$), induced by the vacuum expectation value (VEV) of the SM Higgs boson and $M_N$ is the Majorana mass term for the singlet.

However, it is perhaps fair to say that in its {\em actual} realizations (including details of fitting to the observed neutrino masses), one typically ends up with a tuning of parameters (albeit {\em not} always {\em fine}-tuned, i.e., {\em not} involving large cancellations therein); here, we give some examples of this point. Then, we discuss a {\em natural} version in a warped extra dimensional model [dual to a four dimensional (4D) framework of composite Higgs and partially composite rest of the SM] \cite{Huber:2003sf, Agashe:2015izu}, which is the subject of further study in this paper.

In the original seesaw, the typical choice is that the above Dirac mass term between the two neutrinos
is of order the Higgs VEV, $v$ (or somewhat smaller), and similarly, the Majorana mass term for singlet is close to the UV cut-off scale (denoted by $M_{ \rm UV }$): 
\bea
\hbox{high-scale seesaw}: \; m_D & \lesssim & v \; (\hbox{{\em no} tuning}) \nonumber \\
M_N & \sim & M_{ \rm UV } \; (\hbox{{\em no} tuning, but see below!})
\label{high-scale}
\eea
(Note that in the above and in what follows, $v$ can be replaced by $m_{ \tau }$, i.e., largest of {\em charged} lepton masses with{\em out} qualitative change in conclusions.) Plugging Eq.~(\ref{high-scale}) in Eq.~(\ref{generic}), this results in the SM neutrino mass being much smaller than the electroweak symmetry breaking (EWSB) scale.

However, the observed SM neutrino mass (assuming this is set by the largest of neutrino mass$^2$ differences that have been confirmed, i.e., the atmospheric neutrino oscillations scale) requires that $M_{ \rm UV }$ in Eq.~(\ref{high-scale}) be actually several orders of magnitude smaller than the Planck scale:
\bea
m_{ \nu } \sim 0.1 \; \hbox{eV}
& 
\Rightarrow & 
M_{ \rm UV } \sim 10^{ 14 } \; \hbox{GeV}  \ll M_{ \rm Pl } \sim 10^{ 18 } \; \hbox{GeV} 
\eea
Of course, the latter hierarchy can be technically natural (i.e., radiatively stable), but the point is 
that realizing all this might require additional dynamics.
For example, if this scale corresponds to spontaneous breaking of a gauge symmetry [as in $SU(2)_L \times SU(2)_R \times U(1)_{ B - L }$ or left-right (LR) symmetric models, i.e., $N_R$ is part of a doublet of $SU(2)_R$] by a scalar VEV, then we have to explain why this scalar mass term is much smaller than the Planck scale. % \sh{$\to$ ?} note that weak-scale SUSY (for example) chosen to stabilize the Higgs VEV will of course make $M_{ \rm UV } \ll M_{ \rm Pl }$ technically natural, but still will not help in ``explaining'' this hierarchy.

An alternative is to set the singlet mass scale to be close to the IR (low-scale seesaw), for example, weak scale: % \txtred{\sh{Do we need all this ? Maybe..}(that scale, in turn, being stabilized -- and perhaps even explained -- in the ``standard" ways, for example  weak-scale SUSY or Higgs compositeness\footnote{Although this often requires further model-building, maybe even a coincidence!})}:
\bea
\hbox{low/TeV-scale seesaw}: \; M_N & \sim & M_{ \rm IR } \; (\sim \hbox{TeV}) \; (\hbox{{\em no} tuning})
\eea
but then the tuning is transferred to the Dirac mass term instead:
\bea
m_{ \nu } \ll v & \Rightarrow & m_D \ll v 
\eea
%
%Finally, the so-called ``inverse'' seesaw \cite{inverse} seeks to have natural choices for {\em both} the Dirac mass term between doublet and singlet neutrinos (i.e., $\lesssim v$) {\em and} the mass term for the singlet by itself (that too at the IR/weak scale), but latter one being {\em Dirac} instead, i.e., requiring introduction of {\em another} left-handed (LH) singlet denoted by $S$:
%\sh{Since the previous sentence, to me, was too long, I split it into two as follows. However, we can go back..}
Finally, the so-called ``inverse'' seesaw \cite{inverse} seeks to have natural choices for {\em both} the Dirac mass term between doublet and singlet neutrinos (i.e., $\lesssim v$) {\em and} the mass term for the singlet by itself (that too at the IR/weak scale). However, in the inverse seesaw, the singlet neutrino is {\em Dirac} fermion, requiring introduction of {\em another} left-handed (LH) singlet denoted by $S$:
\bea
M_{ NS } & \sim & M_{ \rm IR } \; (\sim \hbox{TeV}) \; (\hbox{{\em no} tuning})
\eea
In addition the second singlet has a small Majorana mass term denoted by $\mu$ so that the SM neutrino mass formula ends up looking like:
\bea
\hbox{inverse seesaw}: \; 
m_{ \nu }  & \sim & \frac{ m_D^2 }{ M^2_{ N S } } \mu
\label{inverse}
\eea
Of course, tuning is then shifted to the Majorana mass term for $S$:
\bea
m_{ \nu } \ll v & \Rightarrow & \mu  \ll M_{ NS } 
\eea

So, it seems that four-dimensional (4D) models of seesaw might not be entirely satisfactory as far as explaining {\em fully} the small {\em observed} SM neutrino mass.
Recently, it was emphasized \cite{Agashe:2015izu} that 
\begin{itemize}

\item
a {\em natural} realization of seesaw mechanism
occurs in the warped {\em extra} dimensional framework.\footnote{This model was originally proposed in references \cite{Huber:2003sf}, but the basis used in this earlier work obscured the {\em physical} nature of the seesaw mechanism.}  
%
%(which is dual, following the
%AdS/CFT correspondence, to the SM particles being composites of new strong dynamics.
%

\end{itemize}
This framework is dual, following the AdS/CFT correspondence, to varying degree of compositeness of the SM particles. 
%\sh{warped seesaw dual to composite bla bla...}
In a sense, this implementation actually features {\em both} high-scale and inverse seesaw mentioned above.
Namely, from a bottom-up viewpoint, the SM neutrino mass is generated by exchange of pseudo-Dirac singlet states as in inverse seesaw case. Remarkably, 
\begin{itemize}

\item
the smallness of the required Majorana mass term ($\mu$) for the inverse seesaw is itself 
due to a high-scale seesaw:

\end{itemize}
schematically, we have (with $M_{\rm IR} \sim $ TeV as usual)
%\sh{I added ``in'' below b/c, unlike above equations, this equation is not for neutrino mass, but for small Majorana splitting.}
%
%\bea
%\hbox{\txtblue{in} warped/composite (inverse) seesaw}: \; \mu & \sim & \frac{ M_{ \rm IR }^2 }{ M_{ \rm UV } } \; (\hbox{{\em no} tuning})
%\label{mu_warped}
%\eea
%
%\sh{or, we could try something like,}
%
\bea
\hbox{warped/composite seesaw}: \; m_{ \nu }   \sim  \frac{ m_D^2 }{ M^2_{ \rm IR } } \mu, \;\;\; \mu \sim  \frac{ M_{ \rm IR }^2 }{ M_{ \rm UV } } \;\; (\hbox{{\em no} tuning})
\label{mu_warped}
\eea
Note that, even with the above nice feature, we still need (as allued to above) the other hierarchy for getting the observed SM neutrino mass, i.e., $M_{ \rm UV } \ll M_{ \rm Pl }$: this {\em seems} to be a tuning at first sight, but  we will see that this is also explained in warped/composite seesaw.

In detail, the dual CFT picture affords the most transparent understanding of this physics as follows (see more discussion in \cite{Agashe:2015izu} and some using 5D model in Sec.~\ref{sec:5Dmodel} of this paper). The SM Higgs boson arises as a composite of some new strong dynamics which confines at the $\sim$ TeV scale. Rest of the SM (i.e., {\em all} the gauge fields and fermions) start out as elementary degrees of freedom which are external to the strong dynamics, but they ``mix'' with appropriate composites of the latter. 
Thus, the actual SM particles are admixtures of the two sectors.
Such ``partial compositeness'' of the SM fields allows them to couple to the SM Higgs, thus acquiring mass
from its VEV.
In particular, for the case of charged SM fermions, the picture is that external $SU(2)_{\rm L}$ doublet and singlet fermions mix {\em separately} with respective composite ones, {\em starting at the UV cut-off}. Then, only in the far IR, i.e, at $\sim$ TeV scale, these two types of composites (and hence the corresponding external fermions as well) ``connect'' to each other via the Higgs VEV.

For the neutrino sector, the story starts out similarly, i.e., we {\em add} to the SM lepton sector, an external (chiral) SM singlet, denoted by $N_R$, which mixes with an entire composite SM singlet tower from $\sim$ TeV upwards.
However, from then on, there is a departure in the script (vs.~that of charged fermions), again, kind of similarly to the usual seesaw models, but with some crucial difference as follows.
Obviously, this concerns the ``fate'' of the external $N_R$: namely, we assume that the strong dynamics in isolation preserves lepton number so that the composite singlets are purely Dirac to begin with. On the other hand, the {\em external} sector mass terms and interactions need not preserve lepton-number, for example, $N_R$ has a Majorana mass term, $M_N$, which is close to the UV-cut-off, say, $M_{\rm Pl}$.

However, even though lepton-number is violated at the UV cut-off, we can{\em not} write down a SM neutrino mass operator at this stage, since the SM Higgs boson VEV is not ``born'' yet. Instead, the relevant effect of Majorana $N_R$ is that its coupling to strong dynamics will inject lepton-number violation into the strong dynamics also; in particular, integrating out $N_R$ (again, close to the UV cut-off) generates Majorana mass terms for the composite singlet states: note that these Majorana mass terms are for the {\em left} chirality of composite, since that is the one with {\em mass} mixing {\em term} with external $N_R$.

So, we start seeing the ``ingredients'' for a inverse seesaw model, with the seeds being sown in the UV; in particular, it is the two chiralities of the composite singlet who play the role of the $N$, $S$ fields of the usual 4D model of this type!

Thus, we naturally have
\bea
\hbox{warped/composite seesaw}: \; M_{ NS } & \sim & \hbox{TeV/compositeness scale}
\label{MNS_warped}
\eea
Moreover, as already advertised above, we have an explanation for smallness of the Majorana mass term for $S$ [i.e., $\mu$ in Eq. (\ref{inverse})]. Namely, for the TeV mass composites, this mass term will precisely be of the form of $\mu$ in Eq. (\ref{mu_warped}) above, i.e., the ``TeV'' in the numerator there comes from the above-mentioned mass mixing term (between $N_R$ and LH composite) and $M_{ \rm UV }$ in denominator is just the (Majorana) mass term for $N_R$ with itself. We will argue in a bit that this ``effective'' UV scale can actually be naturally smaller than Planck scale.
The final cog in this wheel is the Dirac mass term for the composite singlet with the SM $SU(2)_L$ doublet neutrino: similarly to the case of the charged fermions, this arises from coupling of composite singlet to Higgs VEV and composite doublet, latter mixing with the external SM neutrino. Of course, one difference from charged fermion case is ``absence'' of external leg on the singlet side (since $N_R$ decoupled); so schematically, we get 
\bea
m_D & \sim & \sqrt{ m_{ \tau } \; v } \; \hbox{(\emph{no} tuning)}
\label{mD_warped}
\eea
i.e., with two external fermions, we would have gotten $m_{ \tau }$ vs.~its ``square root'' here with only external doublet present\footnote{For simplicity, we assume here similar degree of compositeness for doublet and singlet {\em charged} lepton.}.
In other words, 
\begin{itemize}

\item
the composite singlets act as a ``bridge'' between EWSB in the IR and lepton-number violation in the UV, 
{\em both} of which are required in order to generate (Majorana) SM neutrino mass.

\end{itemize}

Note that plugging Eqs.~(\ref{MNS_warped}), (\ref{mu_warped}) and (\ref{mD_warped}) into Eq.~(\ref{inverse}), we see that final formula looks like {\em high}-scale seesaw, i.e., using Eq.~(\ref{high-scale}) in Eq.~(\ref{generic})!
In fact, the procedure used in most of the previous literature \cite{Huber:2003sf} for the computation of the SM neutrino mass in this warped extra-dimensional framework reinforces as follows this impression of high-scale seesaw. 
In this 5D model, we have a SM singlet propagating in the bulk, with a Higgs VEV-induced Dirac mass term with the SM lepton doublet field near the IR brane.
In addition, this singlet field has a Majorana mass term on the UV brane, i.e., bulk and IR brane preserve lepton-number.
In the so-called Kaluza-Klein (KK) basis for the singlet 4D states, first the usual mode decomposition is performed by neglecting the above Majorana mass term for the singlet, resulting in zero and massive KK modes.
The effects of the UV brane localized Majorana mass term on these modes are only subsequently taken into account, lifting the (would-be) zero-mode {\em and} mixing them all up.
It turns out that the exchange of only the would-be zero-mode with a super-large Majorana mass term gives rise to the SM neutrino mass, which thus mimics a high-scale seesaw.
However, some of us showed in \cite{Agashe:2015izu} that this is not so in the {\em mass} basis, i.e., it is physically an inverse seesaw (as is clear from the above CFT viewpoint).

In fact, we can further ``exploit'' this process of {\em communication} between the UV (i.e., lepton-number violation) and IR (i.e., EWSB) as follows.
Firstly, it is clear that the lepton-number violating perturbation to the strong dynamics (again, from integrating out the external Majorana singlet, $N_R$) has to be suitably {\em renormalization group} (RG) evolved from the UV scale to IR, i.e., over a large hierarchy. 
Assuming that the strong dynamics is approximately conformal over this hierarchy as would be needed in order to get the observed sizes of SM fermion masses, we see that this transmission can be significantly modulated by the anomalous dimensions of the operators involved.\footnote{This corresponds to profiles for various modes in the extra dimensional dual.}
So, assuming sizable anomalous dimensions,
%\sh{Isn't it better to focus on $c_N > 0.5$ and just mention seesaw scale $\ll$ Planck scale ? I mean we explored the other case for theoretical completeness, but here, being pheno paper, do we need to keep generality of discussion ?}
%
\begin{itemize}

\item
the effective seesaw scale can be much smaller (or larger, depending on {\em sign} 
of the anomalous dimensions!) than the Planck scale: 

\end{itemize}
again, heuristically speaking, 
\bea
M_{ \rm UV } & \sim & M_{ \rm Pl } \times (\hbox{anomalous scaling} \leftrightarrow \hbox{5D profiles}) \nonumber \\
& \sim & 10^{ 12 } \; \hbox{GeV} \; (\hbox{{\em no} tuning})
\eea
where the {\em requirement} of the ``intermediate'' scale in second line corresponds to the choice of $m_D$ in second line of Eq.~(\ref{mD_warped}), using this and Eq.~(\ref{mu_warped}) in Eq.~(\ref{inverse})
and finally setting $m_{ \nu } \sim 0.1$ eV.
Just to be clear, there is {\em no} new dynamics at this scale, cf.~usual, 4D high-scale seesaw.\footnote{
where, for example, this is associated with the breaking of $SU(2)_{\rm R} \times U(1)_{\rm X}$ gauge symmetry down to $U(1)_{\rm Y}$.}
In short, we then have a fully natural seesaw model here, i.e., with no large hierarchies in {\em any} of the {\em fundamental} parameters!

Secondly, because we need the message of lepton-number violation to be brought down to the {\em TeV} scale by {\em particles} beyond the SM, i.e., the composite singlets, we are obviously able to 
%\sh{I am not sure if it'd be ``direct'' probe of the mechanism..}
%
\begin{itemize}

\item
probe the mechanism of generation of SM neutrino mass, namely, by 
producing the lightest of these messengers at the Large Hadron Collider
(LHC)/future colliders (unlike the case of high-scale seesaw).

\end{itemize}
Of course, this is a feature in general of inverse seesaw models so that such signals have been studied before \cite{Chen:2011hc, Das:2015toa}, but (as we will show here) the compositeness of the singlets make a difference!

In a series of papers (this being the first), we initiate the study of LHC signals for the $\sim$ TeV mass singlets in the natural realization of (inverse) seesaw in this warped/composite Higgs setting. We begin here by focussing on a {\em specific}, but {\em well-motivated} model within the above framework. Namely, 
\begin{itemize}

\item
we assume that the strong dynamics has a global symmetry (in the EW sector)
which contains $SU(2)_{\rm L} \times SU(2)_{\rm R} \times U(1)_{\rm X}$ 

\end{itemize}
of which the SM subgroup, i.e.,
$SU(2)_{\rm L} \times U(1)_{\rm Y}$ is gauged by external fields [with $U(1)_Y$ being a combination of
$U(1)_{\rm X}$ and the $U(1)$ contained in $SU(2)_{\rm R}$]\footnote{The warped 5D dual of this scenario is that the {\em bulk} EW {\em gauge} symmetry is extended as above and broken down to the SM subgroup on the UV brane.}. In the canonical case, we would identify ${\rm X} = ( {\rm B} - {\rm L} )$ as in 4D LR models, but in general we could choose other representations under the extra $U(1)$.
The motivation for such an extension of the EW ({\em global}) symmetry in the present context is {\em not} the one for the 4D LR models, i.e., parity restoration at higher energy scales, but rather that it provides a custodial symmetry for suppressing the contributions of the strong dynamics to the EW precision tests, in particular, the $T$ parameter.
Thus, even with the choice of ${\rm X} = ( {\rm B} - {\rm L} )$, there is then {\em no} need for an {\em elementary} (i.e., external to the strong sector) $W^{ \pm }_R$, i.e., charged gauge boson of $SU(2)_{\rm R}$ group, in this model. Similarly, the combination of $U(1)_{ {\rm B} - {\rm L} } $ and $U(1)$ in $SU(2)_{\rm R}$ which is orthogonal to $U(1)_{\rm Y}$ -- often denoted by $Z^{ \prime}$ -- is not gauged, {\em un}like in 4D LR models, i.e., the external sector does not respect the {\em extended} EW symmetry.\footnote{As a bonus, with such a symmetry structure, we {\em automatically} realize the pure Diracness of composite singlets vs.~large, possibly close to UV cut-off, Majorana mass term for the {\em external} singlet.}
We will mostly use the elementary-composite sector picture (called two-site model \cite{Contino:2006nn}, but augmented now by the composite singlet neutrinos) in our actual LHC signal analysis.

Even though we do not have elementary $W_R^{ \pm }$/$Z^{ \prime }$ in this model, given the above global symmetry of strong dynamics, we do have 
\begin{itemize}

\item
{\em composite} $W^{ \pm }_R$ and $Z^{ \prime }$\footnote{We will denote them simply by the same symbols, since there is no chance of confusion with elementary ones in this model. Also, strictly speaking, we have to assume degeneracy of spin-1 composites here in order to classify mass eigenstates in this way: we will consider the case of non-degeneracy in a follow-up paper, where we will give more details of this issue.},
which do couple to singlet neutrinos (cf.~composites of SM gauge bosons obviously do not); 

\end{itemize}
this simultaneous similarity (i.e., ``existence'' of $W^{ \pm }_R$ and $Z^{ \prime }$) {\em and} difference (their compositeness vs.~elementary nature) from 4D LR models will be crucial to the analysis of signals for the present model.
For the lepton sector, we indeed make the canonical choice of fermion representations, but now for the  {\em composites}, since it is that sector which has the $SU(2)_{\rm R}$ symmetry, 
i.e., 
\begin{itemize}

\item
the composite (denoted by $\psi_e$) with which the external RH charged lepton mixes\footnote{called ``electron'' here for simplicity, even though we extend this to the second and third generations also} is part of a doublet of the (global) $SU(2)_{\rm R}$ of strong dynamics, whose other component is the composite RH neutrino (denoted by $\psi_N$), i.e., with which external $N_R$ mixes as mentioned above.\footnote{In detail, one might need {\em two} such $SU(2)_{\rm R}$ doublet composites per generation -- corresponding to two different 5D fields -- in order to obtain the correct SM charged lepton vs.~SM neutrino Dirac mass term, i.e., external charged lepton might actually mix with a {\em different} composite tower than the $SU(2)_{\rm R}$ partners of composite SM singlets associated with the SM neutrino mass. However, this modification does not (qualitatively) affect the present discussion.}

\end{itemize}
Both $\psi$'s have Dirac mass $\sim$ TeV and are vector-like under the SM {\em gauge} and strong dynamics global symmetries.

We begin by considering the production of $\psi_N$ via decays of {\em on}-shell $W^{ \pm }_R$; again such a signal has been studied extensively in the case of usual, 4D LR models \cite{Chen:2011hc}, but the difference here is that $W_R^{ \pm }$ is composite vs.~quarks inside proton being mostly elementary. So, naively, this coupling seems to be negligible (i.e., $\propto$ tiny admixture of composite in SM light quarks or the corresponding Yukawa couplings). Nonetheless, we discuss how 
\begin{itemize}

\item
a significant, albeit still mildly suppressed {\em relative} to SM, light quark-$W_R^{ \pm }$ coupling is induced.

\end{itemize}
This arises by a {\em combination} of elementary-composite mixing for $W^{ \pm}$'s corresponding to $SU(2)_L$ (denoted by $W_L^{ \pm }$)\footnote{Recall that there is no elementary gauge boson mixing {\em directly} with composite $W_R^{ \pm }$.}  {\em and} composite $W^{ \pm }_L-W_R^{ \pm }$ mixing induced by Higgs VEV, with the near degeneracy of these composites in a ``minimal'' model\footnote{This is dual to the 5D model with no IR brane-localized kinetic terms for bulk gauge fields.} amplifying the Higgs VEV effect (see reference \cite{Agashe:2007ki, Agashe:2008jb} for the
5D version of this effect).
(We will consider the case of {\em non}-degenerate spin-1 composites in a follow-up paper.)
In fact, such a mild suppression of production of $W_R^{ \pm }$ (as compared to usual 4D LR models) is perhaps ``welcome'' in the sense that the LHC early run 2 searches are already constraining $2$ TeV $W_R^{ \pm }$ in the usual case, but with compositeness, such low scale for $W_R^{ \rm }$ would then (i.e., given smaller cross-section for the same mass) still be allowed. At the same time, as we will show, the coupling is sizable enough that discovery (for 2 TeV $W^{ \pm }_R$ and $\sim 750$ GeV $\psi_{ e, N }$\footnote{We could contemplate even lighter singlet neutrino, but accomplishing such a hierarchy might require tuning, for example, {\em too} large brane-localized kinetic terms, given that gauge KK cannot be below $\sim2$ TeV due to constraints from EWPT.} such that the above decay is allowed) by the end of run 2 ($\sim$ 300 fb$^{-1}$) would be possible.

Moving onto the relevant decays of $W^{ \pm }_R$\footnote{Other decay channels for $W_R^{ \pm }$ include various components of the Higgs doublet: these were studied in \cite{Agashe:2008jb}, but singlet neutrino was not included there.}, first of all, the largest coupling of $W_R^{ \pm }$ involves {\em composite}
partner of SM $e_R$ and the composite singlet neutrino, i.e., $\psi_e$ and $\psi_N$, cf.~{\em SM} $e_R$
and singlet neutrino in the usual, 4D LR case.
The singlet neutrino decays predominantly (as in 4D LR models) into SM doublet lepton and Higgs doublet (including physical Higgs and longitudinal $W/Z$) via the associated Yukawa coupling\footnote{Note that this coupling is indeed small, given that it involves degree of compositeness of {\em SM} (doublet) lepton, but there is not much of an ``option'' here in terms of decay channel, given that lepton-number is (approximately) preserved.}: the channel we will focus on here (based on smaller background, thus more visibility) is $e_L + W$.
On the other side, $\psi_e$ will similarly decay: we will consider $e_L + Z_{ \rm long}/h$ final state here.
Thus, we see that there is 
\begin{itemize}

\item
an ``extra'' Higgs/$Z$ (vs.~usual, 4D LR models) among the decay products of the $W_R^{ \pm }$,
which, assuming it is tagged, can be used to reduce the SM background.

\end{itemize}
Moreover, it then allows us to possibly reconstruct the full decay of $\psi_e$, thus determining its mass, which is same as that of $\psi_N$ [given the $SU(2)_R$ symmetry].
Including decays of $W$ from $\psi_N$, we then have 
\begin{itemize}

\item
two search channels, i.e., dilepton$+$ dijet (hadronic decay of $W$) and tri-lepton (leptonic decay of W), along with Higgs/$Z$ boson.\footnote{Note that even in the usual, 4D LR  models, one can also get
Higgs/$Z$ boson from singlet neutrino decay, but then we lose lepton(s), i.e., final state with be $lh +$ MET, thereby increasing SM background (for example, SM $Wh$ production will then be relevant), as opposed to our case of Higgs along with di-or-tri-leptons.}

\end{itemize}
We will study both of these and find them to be complementary, for example, rate is larger for the former
(based simply on corresponding branching ratios of $W$),
but so is possibly SM background, given that leptons are typically ``cleaner''.
(Of course, for the case of hadronic decay of the $W$ from $\psi_N$, that side is also fully visible
and hence
can furnish information on $\psi_{ e, N } $ masses.)

Finally, in addition to $W^{ \pm }_R$, we consider production of $\psi_{ e, N }$ pairs 
from decays of {\em on}-shell 
$Z^{ \prime }$. 
Once again, the ``direct'' coupling of quarks inside proton to $Z^{ \prime }$ is negligible;
%(given their elementary vs.~composite, respectively, natures); 
however, mixing does create a larger coupling (just like for the case of $W^{ \pm }_R$ above). 
Note that in usual, 4D LR models, $Z^{ \prime }$ is typically heavier than $W_R^{ \pm }$,
for example, assuming both (being elementary) get their mass from some scalar VEV, just like the case of 
SM $W/Z$. 
Hence, production cross-section of $Z^{ \prime}$ tends to be smaller than that of $W^{ \pm }_R$.
However, in the seesaw model being studied here, 
\begin{itemize}

\item
the $W^{ \pm }_R$ and $Z^{ \prime }$ can be almost degenerate,
since their masses arise from the compositeness scale so that $Z^{ \prime }$ signal
can be comparable to $W_R^{ \pm }$.

\end{itemize}

Here is the outline of the rest of this paper.
We begin in the next Sec.~\ref{sec:5Dmodel} with a brief review of the basic seesaw model in the warped extra dimensional framework
and present details of the implementation in the context of the $SU(2)_{\rm R}$ extension of the SM EW
symmetry mentioned above.
In Sec.~\ref{sec:twosite}, we outline the ``simplified'', i.e., two-site approach \cite{Contino:2006nn} to studying the 5D model
that we will employ in our actual analysis of LHC signals.
We then discuss our main results, starting with production cross-sections and decay branching ratios of various
heavy particles in Sec.~\ref{sec:overview_of_LHC_signals}, followed by computations of SM background and thus the discovery potential for the new particles
in Sec.~\ref{sec:analysis_results}.
Here, we also 
mention/briefly discuss strategies (post-discovery) for distinguishing the composite/warped seesaw model from the usual, 4D LR one.
We conclude and present some directions for future work in Sec.~\ref{conclude}.

%%%%%%%%%%%%%%%%%%%%%%%%%%%%%%%%%%%%%%%%%%%%%%%%%%%%%%%%%%%%%%%%%%%%%%%%%%%%%%%%%%%%%%%%%%%%%%%%%%
%%%%%%%%%%%%%%%%%%%%%%%%%%%%%%%%%%%%%%%%%%%%%%%%%%%%%%%%%%%%%%%%%%%%%%%%%%%%%%%%%%%%%%%%%%%%%%%%%%

\section{5D natural warped seesaw model}
\label{sec:5Dmodel}

%Randall-Sundrum model is a brilliant approach to address both electro-weak hierarchy problem and flavor hierarchy problem. Original RS model (known as RS1) allows all 5D fields  to propagate in the bulk between two branes, namely UV and IR branes, with different profiles in the extra dimension. The zero model of those 5D fields after KK decomposition corresponds to SM fields. Couplings are obtained from overlap of 5D profile of corresponding particles. In this framework, the elementary graviton has a profile peaked at UV brane, which is typically at Planck scale, while Higgs is localized near IR brane. Small overlap between graviton and Higgs field naturally provides TeV scale Higgs mass , thus solving Plank- weak hierarchy problem. Mass hierarchy among charged leptons, known as flavor hierarchy problem, can also be solved by different overlaps of profiles with Higgs . Profiles of leptons are determined by 5D mass. Small difference in 5D mass can lead to hierarchical values near IR/TeV brane. Therefore, mass hierarchy, or equivalently hierarchy among lepton Yukawa couplings with Higgs, is naturally generated by overlap with IR brane localized Higgs field. 

%\sh{I commented a paragraph about general motivation/feature of RS. I believe there are enough discussion along this line in introduction. Also, given that our paper is \emph{specifically} about pheno of already published warped seesaw paper, we may not even need to give one. I open to other option, however.}

\noindent In this section, we provide a brief review of seesaw model in 5D warped extra-dimensions. After discussing general features of warped seesaw, we will focus on a model with the extended bulk gauge symmetry: $SU(2)_{\rm L} \times SU(2)_{\rm R} \times U(1)_{\rm X}$. 
Our 
%
%phenomenological 
%
studies of LHC signals 
are performed using the simplified two-site model of the full 5D warped model. Hence, our discussion about the full 5D model in this
section will be brief, leaving details necessary for the phenomenology to Sec.~\ref{sec:twosite} of the two-site model. 
More details about the 5D results, along
%
%derivations and discussions, 
%
%together 
%
with their 4D CFT dual description, can be found in \cite{Agashe:2015izu}.

%How to explain neutrino masses in this scenario? The first attempt will be adding right handed neutrinos, and couple them to left-handed neutrinos via Yukawa coupling together with Higgs. Just like other charged leptons, SM neutrinos will acquire Dirac mass once Higgs gets a VEV. However, in order to generate super tiny neutrino mass, one needs fine tuning neutrino 5D mass and it is also hard to achieve correct mixing angle among three generations of neutrinos. Another approach is demanding SM neutrinos are Majorana, whose mass are generated via Type I seesaw mechanism in RS model[cite]. This model demands right-handed neutrinos in the bulk, couple to SM left-handed neutrinos via Yukawa coupling and also acquire large Majorana mass on the UV brane. The 5D Lagrangian contains

%\noindent \sh{I also commented discussion about Dirac vs Majorana neutrino mass etc. I found we have lots of discussion about it in the introduction. I am okay with having such discussion, but not too much repetition.}

\noindent We begin our discussion by taking usual Randall-Sundrum framework with all SM fermions and gauge bosons propagating the bulk of a slice of $AdS_5$. For concreteness, we consider SM Higgs to be localized on the IR brane. %As indicated above, the bulk gauge symmetry is extended to $SU(2)_{\rm L} \times SU(2)_{\rm R} \times U(1)_{\rm X}$ and the Higgs is bi-doublet under this gauge group. In this way, the electroweak sector has a built-in custodial symmetry and this provides a protection mechanism for large corrections to T-parameter of the electroweak precision measurements. 
The 5D SM gauge singlet field, $N$, which is the analog of the 
%
%corresponding to 
%
the right-handed neutrinos of the usual, 4D seesaw models, propagates the bulk.
%
%\ka{I added the following.}
%
Like all 5D fermion fields, $N$ can be decomposed into {\em both} left ($L$( and right ($R$) chiralities (denoted
by $N_{ L, R }$, respectively) from the
4D viewpoint.
$N_R$ couples to SM $SU(2)_{\rm L}$ lepton doublet, in particular left-handed neutrinos, and the Higgs on the IR brane with 5D Yukawa coupling $y_{\rm 5D}$. In addition, $N_R$ acquires large Majorana mass, which is taken to be localized on the UV brane. %This localization of Majorana mass on the UV brane is well-explained with the extended gauge group. Namely, by taking $U(1)_{\rm X} \to U(1)_{\rm B-L}$,  
These can be summarized in the following 5D Lagrangian 
\bea
\mathcal L_{\rm 5D} \ni y_{\rm 5D} L H N + c_N k \bar{N} N + \delta (z-z_h)  \frac{1}{2}\frac{m_N}{k} N_R N_R,
\label{eq:warped_seesaw_5D_Lagrangian}
\eea
%
%
%\bea
%\mathcal L_{\rm 5D} \ni y_{\rm 5D} L H N + c_N k \bar{N} N + \frac{\delta (z-z_h)}{z_h/z}  \frac{1}{2}\frac{m_N}{k} N_R N_R,
%\label{eq:warped_seesaw_5D_Lagrangian}
%\eea
%
%
where since $N$ is 5D fermion field, it is four component spinor, containing $N_L$ and $N_R$ 4D Weyl spinors. $c_N k $ is 5D mass parameter for $N$ (in units of the AdS curvature scale, $k$) and $m_N$ is Majorana mass of $N_R$. UV(IR) brane is at $z=z_h (z_v)$.

%\noindent \sh{(i) I commented two sentences saying the above model is \textit{natural warped seesaw} model. I am okay to put it back with some changes. I just thought we have detailed discussion in the introduction and with just lagrangian, it is not explaining why it is so. (ii) do we need $z_h/z$-factor in the denominator ? With $\delta$-function, it is 1 anyway ?}
%This model is naturally in the sense that all parameters are $O(1)$ in this natural size, thus no tuning, and correct neutrino mass and large mixing among neutrinos can be achieved [cite].  This is what we call \textit{natural warped seesaw}.

\noindent The above model was studied in \cite{Huber:2003sf} using so-called KK-basis where KK decomposition was done without taking into account the large Majorana mass term from the beginning. The effects of the Majorana mass was added as a posteriori process and this leads to large Majorana masses for zero- and KK-modes and large mixing among all modes. Hence, although analysis using KK-basis produces correct neutrino mass formula, using a basis that is vastly different from the \emph{mass} basis obscures the physical
%
%dynamical 
%
picture. In particular, the results from KK-basis 
{\em naively} suggest (or give the misleading impression)
%
%\ka{I avoided using the word ``wrong" here.}
%
%created a wrong impression 
%
that the above 5D warped 
%
%Type I 
%
seesaw model is indeed of Type I 
%
%seesaw 
%
in the sense that the SM neutrino mass is generated by the \emph{dynamical} exchange of a super-heavy singlet mode, i.e.,
at the (effective) seesaw scale (for more discussion of this point, see \cite{Agashe:2015izu}). 

%In order to show SM couplings, one needs find the zero mode and its profile after KK decomposition. Previously, people use normal KK decomposition using pure Dirichlet or Neumann boundary conditions on UV and IR branes, ignoring UV brane Majorana mass for the moment. After getting profiles for all KK modes, Majorana mass contribution was added to get final spectrum.  It is clear that the would-be zero mode will obtain large Majorana mass, which corresponds to 4D seesaw scale. All KK modes will not contribute to SM neutrino mass due to equation of motion. So the remaining physics will be just like 4D Type I seesaw, with $O(1)$ Yukawa coupling and $10^{12}$ GeV Majorana mass. That's why people believe that such 5D Type I seesaw will not have any new particle within LHC reach.

%\ka{I tried to avoid too much repetition of ``dynamical" here!}

\noindent However, as shown in \cite{Agashe:2015izu}, analysis based on the mass basis, including the Majorana mass term from the beginning, reveals very {\em different} dynamical picture. The mass eigenstates of 4D effective theory (after KK-decomposition) of Eq.~(\ref{eq:warped_seesaw_5D_Lagrangian}) is a tower of pseudo-Dirac singlet fermions with tiny Majorana splitting. For the choice of $c_N \sim -0.3$ that renders correct SM neutrino mass, dominant contributions to the SM neutrino masses come from the
%
%\emph{dynamical} 
%
exchange of a few low lying mass eigenstates (cf.~super-heavy modes in the KK basis). Namely, the SM neutrino mass is generated not by an exchange of super-heavy Majorana singlet mode, but by exchanges of $O({\rm TeV})$ pseudo-Dirac singlet modes. Therefore, the dynamical nature of the warped seesaw is \emph{inverse seesaw} \cite{inverse}, not Type I. Moreover, it is indeed very \emph{natural} realization of it, because the SM neutrino mass is obtained with all dimensionful parameters taken to be near the cut-off scale and all dimensionless parameters to be $O(1)$. 
This new finding, then, 
{\em re}-focuses attention on LHC signals from the 
$O({\rm TeV})$ scale singlet pseudo-Dirac fermions that arise in this model.
%
%\ka{I mean these states ``existed/were accessible" always!}
%
%opens otherwise impossible, very exciting chances that such $O({\rm TeV})$ scale singlet pseudo-Dirac fermions could be discovered at LHC. 
%
Since the production and decay channels depend on details of the model, now we describe a concrete model based on the extended bulk gauge symmetry, whose simplified two-site version (presented in next section) will be used for our collider studies in Sec.~\ref{sec:analysis_results}. \\

%However, as pointed out in [cite], previous KK decomposition is good to get correct SM neutrino mass but obscures dynamics of right-handed neutrinos. The simply reason is that such basis is not (actually far away from )mass eigen-basis. After correctly solving the spectrum using correct boundary condition on the branes, namely taking Majorana mass into account at the beginning, the mass eigenstate of $N$ are $O(\rm TeV)$ pseudo-Dirac pairs with tiny Majorana splitting.  For the parameter space which could address correct neutrino mass as well as other phenomenology, SM neutrino masses are dominantly coming from effects of a few low lying KK modes of $N$.  Remarkably, this turns out to have Inverse seesaw feature, where SM left-handed neutrinos have Yukawa coupling with right-handed neutrino, which obtain Dirac mass with its left-handed partner, and left-handed partner having tiny Majorana mass.  This interpretation of the same model, reveals its true dynamical particles are $O(\rm TeV)$ particles, which could give new interesting signals at LHC.

\noindent \textbf{Natural realization with custodial symmetry
%
%A model with left-right symmetry
%
}

%The above discussion is rather general, which does no depend on the details of RS model and the gauge structure of  $N$ . Now we will give an explicit model with is testable at LHC.
 
\noindent In order to have sizable signal production of the new particles in the 5D model (i.e., the KK excitations of
SM) at the LHC, a KK scale of the order $O(1)$ TeV is desirable; of course naturalness of the EW scale
also prefers such a low scale. 
%
%\ka{I added last phrase in preceding sentence!}
%
However, minimal RS model with only the SM gauge symmetry in the bulk is in tension with EW precision tests, both oblique and non-oblique (from $Z \to b \bar b$ coupling) corrections, and consistency requires KK scale of $\gtrsim O(10)$ TeV. This bound can be relaxed by extending the bulk EW\footnote{Since 
QCD gauge group int he bulk will not play any role in our study, we will simply drop it from hereon.} gauge group to $SU(2)_{\rm L} \times SU(2)_{\rm R} \times U(1)_{\rm X}$. In particular, the extended gauge group provides custodial symmetry for both T-parameter and $Z \to b \bar b$ coupling, and KK scale as low as $O(1)$ TeV is allowed \cite{Agashe:2003zs,Agashe:2006at}.\footnote{In fact, 
even with the extended bulk gauge group, KK scale is constrained {\em generically} to be $\gtrsim O(3)$ TeV. Thus,
special regions of parameter space and/or additional contributions to these observables (perhaps
from further model building) will be needed in order to have KK scale as low as $O(1)$ TeV. Given that resonances with mass $O(3)$ TeV or heavier is slightly beyond the LHC reach, having new colliders with higher energy reach are required and hence motivated for a better test.}
There are also constraints from flavor/CP tests which generically require $\gtrsim O(10)$ KK scale, but
here we assume addition flavor structure (for example, flavor symmetries) in order to ameliorate those bounds \cite{Csaki:2008zd} .

%Fortunately, RS model with extended gauge symmetry, $SU(2)_L \times SU(2)_R \times U(1)_X$ , in the bulk and on the IR brane can pass  EW precision tests, as well as bring KK scale to $O(1)$ TeV, with the help of custodial symmetry. In this model, gauge symmetry breaks down to  $SU(2)_L \times  U(1)_Y$ by boundary condition on the UV brane. On the IR brane, $SU(2)_L\times SU(2)_R$ breaks down to $SU(2)_V$ by Higgs VEV.  By choosing the appropriate representations of SM fields, this model can also be safe from $Z \to b \bar b$[cite]. One canonical choice of $U(1)_X$ could be $U(1)_{B-L}$, which ensures lepton number is conserved in the bulk and on the IR brane, where $U(1)_X$ is unbroken. This forbids Majorana mass term of $N$, which is obviously lepton number violating,  in the bulk and on the IR brane. Remarkably, RS model with extended gauge symmetry, turns out to be a more natural model to embed Type I seesaw with Majorana mass on the UV brane.

%\ka{I have done some rearrangements in what follows: the original versions are still in old tex file}

\noindent On the UV brane, the gauge symmetry is broken down from 
$SU(2)_{\rm R} \times U(1)_{\rm X}$ to  $U(1)_{\rm Y}$ by choice of boundary conditions (BC). 
Specifically, 
the gauge fields associated with the broken generators $\left( SU(2)_{\rm R} \times U(1)_{\rm X} \right) / U(1)_{\rm Y}$ will have 
Dirichlet BC, denoted henceforth as ``$-$'', whereas $U(1)_Y$ 
and $SU(2)_L$ has Neumann ($+$).
All gauge fields are taken to be $+$ on IR brane.
In particular, only fields with $(++)$ BC have zero-modes up on KK-decomposition, i.e., only gauge fields for SM gauge group in this case.
We use $W_R$ and $Z'$ to denote the extra gauge fields, i.e., for charged $SU(2)_{\rm R}$ and $\left( U(1)_{\rm R} \times U(1)_{\rm X} \right) / U(1)_{\rm Y}$, respectively:
these have $(-+)$ BC and hence only have massive/KK
%
%have no-zero 
%
modes.

%Gauge fields in this model are SM gauge fields and its composite modes, together with $W_R$ $Z'$. $W_R$ is the gauge boson from $SU(2)_R$. Linear combinations of $T^3_R$, the diagonal generator in $SU(2)_R$ and $U(1)_X$ give $U(1)_Y$ and $U(1)_Z'$, whose gauge boson is $Z'$. Since both $W_R$ and $Z'$ is broken on the UV brane and unbroken on the IR, the boundary condition for these two fields are $(-\,+)$. Other gauge fields have (+\,+) boundary condition. Interestingly, after EWSB, there exists \textit{maximal mixing} among $W^{(1)}_L$ and  $W^{(1)}_R$[cite], which provide a way to produce $W^{(1)}_R$ from light quarks. We will make this point more clear in Sec.~\ref{sec:gaugemixing}.

\noindent Higgs field, which we choose to be localized on the IR brane, is a bi-doublet of $SU(2)_{\rm L}\times SU(2)_{\rm R}$, with zero charge under $U(1)_X$:
%
%Higgs field is chosen to be localized on the IR brane in this model. In order to achieve custodial symmetry protection, Higgs needs to be bi-doublet under $SU(2)_L\times SU(2)_R$. So it has representation
%
\bea
H \in (2,2)_0.
\eea 
This representation 
results in a custodial symmetry, i.e., the Higgs VEV breaks $S(2)_L \times 
SU(2)_R$ down to $SU(2)_V$, which suppresses contributions from the gauge sector to the $T$-parameter: note that $U(1)_X$ remains unbroken in this process.
The Higgs VEV will also generate mixing between various modes of $W_R$ and $W_L$, an effect which can be treated perturbatively and which 
will be very important for LHC signals for the singlet neutrinos. We will make this point clearer in Sec.~\ref{sec:twosite}.

\noindent Moving onto representation of fermions under the extended gauge group, first note that (just like for gauge fields)
SM fermions will arise as zero-modes of 5D fields with $(++)$ BC.
We begin with the leptons,
where we choose the simplest possibility, i.e., $X$ is same as $(B - L)$
in this sector. Thus, we take $L$, i.e., the 
SM $SU(2)_{\rm L}$ lepton doublet, to be a singlet of $SU(2)_R$,
while the right-handed charged lepton (denoted by $l$) is promoted to be a doublet of $SU(2)_R$, denoted by $L_R$ [as in the canonical, 4D (gauged) left-right (LR) symmetric models]:
\bea
L\in (2, 1)_{-\frac{1}{2}}~~~~~ L_R,\tilde L_R\in (1, 2)_{-\frac{1}{2}}.
\eea
where numbers in the parenthesis denote representation under $SU(2)_{\rm L}$ and $SU(2)_{\rm R}$, while representation of $U(1)_X$ is shown as a subscript [we will explain momentarily why there are {\em two} $SU(2)_R$ doublets].
Remarkably, akin to usual, 4D LR symmetric models, we see that the 
%\pd{Question$\to$}
$SU(2)_R$ partner of $\ell$ has the precisely the characteristics to play the role of the $N$ field mentioned above, i.e., it is a (i) singlet under SM gauge group;
(ii) it has a Yukawa coupling with lepton doubleton the IR brane and (iii)
a Majorana mass term for it can be written only only the UV brane, since that is the only location where $SU(2)_R \times U(1)_X$ (under which it is charged) is broken.
As a by-product, such a choice gives rise to a way to produce $N$ via decay of $W_R$. In fact, this will be the production channel for our signal process.

\noindent In more detail, note that we will actually need {\em two} $SU(2)_{\rm R}$ lepton doublets, namely:
%
% as follows:
%
%
\bea
\tilde L_R=\left ( \begin{array}{ll}
N (+ + ) \rightarrow (-+) \\
~\\
\tilde{\ell} (-+)
\end{array}
\right)_R ~~~~~~~~~~~L_R=\left ( \begin{array}{ll}
\tilde N (-+) \\
~\\
\ell (++)
\end{array}
\right)_R
\label{lepton_content}
\eea 
Here the SM lepton ($\ell$) is obtained as the zero-mode from the 2nd multiplet above, i.e., with $(++)$ BC;
its $SU(2)_R$ partner (denoted by $\tilde{N}$) is chosen to be $-$ on the UV brane (thus having no zero-mode at all):
this is 
consistent with the bulk gauge symmetry since $SU(2)_{\rm R} \times U(1)_{\rm X}$ is broken on UV brane to $U(1)_{\rm Y}$ (i.e. different BC's for
two components of doublet are allowed), while this symmetry is unbroken on IR brane (i.e. it should be same BC for both fields, which is $+$ in this case).
Note that $\tilde{N}$ then plays no role in the seesaw for the 
SM neutrino mass (hence will be dropped from now on). 
On the other hand, the BC's are ``switched'' in the 1st doublet, i.e., $\tilde{\ell}$ has no zero-mode, whereas
the $N$ here will be driving the SM neutrino mass seesaw
(thus will be denoted as {\em the} singlet neutrino henceforth). 
Note that $N$ has $(++)$ BC to ``begin with'', but adding a UV brane localized Majorana mass term ``repels'' $N$ profile away from UV brane, resulting in effective boundary condition of the form $(-\,+)$ and hence removing the corresponding 
zero-mode.

The simple reason for having two $SU(2)_{\rm R}$ doublets, instead of housing $N$ and SM right-handed lepton in a single $SU(2)_{\rm R}$ doublet, is the following.
The $N$ and SM right-handed lepton, i.e., $\ell$, fields should have different 5D bulk mass parameters in order to produce correct masses for charged lepton and neutrino \cite{Huber:2003sf}, i.e.,
we require $c < -0.5$ for the field giving charged lepton zero-mode so that 
this mode is localized near the UV brane \footnote{Such a profile also needs to be chosen for the $L$ zero-mode.}, thus giving the observed charged lepton mass, whereas we need $c \sim -0.3$ for $N$ (as mentioned above), i.e.,
that would-be zero-mode should be peaked near the IR brane instead.  
However, 
%
%On the other hand, 
%
by $SU(2)_{\rm R}$-invariance, fields in a doublet should have a common 5D mass parameter.
Thus, we need to ``split'' the SM charged lepton and singlet neutrino
multiplets as shown above. 
%\sh{Similarly, I commented the first reason and kept only the second one. We can put it back if the first reasoning is correct. I need to think.}

\noindent Following \cite{Agashe:2006at}, i.e., in order to suppress corrections to the $Z b \bar{b}$ coupling, we choose the representations of 
the quarks to be somewhat non-minimal
%
%all other SM fermion fields 
%
as follows. %Following \cite{•}, we choose left-handed quarks doublets to be in bi-doublet representation under $SU(2)_L \times SU(2)_R$. Right-handed up (down) quarks are in singlet (triplet) representation of $SU(2)_R$. The representation for quarks and leptons are:
\bea
Q_L\in (2 , 2)_{\frac{2}{3}} ~~~~~u_R \in (1,1)_{\frac{2}{3}} ~~~~d_R\in(1,3)_{\frac{2}{3}} 
\eea
Here, $Q_L$ denotes the
%
%bi-doublets under $SU(2)_{\rm L} \times SU(2)_{\rm R}$ which contains 
%
SM left-handed quarks doublet and $u_R$, $d_R$ are the $SU(2)_L$ singlets. 
For the ``extra"'
%
%other BSM particles 
%
fields 
in $SU(2)_{\rm R}$ doublet or triplet representations above, we take Dirichlet-Neumann $(-\,+)$ boundary condition in order to remove zero-mode
(just like was done for leptons above).
As usual, $t_R$ zero-mode is taken to be localized near the IR brane, while
$(t,b)_L$, i.e., $Q^3_L$, has a (roughly) flat profile and rest of the quarks are peaked near the UV brane (just like the SM leptons).

%The reasons for adding two doublets are following: (1) $SU(2)_R $ is broken on the UV brane, so we can not have both Neumann boundary condition for two partners in the same $SU(2)_R$ doublet. A practical model needs both $N$ and $\ell$  having zero modes, which can not come from the same $SU(2)_R$ doublet. (2 )$N$ and $\ell$ should have different 5D bulk masses in order to achieve correct SM spectrum[cite]. This means we can not put them in the same doublet, which has the same 5D bulk mass.  

%As pointed out in [warped seesaw], although $N$ have $(+\,+)$ boundary condition, they will become pseudo-Dirac in mass eigenstates in this warped seesaw model. Large Majorana mass term on the UV brane effectively repels $N$ profile away from UV brane, as if it has  $(-\,+)$ boundary condition. Since the Majorana splitting ($O(\rm MeV)$) is so tiny comparing to its Dirac mass ($O(\rm TeV)$), for the signals studied in this paper, we treat $N$ as \textit{pure Dirac} with $(-\,+)$ boundary condition, which will have the same mass as its $SU(2)_R$ partner $\tilde \ell$.

\noindent We mentioned that the spectrum of $N$ in 4D effective theory is a tower of pseudo-Dirac fermions. Since the Majorana splitting ($O(\rm MeV)$) for these pseudo-Dirac pairs is very tiny comparing to its Dirac mass ($O(\rm TeV)$), we are unlikely to be able to probe
any effects from such Majorana splitting. Moreover, as far as investigating the discovery potential of the lightest pseudo-Dirac singlet mode is concerned, the existence of tiny Majorana splitting will not make any difference. For this reason and for simplicity, in our collider study, we ignore Majorana splitting and treat $N$ as \textit{pure Dirac} with $(-\,+)$ boundary condition, which will have the same mass as its $SU(2)_{\rm R}$ partner $\tilde \ell$.

\begin{figure}
\center
\includegraphics[width=0.8\linewidth]{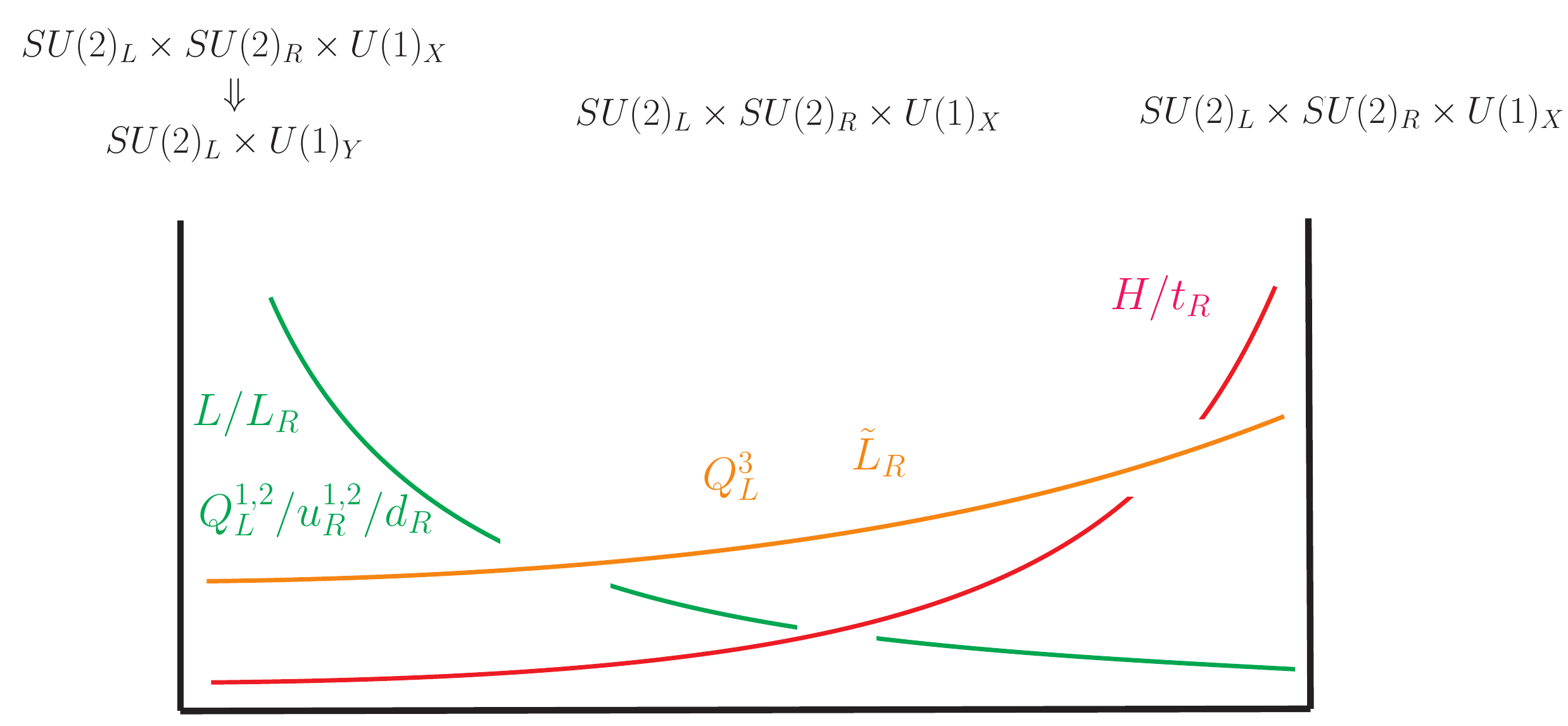}
\caption {RS model with extended bulk gauge symmetry $SU(2)_{\rm L} \times SU(2)_{\rm R} \times U(1)_{\rm X}$ and singlet neutrino. Gauge symmetries together with its breaking pattern are shown on the relevant position along the extra-dimension. The position of the fields shows where the zero-mode profile of the corresponding 5D fields is localized. %the more close to UV(IR) brane , the more this field is peaked at UV(IR) brane. 
For fields with (close to) flat zero-mode profile, they are located in the middle of the bulk. }
%\sh{Some improvement is needed for this plot.}}
\label{fig:5Dsetup}
\end{figure}

The 5D fields discussed thus far are summarized in Fig.~\ref{fig:5Dsetup}. The position of the fields shows where the zero-mode profile of the corresponding 5D fields is localized. Fields that are closer to the UV (IR) brane signifies that their zero-mode profiles are peaked near the UV (IR) brane. For fields with (close to) flat zero-mode profile, they are located in the middle of the bulk. \\
%
%

%\ka{The following paragraph is not really needed: it's more for the sake of completeness. So, you can remove it if you wish.}

\noindent\textbf{Couplings of KK modes}
The couplings among various 4D particles are 
%
%given by 
%
proportional to the overlap of their respective profiles in the extra dimension.
Now the light
quarks are localized near the UV brane, while the KK modes are near the IR brane.
However,  the non-zero (Neumann BC)
profile of KK of {\em SM} gauge bosons at the UV brane does induce a significant 
%
%sizable 
%
coupling 
to the light quarks.
On the other hand, KK $W^{ \pm }_R$ and $Z^{ \prime }$ vanish at the UV brane (Dirichlet BC), rendering such a coupling to be
negligible.
Nonetheless, as we discuss in Sec.~\ref{sec:twosite}, 
EWSB mixing among KK $W_L$ and $W_R$ does induces a sizable coupling of
KK $W^{ \pm }_R$ to light quarks provided there is degeneracy between KK $W^{ \pm }_R$ and KK $W^{ \pm }_L$
,similarly KK $Z$ and KK $Z^{ \prime }$.
This coupling can then be 
used in production of KK $W^{ \pm }_R$ and $Z^{ \prime }$.
Once produced, their 
decay is dominantly to modes localized near IR brane such as (light) KK fermions and/or top quark/Higgs boson,
since those couplings are the largest. \\

\noindent\textbf{Spectrum of KK modes}

\noindent Mass of KK gauge boson is dictated by boundary condition of corresponding 5D gauge field. The mass of first KK mode of gauge fields with $(+\,+)$ boundary condition is typically $O(1) \times $ warped-down $k$ 
%
%KK scale 
%
and we denote it as $m_{\rm gauge}$. On the other hand, first KK mode of gauge fields with $(-\,+)$ boundary condition has slighter smaller mass than $m_{\rm gauge}$.

%Mass of KK gauge boson is dictated by boundary condition. The mass of first KK mode of  gauge fields with $(+\,+)$ boundary condition is typically $O(1)$ KK scale, call it $m_{\rm gauge}$. First KK mode of gauge fields with $(-\,+)$, is slighter smaller than $m_{\rm gauge}$.

%
\noindent %\sh{I added superscript $(1)$ to specify it is first KK and to make clear distinction between 5D field and 4D KK-mode.} 
KK fermion masses are determined by boundary condition and 5D mass $m_5$, or $c=\frac{m_5}{k}$. For fermion fields with
$c$ chosen such that the corresponding (would-be) zero-mode is localized near the UV brane, we find that 
%
%$c <  -0.5$, 
%
the KK mass is larger than KK gauge mass $m_{\rm gauge}$, regardless of its boundary condition
[assuming brane localized kinetic terms (BKT) are negligible].
%
%very small]. 
%
This is the case for all {\em leptonic}
%
%fermion 
%
fields, except for $\tilde L_R$. In order to produce SM neutrino mass, $\tilde L_R$ has $c \sim -0.3$ and resulting KK mass is naturally smaller than $m_{\rm gauge}$. However, in the minimal setup, its mass is still bigger than $\frac{1}{2}m_{\rm gauge}$, preventing the decay of $W_R^{(1)}$ into $N^{(1)}$ and $\tilde{\ell}^{(1)}$. As is well-known, turning on BKT's could lower the mass of corresponding KK modes. 
%
%In Appendix.~\ref{sec:Appendix}, 
%
We can show $O(1)$ BKT on the IR brane for $\tilde L_R$ can result in mass of $N^{(1)}$ and $\tilde{\ell}^{(1)}$ smaller than $\frac{1}{2}m_{\rm gauge}$. % allowing on-shell decay of $W_R^{(1)}$ into $N^{(1)}$ and $\tilde{\ell}^{(1)}$. 
%
%\sh{Did someone write down Appendix ?} 
%\ka{In the light of lack of time, I suggest that we postpone appendix A to v2!}
%
Another interesting fact about BKT is that, both $N_R^{(1)}$ and $N_L^{(1)}$ can have 
%
%$O(1)$ 
%
similar 
coupling to gauge field $W_R^{(1)}$. In the absence of BKT, the coupling of $N_L^{(1)}$ to $W_R^{(1)}$ is mildly suppressed, i.e.,
by $O(1)$,  as compared to $N_R^{(1)}$, since $W_R$ is peaked near IR brane where $N_L^{(1)}$ has vanishing boundary condition. 
%
%\sh{Is the previous claim true ? I understand that $N_L^{(1)}$ has vanishing boundary condition on the IR brane simply because of $\partial_5$ term, enforcing opposite B.C. between RH and LH. Still, I think most of its normalization comes from region near the IR brane, and hence still large overlap ? Or, what am I missing ? Vanishing B.C. on IR brane can lead to significant drop in the overlap with other KK modes, e.g. $W_R^{(1)}$ ?} 
%
%\ka{I modified as above}
%
Combination of these two features opens new decay channels for KK $W_R^{(1)}$. Namely, the decay of $W_R^{(1)}$ into pair of $N_R^{(1)}$ and $\tilde{\ell}_R^{(1)}$ together with $N_L^{(1)}$ and $\tilde{\ell}_L^{(1)}$, which are our signal channels. 

%
%\ka{More points for the sake of completeness.}
%
A similar analysis can be applied to the quark sector: we find that, in the absence of BKT's, the only KK fermions which are a bit lighter than
the $W^{(1)}_R$ (but still heavier than $1/2 \; m_{ \rm gauge }$) are the $SU(2)_R$ partners of $Q^3_L$ (like the case of $\tilde{L}_R$ above).
We assume that BKT's for these states are {\em not} turned on ({\em un}like for $\tilde{L}_R$) so that
KK $W_R$ can{\em not} decay into pairs of these extra fermions. 
The decay channel for $W^{(1)}_R$ into SM $Q^3_L$ and the above extra fermions is kinematically open; however,
given the (roughly) flat profile of $Q^3_L$, this coupling is nonetheless suppressed compared to the coupling to
$N^{(1)}$ and $\tilde{\ell}^{ (1) }$ so that this decay mode can be neglected.

%KK fermion masses are determined by boundary condition and 5D mass $m_5$, or $c=\frac{m_5}{k}$. For fermions with $c <  -0.5 $, no mater what the boundary condition is, its mass is heavier than $m_{\rm gauge}$. This is the case for all beyond SM KK particles expect $\tilde L_R$. $\tilde L_R$ has $c \sim -0.3$ , which is naturally smaller than $m_{\rm gauge}$. However, its mass is still bigger than $\frac{1}{2}m_{\rm gauge}$. As is well-known, brane localized kinetic (BKT) term could lower the mass of corresponding KK modes. In Appendix. \ref{App:BKT}, we show $O(1)$ BKT on the IR brane for $\tilde L_R$ could result in mass of $N$ and $\tilde \ell$ being smaller than $\frac{1}{2}m_{\rm gauge}$. Another interesting fact with BKT is that, $N_R$ and $N_L$ could  both have $O(1)$ coupling to gauge field $W_R$. On the contrary, with absent of BKT, $N_L$ has suppressed coupling to $W_R$ due to $W_R$ being peaked at IR brane where $N_L$ has vanishing boundary condition.   Combination of these two features open new decay channels of KK $W_R$  to pair of $N_R$ and $\tilde \ell_R$ together with $N_L$ and $\tilde \ell_L$, which are our signal channels.

\noindent In this paper, we study the on-shell production of KK gauge bosons $W_R^{(1)}$ and its decays to $N^{(1)}$-$\tilde{\ell}^{(1)}$ pair. Particles heavier than $W_R^{(1)}$ are dropped for simplicity of study. Below, we summarize the spectrum of particles of interest: 
\bea
m_{\rm gauge} > 2 m_{\tilde L_R} \gg \textrm {mass of SM particles}
\eea
where KK gauge bosons included in our phenomenological study are $W^{(1)}_L$, $W^{(1)}_R$, and $Z^{(1)}$, $Z'^{(1)}$.

\section{Two site approach to natural warped seesaw}
\label{sec:twosite}

Full 5D warped model contains all the degrees of freedom with perturbative couplings. In this sense, it is fully calculable 5D effective theory and any relevant questions can be answered by explicit computation. However, for a specific phenomenological search, only a finite subset of degrees of freedom and related couplings are involved and a simplified model consisted of only relevant particles and couplings will be much more efficient in practice. Two site model of \cite{Contino:2006nn} provides one way to obtain a simplified 4D effective theory by a consistent truncation of a full 5D warped model to the first KK modes. This approach not only simplifies phenomenological studies, but also can encompass phenomenology of broader class of 5D warped models, or its 4D composite models, thereby allowing more inclusive/systematic searches.  

\noindent Two site model, as the name suggests, consists of two sectors/sites: the elementary sector and the composite sector. 
%
%\sh{Previously, there was phrase like "Motivated by AdS/CFT duality"... Is it ? Or it is more like along the line of deconstruction of extradimensions. I am not sure. I remember we had a discussion about it. Since I am not sure, I just simply state that there is composite and elementary sector in two site model, without mentioning exactly where it comes from.}%
%
%\ka{I'm Ok with this, as long as we cite original 2-site paper, which w do already!}
%
The composite sector represents strong dynamics which confines at $O({\rm TeV})$ scale, the scale where the scale invariance is spontaneously broken and composite resonances are ``born''. In principle, there will be towers of infinite resonances. However, as a phenomenological simplified model, only the lightest resonances, the relevant particles for the collider searches, are kept. Elementary sector, on the other hand, exhibits physics external to strong dynamics, but with couplings to the composite sector. These couplings induce mixing between elementary and composite states and upon diagonalization, this leads to massive mass eigen-modes, dual to first KK modes in 5D, and massless modes, dual to zero mode, i.e. SM fields. In this way, it is easily seen that both SM fields and the first KK modes of 5D model are generically the admixture of elementary and composite states, the amount of compositeness being determined by the size of the mixing at the $O({\rm TeV})$ scale. Such a feature is known as \emph{Partial Compositeness} in 4D strong dynamics, a robust mechanism that solves flavour hierarchy problem of the SM. 5D dual of partial compositeness is the localization of the zero-mode profile along the extra-dimension, localization near the IR (UV) brane being dual to more composite (elementary).   

%5D model is good to visualize the coupling via overlap of profiles, but is not simple to work in details.  Instead, as pointed out in [two site], a simplified effective 4D theory, namely two site model, can represent a class of 5D RS and 4D composite models. Motivated by AdS/CFT duality, two site mode has one elementary sector and one composite sector.
%, dual to physics at UV brane and one composite sector, dual to physics in the bulk with IR brane.  
%Composite sector represents strong dynamics with spontaneous symmetry breaking. In principle there will be infinite tower of resonances in such strong dynamics. However, composite sector in two site model only contains the lightest resonance for simplicity. Elementary sector shows physics external to strong dynamics with specific coupling to the composite sector. After diagonalization, the massive modes are dual to first KK modes in 5D, whereas massless mode are dual to zero mode, which are SM fields. It is clear that SM field generally will be admixture of parts from elementary and composite sectors.  Such feature is known as Partial Compositeness in 4D strong dynamics, which could reveal properties of profiles of 5D fields. 

%Natural warped seesaw can be understood in two site language as follows. $N_R$ has large Majorana mass term in elementary sector, while composite $\chi _L $ and $\chi_R$ have $O(TeV)$ Dirac mass. Also $N_R$  has  $O(TeV)$ mass mixing with $\chi _L$. 

\noindent Two site model of the natural warped seesaw that we reviewed in Sec.~\ref{sec:5Dmodel} can be described as follows. We begin by discussing the singlet neutrino $N_R$. In the elementary sector, there is elementary field $N_R$ that has large Majorana mass term $m_N$. In the composite sector, as already mentioned in the introduction, there is a composite singlet Dirac fermion $(\chi_L, \chi_R)$ with $O({\rm TeV})$ Dirac mass. Finally, there is mass mixing between $N_R$ and $\chi_L$, i.e. they have the same quantum number, with the size of the mixing being characterized by the relevant scale, i.e. of the order of $O({\rm TeV})$. These can be summarized by the following Lagrangian (dropping kinetic terms for simplicity):
\bea
\mathcal L_{\rm seesaw}&=&\mathcal L_{\rm elementary}+\mathcal L_{\rm composite}+\mathcal L_{\rm mixing} \nonumber \\
&=& \frac{m_N}{2} N_R N_R+ (m_D\bar{\chi }_L\chi_R+\Delta \bar{\chi}_L N_R + {\rm h.c.})
\eea
where $m_N$ ($m_D$) is Majorana (Dirac) mass for elementary (composite) states and $\Delta$ is the mass mixing between the two. Both $m_D$ and $\Delta$ are $O$(TeV), while $m_N\gg {\rm TeV}$. Largeness of the Majorana mass $m_N$ allows us to integrate out $N_R$, i.e. use equation of motion for $N_R$, to get 
\bea
\mathcal L_{\rm seesaw}= (m_D \bar{\chi}_L \chi_R + {\rm h.c.})+\frac{m_D \Delta}{m_N} \chi_L \chi_L.
\eea
%
%
%\sh{Previously, it was $\frac{\Delta m_D}{m_N}$, but I flipped the order and rewrite it as $\frac{m_D \Delta}{m_N}$ to avoid the possible confusion viewing $\Delta m_D$ as some kind of mass difference/gap.}
%
%\pd{I modified below}
\noindent Notice that integrating out $N_R$ generates the Majorana mass for \emph{left-handed} $\chi_L$ of the composite singlet fermion, that is, it transmits lepton-number violation into the composite sector. Since $\frac{m_D \Delta}{m_N}\ll m_D$, it is clear that the composite fermion $(\chi_L , \chi_R)$ becomes pseudo-Dirac and the exchange of this pseudo-Dirac singlet fermion between the two left-handed SM neutrinos then is the dynamical origin of the SM neutrino mass.
%
% generation. 
%
Namely, it is the inverse seesaw for SM neutrino mass generation. Notice, however, that the way the small Majorana splitting %for the pseudo-Dirac fermion 
is generated is by the ``exchange'' of super-heavy $N_R$, which can be viewed as Type I seesaw. As mentioned in Sec.~\ref{sec:5Dmodel}, since the Majorana splitting is much smaller than Dirac mass, we simply drop it and treat $(\chi_L , \chi_R)$ as a pure Dirac fermion for our collider analysis presented in Sec.~\ref{sec:analysis_results}. For the rest of the study, we simply use ($N^{(1)}_L$, $N^{(1)}_R$) to denote  $(\chi_L , \chi_R)$ and put them and their $SU(2)_{\rm R}$ partner, denoted as ($\tilde \ell^{(1)}_L$, $\tilde \ell^{(1)}_R$), in the doublet $\tilde L_R$.

\noindent For the rest of the model, following \cite{Contino:2006nn}, we consider an elementary sector with elementary gauge group $[SU(2)_{\rm L} \times U(1)_{\rm Y}]^{\rm elem}$ and a composite sector with global symmetry $[SU(2)_{\rm L} \times SU(2)_{\rm R} \times U(1)_{\rm X}]^{\rm comp}$. Focusing on the gauge sector first, there will be mixing terms between elementary gauge bosons and corresponding composite vector mesons, i.e. composite vector bosons associate with $[SU(2)_{\rm L} \times U(1)_{\rm Y}]^{\rm comp}$ subgroup of the full global symmetry of the composite sector. These mixing terms between elementary and composite vector bosons break both elementary and composite symmetries. However, it does so in a way that only one linear combination of the elementary gauge boson and composite vector meson gets a mass, leaving the other orthogonal combination being still massless. Namely, there is unbroken gauge symmetry which we identify as the SM gauge group $[SU(2)_{\rm L} \times U(1)_{\rm Y}]^{\rm SM}$. These massless (massive) mass eigenstates are dual to zero-(KK-)mode SM gauge boson arising in the 5D model. In this way, we understand that there is mixing between elementary and composite vector bosons, allowing the coupling between elementary fermions and composite vector mesons. 

On the other hand, since there exist no associated elementary gauge bosons, the charged vector mesons for $SU(2)_{\rm R}$ ($W^{(1)}_R$) and the one for $\left( U(1)_{\rm R} \times U(1)_{\rm X} \right) / U(1)_{\rm Y}$ ($Z'^{(1)}$), i.e. orthogonal to $U(1)_{\rm Y}$, do not have mixing with elementary gauge bosons, i.e. purely composite. This feature is dual to the fact that the corresponding 5D gauge bosons have odd boundary condition on the UV brane and have no zero-mode. 
\noindent SM fermion fields are admixture of elementary and composite states (resulting from presence of mass terms
along the lines of what was discussed for singlet neutrino above).
%
%the amount of compositeness depending on the size of mixing. 
%
In this study, just for simplicity, we treat all SM fermions to be purely elementary, except $(b_L,t_L)$ and $t_R$. As we discussed in Sec.~\ref{sec:5Dmodel}, the 
%KK scale 
mass for KK modes of all SM fermions are higher than gauge KK;
%{I changed KK scale to mass of KK modes of SM fermion...}
however, the KK modes from the $\tilde{L}_R$ multiplet in
Eq.~(\ref{lepton_content}) are taken to be lighter. This is mapped into the two site model by the fact that all ``excited'' composite modes of the SM fermions\footnote{Note that this also applies to the composites with which the external RH charged lepton mixes, i.e., corresponding to the 
field $\ell$ from the $L_R$ multiplet in Eq.~(\ref{lepton_content}).} are heavier than composite vector mesons, thus for 
%
%further 
%
simplicity, we neglect them in what follows. 
However, the composite $SU(2)_R$ doublet containing the singlet neutrino (discussed above) is light.
Higgs is chosen to be pure composite state as a standard choice. %
The diagonalized Lagrangian before EWSB (see next section for this effect) containing all these degrees of freedom is given by
\bea
\mathcal L= \mathcal L_{\textrm{gauge}}+\mathcal L_{\textrm{fermion}}+\mathcal L_{\textrm{Higgs}}.
\eea
where we provide each part below one by one. First of all, $\mathcal L_{\textrm{gauge}}$ is given by
\bea
\mathcal L_{\textrm{gauge}}=-\frac{1}{4}F_{\mu\nu}^2+\frac{1}{2}\left(D_\mu\rho_{\nu}D_\nu\rho_{\mu}-D_\mu\rho_{\nu}D_\mu\rho_{\nu}\right)+\frac{m^2_{\star}}{2}\tilde{\rho}^2_{\mu}+\frac{m^2_{\star}}{2 \cos^2{\phi}}\rho^{2}_{\mu }+\frac{ig}{2}F_{\mu\nu}[\rho_{\mu},\rho_\nu],
\eea
%
%
%\sh{the super script on $\rho$ in the fourth term is typo ?}
where $\rho _\mu=(W^{(1)}_{L\mu}, B^{(1)}_\mu)$ (using the 5D notation, i.e., 
KK of SM gauge fields), $\tilde \rho _\mu=(W^{(1)}_{R \mu}, Z'^{(1)}_\mu)$ 
({\em non}-SM gauge bosons) and $A_\mu=(W^{(0)}_{L\mu}, B^{(0)}_\mu)$ (the SM gauge bosons), and we have dropped gauge indices to avoid notational clutter. %\sh{I removed gauge indices and wrote as above, but I am okay with putting them back.} 
$F_{\mu \nu} $ is the field strength of $A_\mu$. All covariant derivatives in the Lagrangians are with respect to the unbroken SM gauge group, namely $D\mu=\partial_\mu-igA_\mu$.
Gauge couplings are SM gauge couplings $g=(g_W,g_Y)$ and composite gauge couplings $g_\star=(g_{\star W},g_{\star Y}) $, and $\tilde g_\star=(g_{\star R}, g_{\star Z'})$, where $g_{\star Y}=\frac{g_{\star R} g_{\star X}}{\sqrt{g^{2}_{\star R}+g^{ 2}_{\star X}}}$. %
%
%where $\rho _\mu=(W^{(1)a}_{L\mu}, B^{(1)}_\mu)$ and $\tilde \rho _\mu=(W^{(1)a}_{R \mu}, Z'^{(1)}_\mu)$ and $A_\mu=(W^{(0)a}_{L\mu}, B^{(0)}_\mu)$. $F_{\mu \nu} $ is the field strength of $A_\mu$, $D_\mu=\partial_\mu+i g A_\mu$, and $g=(g_W,g_Y)$, $g_\star=(g_{\star W},g_{\star Y}) $ $\tilde g_\star=(g_{\star R} g_{\star Z'})$. Here $g_{\star Y}=\frac{g_{\star R} g_{\star X}}{\sqrt{g^{2}_{\star R}+g^{ 2}_{\star X}}}$
%
The elementary-composite mixing angle $\phi=(\phi_W,\phi_Y)$ is defined as $\sin \phi = \frac{g}{g_\star}$. 
Finally, $m_{ \star }$ denotes the composite spin-1 mass {\em before} mixing with elementary states, hence this is also the
mass for $\tilde{ \rho }$'s (i.e., $W^{ \pm }_R$ and $Z^{ \prime }$) who do not have such mixing. Whereas, for composite partners of {\em SM} gauge bosons, i.e., $\rho$'s, the mass is modified by this mixing
as indicated above.

Note that we are providing phenomenologically most relevant terms only, dropping terms with 3 or more  $\rho$'s or $\tilde \rho$'s. This is valid approximation since we are working to leading order and, at the leading order, only two body decays of the heavy particles, e.g. $\rho$'s or $\tilde \rho$'s, are relevant. 

%

%
%Next, 
%
Moving onto the fermion sector, $\mathcal L_{\textrm{fermion}}$ (for the fields relevant for our collider study) is given by
%
%
%Note here the terms with 3 and more than 3 $\rho$s or $\tilde \rho$s are omitted, since we are working at leading order and only two body decay of heavy particles, which is the dominant decay channels in composite models. And $\phi=(\phi_W,\phi_Y)$ is defined as $\sin \phi = \frac{g}{g_\star}$
%
%
%\bea
%\mathcal L_{\textrm{fermion}}&=&\bar \psi_{\rm SM} i \slashed{D} \psi_{\rm SM} + \bar {\tilde L} (i \slashed D - m_N ) \tilde L  \\
%&~&+g (\cos^2 \phi_{\rm SM} \cot \phi - \sin^2 \phi_{\rm SM}  \tan \phi )\bar \psi_{\rm SM} \rho^\star_\mu \gamma^\mu \psi_{\rm SM}\\
%&~& + \tilde g_\star \cos^2 \phi_{\rm SM} \bar \psi_{\rm SM} \tilde \rho_\mu \gamma ^{\mu} \psi_{\rm SM} \\
%&~&+ \tilde g_\star \bar{\tilde L} \tilde \rho_\mu \gamma^{\mu} L+  g_\star \bar{\tilde L} \rho_\mu \gamma^{\mu} \tilde L
%\eea
%where $\psi_{SM}$ denotes all SM fermions. All couplings here should multiply  by appropriate charges to get final coupling. 
%
%Here we assume all excited composite modes of SM fermions, even one's in the same multiplets of SM fermions, are heavier than first KK mode of gauge bosons, so we neglect couplings of those modes for our study. 
%
%Also, we assuming mass of $\tilde L$ doublet is smaller than half of KK gauge bosons, so that KK gauge bosons can decay to pair of these two heavy fermions. 
%
\bea\label{eq:Lfermion}
\mathcal L_{\textrm{fermion}}&=&\bar \psi_{\rm SM} i \slashed{D} \psi_{\rm SM} + \bar {\tilde L}_R (i \slashed D - m_D ) \tilde L_R  \nonumber \\
&~&-g  \tan \phi \; \bar \psi_{\rm light} \rho_\mu \gamma^\mu \psi_{\rm light} \nonumber \\
&~&+g (\cos^2 \phi_{Q^3_L} \cot \phi - \sin^2 \phi_{Q^3_L}  \tan \phi ) \; \bar Q^3_L \rho_\mu \gamma^\mu Q^3_L+g_{\star Y} \; \bar t_R B^{(1)}_\mu \gamma ^{\mu} t_R \\
&~& + \tilde g_\star \cos^2 \phi_{Q^3_L} \; (\bar b_L Z'_\mu \gamma ^{\mu} b_L +\bar t_L Z'_\mu \gamma ^{\mu} t_L) +g_{\star Z'} \; \bar t_R Z'_\mu \gamma ^{\mu} t_R  \nonumber \\
&~&+ \tilde g_\star \; \bar{\tilde L}_R \tilde \rho_\mu \gamma^{\mu} \tilde L_R +g_{\star Y} \; \bar{\tilde \ell} B^{(1)}_\mu \gamma^\mu \tilde \ell \nonumber
\eea
%
%
%\sh{Again, I see $\rho^\star$ here.. I am confused about notation now.. Also, do we need to give explicit expression for covariant derivatives ? .. Peizhi ?}
where $\psi_{SM}$ denotes all SM fermions and $\psi_{\rm light}=\psi_{\rm SM}-\{Q^3_L, t_R\}$, i.e. light SM fermions. 
%
%\sh{I've changed $\psi_{\rm SM}/\{Q^3_L, t_R\}$ to $\psi_{\rm SM}-\{Q^3_L, t_R\}$. To me, $A/B$ is quotient operation. If I am wrong, let me know.} 
%
%\ka{I added the following: we should then {\em try} to use $N^{ (1) }$ and $\tilde{l}^{ (1) }$ later on, for example, in Feynman diagrams etc. also.}
%\pd{I moved the statements up}
%The only purely composites fermions relevant for our study are the $SU(2)_R$ doublet $\tilde{L}_R$ (which is vector-like),
%containing the singlet neutrino and its charged lepton partner: these were denoted by
%$\chi^{ N, \; l }$ earlier in this section, but, for simplicity, henceforth we will instead use the 5D notation 
%[see Eq.~(\ref{lepton_content})] for these fields, i.e., $N^{ (1) }$ and $\tilde{\ell}^{ (1) }$ (even though we are following the two-site approach)
%
%components
%

%
It is 
%
%implicitly 
%
understood that all couplings should be multiplied by appropriate charges to get final coupling, which we do not show explicitly. The mixing angle between elementary and composite states of the associated fermion $\psi$ is denoted by $\phi_{\psi}$, with $\sin  \phi_{\psi} =1 (0)$ corresponds to pure elementary (composite). The specific representations of fermions are discussed in Sec.~\ref{sec:5Dmodel}. As mentioned earlier, here we assume that light SM fermions are purely elementary, i.e. $\sin \phi_{\psi_{\rm light}}=1$ (corresponding to the zero-modes
being localized near the UV brane in the 5D model), and $Q^3_L$ is slightly composite (roughly flat profile) and $t_R$ is fully composite
(localized near the IR brane), i.e. $\sin \phi_{t_{R}}=0$. 
Finally, $\mathcal L_{\textrm{Higgs}}$ has the form %
%
%
%where $\psi_{SM}$ denotes all SM fermions. All couplings here should multiply  by appropriate charges to get final coupling. $ \phi_{\psi}$ is the mixing angle of the associated ferimion $\psi$, and $\sin  \phi_{\psi} =1(0)$ corresponds to pure elementary (composite). The specific representations of fermions are discussed in Sec.~\ref{sec:5Dmodel}. Here we just assuming light fermions have $\sin \phi_{\psi_{\rm light}}=1$, where $\psi_{\rm light}=\psi_{\rm SM}/\{Q^3_L, t_R\}$ and $Q^3_L$ is slightly composite and $t_R$ is fully composite, meaning $\sin \phi_{t_{R}}=0$.
%
%
\bea\label{eq:Lhiggs}
\mathcal L_{\textrm{Higgs}}&=&\left|D_\mu \textbf{H} + i g \cot \phi \; \rho_\mu  \textbf{H} - i \tilde g_\star \; \textbf{H} \tilde \rho_\mu \right|^2 + V(\textbf{H}) -y \bar L \textbf{H}  \tilde L_R,
\eea
%
%\sh{(i) I made the above eq. one line, instead of two lines given previously. I am okay with going back to two-line form. (ii) $D_\mu$ contains only SM gauge connection ?}
%
%
%\bea\label{eq:Lhiggs}
%\mathcal L_{\textrm{Higgs}}&=&\left|D^\mu \textbf{H} + i g \cot \phi \; \rho_\mu  \textbf{H} - i \tilde g_\star \; \textbf{H} \tilde \rho_\mu \right|^2 + V(\textbf{H}) \\
%&~& -\bar L Y H \tilde L
%\eea
%
%
where $\textbf{H}$ denotes Higgs bi-doublet $\textbf{H}=(i\sigma_2 H, \,  H)$ and $H$ denotes SM Higgs doublet. $L$ is SM lepton doublet and $y$ is the Yukawa coupling constant. %$\textbf{H}$ transforms under $[SU(2)_L\times SU(2)_R]^{\rm comp}$.
All the vector fields in the Lagrangians shown above are in matrix forms. Final results can be obtained by taking traces of the above Lagrangians with appropriate normalization. 

%\ka{Again, for the sake of completeness (in order to match what was done in 5D model section), I feel we should point out which couplings are relevant for production vs.~decay. So, I added the sentences below. This can be simply removed if it is too much repetition.}

Note the sizable couplings of light quarks to $\rho$'s, i.e., excited SM gauge bosons, in 2nd line of 
Eq.~(\ref{eq:Lfermion});
%
%above equation
%
these are nonetheless
suppressed compared to the SM gauge couplings by the smallness of the elementary-composite mixing factor and correspond to the profile of the gauge KK modes
at the UV brane in the 5D picture.
In any case, it is these couplings that will
will be relevant for production of the spin-1 states at the LHC.
On the other hand, the coupling of light quarks to {\em non}-SM gauge bosons, i.e., $W^{ \pm(1) }_R$ and $Z^{ \prime(1) }$
(denoted collectively by $\tilde{ \rho }$)
%
%in above equation) 
%
is negligible, due to the absence of the elementary counterparts (and dual to profile of those gauge KK vanishing on the UV brane).
However, as we will see below, even ere a sizable coupling will be generated due to EWSB effects.
As far as decay of spin-1 states is concerned, 
it is the couplings in last line of Eq.~(\ref{eq:Lfermion}), i.e., to
composite leptons,
to top quark in line above it and to Higgs particles from Eq.~(\ref{eq:Lhiggs}) which dominate.
%
%above equation 
%

\subsection{Higgs induced gauge mixing}
\label{sec:gaugemixing}

%Given above Lagrangians, after EWSB, the original basis is not a mass eigenstates anymore. In this section, we shall discuss the diagonalization of mass eigenstates, which will result in maximally mixing among $W_L^{(1)}$ and $W_R^{(1)}$. $g_{\star W}$ and $g_{\star R}$ are chosen to be the same for the rest of the results.

\noindent When Higgs gets a VEV and the electroweak symmetry is spontaneously broken, it generates mixing among gauge bosons and fermions. In order to obtain mass spectrum, then, mass matrices should be diagonalized. In this section, we shall discuss the diagonalization of mass matrices and show, in particular, that the mass eigenstates of massive vector bosons consist of $O(1)$ components of both $W_L^{(1)}$ and $W_R^{(1)}$. That is, EWSB induces a significant mixing between $W_L^{(1)}$ and $W_R^{(1)}$, and this will be the main production channel for $W_R^{(1)}$
(as mentioned earlier). We choose $g_{\star W}$ and $g_{\star R}$ to be the same for our benchmark points. 

\noindent The mass matrix for charged vector bosons is given by
\bea
 \left (W_L^{+(0)}~~W_L^{+(1)}~~W_R^{+(1)}\right) \mathcal M^2  \left (W_L^{-(0)}~~W_L^{-(1)}~~W_R^{-(1)}\right)^{T}
\eea
where
\bea
\mathcal M^2 =\frac{1}{4} \left ( \begin{array}{ccc}
g_W^2 v^2  & g_W^2 v^2 \cot \phi_{W} & -g_W g_{\star W }v^2 \\
g_W^2 v^2 \cot \phi_{W} & 4 \frac{m_\star^2}{\cos^2 \phi_W}+(g_W \cot \phi_{W} v)^2&   -g_W \cot \phi_{W}  g_{\star W} v^2 \\
-g_W g_{\star W} v^2 & -g_W \cot \phi_{W}  g_{\star W} v^2& 4 m_\star^2+(g_{\star W} v)^2\\
\end{array}\right)
\label{eq:mass_matrix_gauge_bosons}
\eea
%
%we can find the mass splitting of $m_{W^{(1)}_L}$ and $m_{W^{(1)}_R}$:
%\bea
%m^2_{W^{(1)}_L}-m^2_{W^{(1)}_R}= \tan^2 \phi_W m^2_\star
%\eea 
%
%In the following study, we assume $g\ll g_\star~\tilde g_\star$, so we can use approximation $\cos \phi \approx 1$ in most of the cases, except the mass splitting mentioned above.
%So the mass matrix is slightly simplified to 
%
%\ka{For the sake of completeness, I think we need to emphasize the following.}
%
Note that we assume the {\em same} purely composite mass $m_\star$ for all composite gauge fields, i.e., before mixing with elementary states; this mixing does
perturb the mass for excited {\em SM} gauge bosons as seen above.
We will return to the more general case of {\em non}-degenerate composites in a follow-up paper.

\begin{figure}
\center
\includegraphics[width=0.6\linewidth]{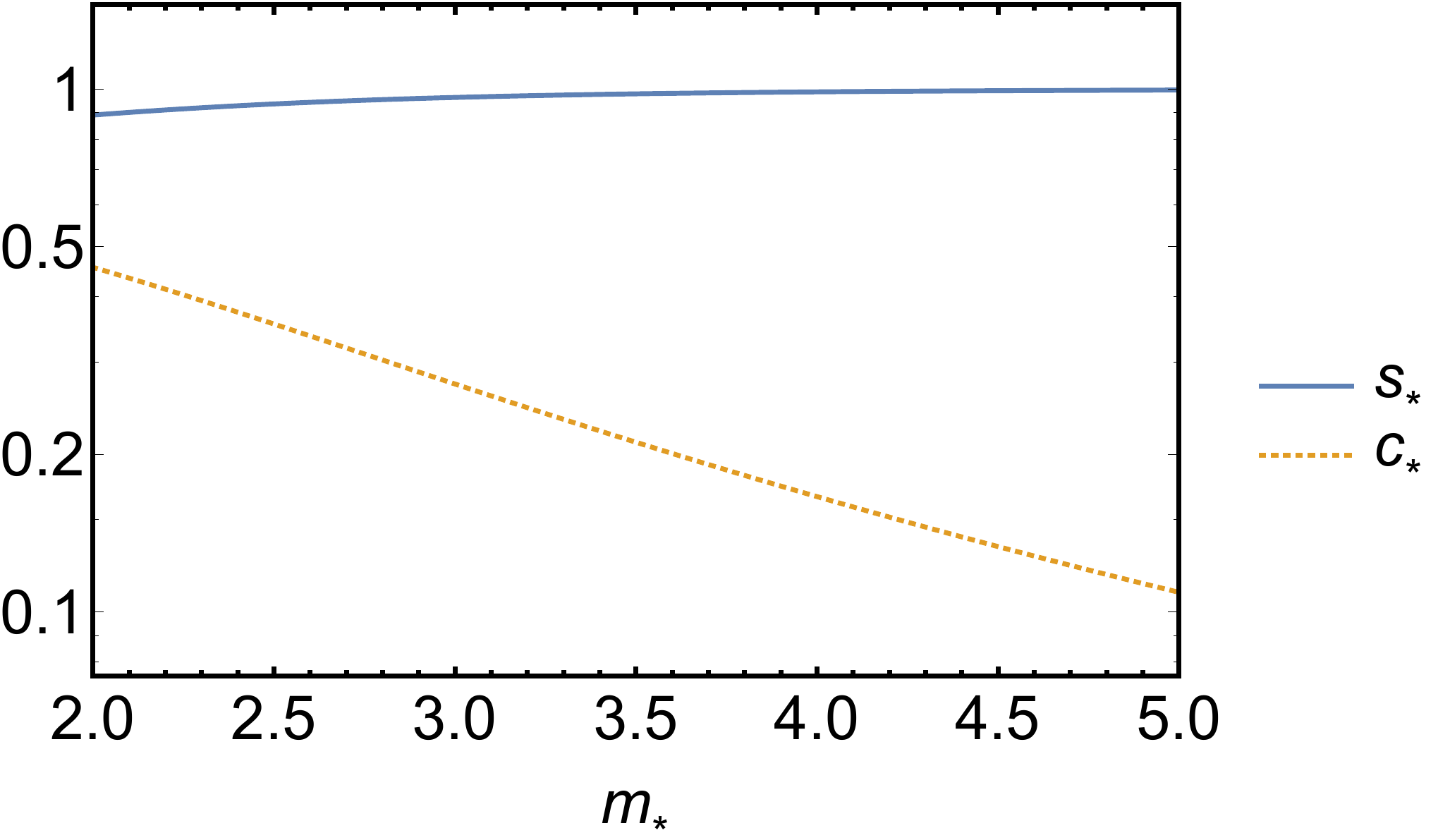}
\caption{The figure denotes sine and cosine of mixing angle($s_\star$,$c_\star$) between $W^{(1)}_L$ and $W^{(1)}_R$ as a function of $m_\star$, with $g_{\star W}=g_{\star R}=3$. SM parameters are chosen to be standard values $g_W=0.65$and $v=246\textrm{GeV}$.}
\label{fig:LRMixing}
\end{figure}

\noindent Performing explicit diagonalization of the above matrix analytically can be quite challenging. However, we can use the following method to get an approximate result. Our procedure will be valid
%
%construct a good approximation results as long 
%
as the following relations hold:
%
%
%The general diagonalization  of  the mass matrix is complicated. However, we could use the following method to get the approximated results. This result is a good approximation when 
%
%
\bea
\frac{1}{4}g_\star^2 v^2 \ll m^2_\star~ \textrm{and}~ g\ll g_\star,~\tilde g_\star
\label{eq:approx}
\eea
We demand that the mass matrix can be fully diagonalized by the following transformation by $U$.
\bea
U^{ \dagger } \mathcal M^2 U & = &\mathcal M^2_{ \rm diag }
\eea
where 
\bea
U & = & U_{ 12 } U_{ 13 } U_{ 23 } 
\eea
with
\bea
U_{ 12 } & = & %
\left( 
\begin{array}{ccc}
c & -s & 0 \\
s & c & 0 \\
0 & 0 & 1 
\end{array}
\right) \nonumber
\\
U_{ 13 } & = & 
\left( 
\begin{array}{ccc}
C & 0 & - S \\
0 & 1 & 0  \\
S & 0 &  C
\end{array}
\right)
 \\
U_{ 23 } & = & 
\left( 
\begin{array}{ccc}
1 & 0 & 0 \\
0 & c_{ \star } & - s_{ \star }  \\
0 & s_{ \star } & c_{ \star }
\end{array}
\right). \nonumber
\eea
Here $s,~S,~s_\star$ represent the sines of $\theta_{12},~\theta_{13},$ and $~\theta_{23}$, whereas $c,~C,~c_\star$ represent cosines of associated angles. Making use of the approximations of Eq.~(\ref{eq:approx}), one readily finds that
\bea
\tan 2 \theta_{ 12 } & \approx & \frac{ 2 g^{ 2}_{\star W} v^2   } {4 m_\star^2+g^{2}_{\star W} v^2  } \nonumber \\
\tan 2 \theta_{ 13 } &\approx& \frac{ -2 g^{2}_{\star W} v^2  } { 4 m_\star^2+g^{2}_{\star W} v^2}\\
\tan 2 \theta_{ 23 } & \approx & \frac{ -2 g_{\star W}^2 v^2   } { 4 \sin^2 \phi _W m^2_\star} \nonumber
\eea
The last formula corresponds to the mixing between two composite vector bosons, $W^{(1)}_L$ and $W^{(1)}_R$. Since $\frac{1}{4}g_\star^2 v^2 \ll m^2_\star$, we would naively think that this mixing is small. However, using $ \sin^2 \phi _W =  \frac{g_W^2}{g^2_{\star W}}$, we get
\bea
\tan 2 \theta_{ 23 } & \approx & \frac{ - g_{\star W}^2 v^2   } {  2(\frac{g^2_W}{g^2_{\star W}}m_\star)^2}
\label{eq:mixing_WL_WR_1}
\eea
and, as far as $g \ll g_\star$, this mixing angle is $O(1)$ ! That is, in most of the parameter space of interest, we get a \emph{significant mixing} between $W^{(1)}_L$ and $W^{(1)}_R$. (see Fig.~\ref{fig:LRMixing}) % which contradicts to naive impression of mixing angles among composite sectors.  
The same feature was pointed out in \cite{KA 08} using full 5D model, instead of two site model presented here. %
%\sh{I added the following..} %
%
The origin of the above large mixing can be understood as follows. From the mass matrix Eq.~(\ref{eq:mass_matrix_gauge_bosons}), one can find
\bea
m^2_{W^{(1)}_L}-m^2_{W^{(1)}_R}= \tan^2 \phi_W m^2_\star
\eea 
and using $ \sin^2 \phi _W \approx \tan^2 \phi_W \approx \frac{g_W^2}{g^2_{\star W}} \ll 1$, Eq.~(\ref{eq:mixing_WL_WR_1}) can be rewritten as follows.
\bea
\tan 2 \theta_{23} \approx \frac{-\frac{1}{2} g_{\star W}^2 v^2}{\left( M^2_{W_L^{(1)}}-M^2_{W_R^{(1)}} \right)}
\label{eq:mixing_WL_WR_2}
\eea
%
%
%This expression manifests the fact that the large mixing arises when the mass gap $m^2_{W^{(1)}_L}-m^2_{W^{(1)}_R}$ is suppressed, i.e. masses are degenerate, compared to $g_{\star W}^2 v^2$.
%\pd{I rephrase the sentence below.}
This expression manifests the fact that the large mixing arises when $W^{(1)}_L$ and  $W^{(1)}_R$ have almost degenerate masses, i.e. the mass gap $m^2_{W^{(1)}_L}-m^2_{W^{(1)}_R}$ is suppressed or comparable to $g_{\star W}^2 v^2$.

\noindent The relation with the mass basis denoted by $W$, $W_L$ and $W_R$ is given by 
\bea
\left( 
\begin{array}{c}
W \\
W_L \\
W_R
\end{array}
\right)
& = & U_{ 23 }^{ \dagger } U_{ 13 }^{ \dagger } U_{ 12 }^{ \dagger } 
\left( 
\begin{array}{c}
W_L^{(0)}\\
W_L^{(1)} \\
W_R^{(1)}
\end{array}
\right).
\eea
Since $\frac{1}{4}g_\star^2 v^2 \ll m^2_\star$ , $\theta_{ 12 } \approx \theta_{ 13 } \ll 1 $, we can approximate $s=S$. After dropping all terms with two or more s or S, we can get
\bea\label{eq:W_mass_basis}
W & = & C (c  \;  W_L^{ (0) } +  s \;  W_L^{(1)}) + S \;  W_R^{(1)} \nonumber \\
W_L & \approx & -  s  W_L^{ (0) }  + c ( \;  c_{ \star }   W_L^{(1)} +  s_{ \star }  W_R^{(1)} ) \\
W_R & \approx& -  S   W_L^{ (0) } +C ( -s_{ \star }  W_L^{(1)}  + c_{ \star }  \;W_R^{(1)}).  \nonumber
\eea
\noindent The typical size of these mixing angles are 
\bea
s\approx |S|\sim \frac{  g^{ 2}_{\star W} v^2   } {4 m_\star^2  } \ll 1 \nonumber \\
s_\star \sim \frac{ - g^{2}_{\star W} v^2   } {  4(\frac{g^2_W}{g^2_{\star W}}m_\star)^2} \sim 1.
\eea
%
%\ka{Again, I am trying to be complete by writing the following.}
%
Given the above {\em large} mixing between $W^{ (1) }_{ L, \; R }$ induced by the Higgs VEV, it is clear that
light quarks will now
couple similarly (and significantly) to {\em both} the mass eigenstates, cf.~in the basis prior to EWSB,
the coupling to one of the states, i.e., $W_R^{ (1) }$, was negligible.

\noindent The masses of the physical 
%
%eigen
%
states will also be perturbed due to the EWSB effects. %\sh{Please check the following sentence if it conveys the intended meaning.} 
Here, for simplicity, we kept only two massive states $W_L^{(1)}$ and $W_R^{(1)}$ to obtain the mass splitting, assuming that the small fraction of $W^{(0)}$ in the mass eigenstate does not make any difference. With such mass splitting, the physical mass for $W_L$ and $W_R$ are given by
\bea
m^2_{W_{L/R}}\approx m^2_\star+\frac{1}{4}g_{\star W}^2 v^2 \pm \sqrt{\frac{g_W^4}{4 g_{\star W}^4} m^4_\star+\frac{1}{16} g^4_{\star W} v^4}
\eea
where the $+(-)$ sign is for $m_{W_L}(m_{W_R})$. 

\noindent Similar analysis can be done for neutral gauge bosons. The mass matrix is given by
\bea
 \left (Z^{(0)}~Z^{(1)}~Z'^{(1)}\right) \frac{1}{2}\mathcal M^2   \left (Z^{(0)}~Z^{(1)}~Z'^{(1)}\right)^{T}
\eea
where
\bea
\mathcal M^2 =\frac{1}{4} \left ( \begin{array}{ccc}
g_Z^2 v^2  & g_Z^2 v^2 \cot \phi_{Z} & -g_Z g_{\star Z'}c'^2 v^2 \\
g_Z^2 v^2 \cot \phi_{Z} & 4 \frac{m_\star^2}{\cos^2 \phi_Z}+(g_Z \cot \phi_{Z} v)^2&   -g_Z \cot \phi_{Z}  g_{\star Z'} c'^2 v^2 \\
-g_Z g_{\star Z'}c'^2 v^2 & -g_Z \cot \phi_{Z}  g_{\star Z'} c'^2v^2& 4 m_\star^2+(g_{\star Z'} c'^2 v)^2\\
\end{array}\right).
\eea
Here $c'=\sqrt{1-\tan^2\theta _W}$ and $\theta _W$ is Weinberg angle in the composite sector. 
%
%\sh{How good/plausible is the following assumption ?} 
%
%\ka{In 5D language, I think this is valid at tree-level and with no BKT's so that 4D gauge coupling for all gauge fields here 
%is simply volume-suppressed as
%compared to 5D.}
%
We assume that the composite sector has the same Weinberg angle as the SM. Mass eigenstates are denoted by $Z$, $Z_1$ and $Z'$, and are related to gauge basis fields by 
\bea
Z & = & C (c  \;  Z^{ (0) } +  s \;  Z^{(1)}) + S \;  Z'^{(1)} \nonumber \\
Z_1 & \approx & -  s  Z^{ (0) }  + c ( \;  c_{ \star }   Z^{(1)} +  s_{ \star } Z'^{(1)} )\\
Z' & \approx& -  S   Z^{ (0) } +C ( -s_{ \star }  Z^{(1)}  + c_{ \star }  \;Z'^{(1)}) \nonumber
\eea
\noindent The typical size of mixing angles are 
\bea
s\approx |S|\sim \frac{  g^{2}_{\star Z} v^2   } {4 m_\star^2  } \ll 1 \nonumber \\
s_\star \sim \frac{ - g^{2}_{\star Z} v^2   } {  4(\frac{g^2_Z}{g^2_{\star Z}}m_\star)^2} \sim 1
\eea
And the spectrum of the mass eigenstate is 
\bea
m^2_{Z_1/Z'}\approx m^2_\star+\frac{1}{4}g_{\star Z}g_{\star Z'}c'^2 v^2 \pm \sqrt{\frac{g_Z^4}{4 g_{\star Z}^4} m^4_\star+ \frac{1}{16}g^2_{\star Z}g^2_{\star Z'}c'^4v^4}.
\eea

\subsection{Lepton Mixing}

%Apart from mixing in the gauge sector, EWSB also induces mixing in the fermion sector. As discussed earlier,  Since we assume all composite modes (dual to 5D KK modes) of SM particles are heavy and neglected, the mixing only appear in SM lepton doublet $L$ and $SU(2)_{R}$ doublet $\tilde L_R$. The Lagrangian contains Yukawa coupling of $L$ and $\tilde L_R$ in two site model has:

Apart from mixing in the gauge sector, EWSB also induces mixing in the fermion sector. As discussed earlier, composite ``excited'' modes for SM particles are neglected because they are heavier than composite vector bosons and composite states of singlet neutrino. For this reason, we focus on the mixing among SM lepton doublet $L$ and the composite $SU(2)_{R}$ doublet $\tilde L_R$. The relevant parts of the Lagrangian containing Yukawa coupling of $L$ and $\tilde L_R$ are as follows:
\bea
\mathcal L \ni y L_i \textbf{H}\tilde L_{Ri}+ m_D \bar{\tilde L}_{Ri} \tilde L_{Ri}
\label{eq:Lagrangian_lepton_mixing}
\eea
where $y$ is the Yukawa coupling and $i$ denotes the generation index of leptons, $i=\{e,~\mu,~\tau\}$. $m_D$ is the Dirac mass for composite $\tilde L_i$. The elementary (composite) $SU(2)_{\rm L}$ ($SU(2)_{\rm R}$) doublet $L$ ($\tilde{L}_R$) is defined as
\bea
L_e=(\nu^{(0)}_{e\,L},  \, e^{(0)}_L) \nonumber \\
\tilde L_{Re}=(N^{(1)}_{e},  \, \tilde e^{(1)}).
\eea
When the Higgs field gets a VEV, the Lagrangian Eq.~(\ref{eq:Lagrangian_lepton_mixing}) generates \emph{neutrino Mixing} as can be seen from
\bea
&~&\frac{y v}{\sqrt{2}} \bar \nu_L^{(0)} N_R^{(1)} + m_D \bar N_L^{(1)} N_R^{(1)} \nonumber \\
&~&  \nonumber \\
&=& m_D (\bar N_L ^{(1)}+ \frac{y_{4D} v}{\sqrt{2} m_D} \bar \nu^{(0)}_L) N_R^{(1)}
\eea
From this, we can obtain physical mass eigenstates denoted as $N_L$, $N_R$, and $\nu_L$:
\bea\label{eq:N_nu_mass_basis}
N_L &\approx& N^{(1)}_L + V_{\ell N} \nu^{(0)}_L \nonumber \\
N_R &=& N_R^{(1)}\\
\nu_L &\approx& \nu^{(0)} _L - V_{\ell N} N^{(1)}_L \nonumber
\eea
where the mixing is given by $V_{\ell N}=\frac{y v}{\sqrt 2 m_D}$. %
The same Lagrangian also introduces \textit{electron mixing} after EWSB
(we can safely neglect the SM electron Yukawa coupling or mass term here as compared to the others):
\bea
&~&\frac{y_L v}{\sqrt{2}} \bar e_L^{(0)} \tilde e_R + m_D \bar {\tilde e}_L^{(1)} \tilde e_R^{(1)} \nonumber \\
&~& \nonumber \\
&=& m_D (\bar{ \tilde e}_L^{(1)} + \frac{y_{4D} v}{\sqrt{2}m_D} \bar e_L^{(0)}) \tilde e_R^{(1)}.
\eea
Again, from this, we can obtain physical mass eigenstates denoted as $\tilde e_L$, $\tilde e_R$, and $e_L$:
\bea\label{eq:charged_lepton_mass_basis}
\tilde e_L &\approx& \tilde e^{(1)}_L + V_{\ell N} e^{(0)} _L \nonumber \\
\tilde e_R &=& \tilde e_R^{(1)}\\
e_L &\approx& e^{(0)} _L - V_{\ell N} \tilde e^{(1)}_L. \nonumber
\eea
%
%
%\sh{I need to understand better this last paragraph.. I have not touched it yet. To be honest, I am tired..}
%
%\ka{I shortened it, by removing the sentences about neglecting electron mass -- which is already mentioned above - and at the end about another multiplet, which is getting into too many details etc.}
%
%In both cases, the mixing only appears in the left-chirality, 
%
%since they are massless (chiral) in SM.  
%
In principle,
there is a similar effect from mixing of SM $SU(2)_L$ {\em singlet} charged lepton (after EWSB) with
composite $SU(2)_L$ doublets; however, since we assumed that such composites are heavy, we
can neglect it.
Moreover, electrons and neutrinos have the same mixing $V_{\ell N}$. This is because 
%
%(1) %Electron mixing, in principle, is not as simple as neutrino case, because electron is massive in SM. 
%from phenomenology point of view, electron can be treated as massless in high energy colliders, just as neutrinos.  
%
(1)
$N$ and $\tilde \ell$ are in the same $SU(2)_R$ doublet with the same mass $m_D$, together $\nu_L$ and $l_L$ being in the same $SU(2)_L$ doublet and (2) these two mixings originate from the same Yukawa coupling.  
%
%Though $\tilde \ell^{(1)}$ looks unfamiliar to people, it just acts like $\ell^{(1)}_R$ in the case with no seesaw sector, where KK modes of right-handed SM lepton mixes with SM ones. In our study, decoupling $\ell^{(1)}_R$ will effectively let $\tilde \ell^{(1)}$ takes its role.

\section{Overview of LHC signals}
\label{sec:overview_of_LHC_signals}

%As mentioned above, natural warped seesaw provide a way to produce composite $W_R$ in LHC via maximal mixing between $W^{(1)}_L$ and $W^{(1)}_R$ . Also, since both $W_R$ and $\tilde L_R$ are pure composite, $W_R \to N \tilde \ell$ channels dominant the branching ratio of $W_R$. Combination of these two feature achieves significant production of $N$ in LHC. The final states we are focused on is $N$ decaying to $W \ell$ and $\tilde \ell$ decaying to $H \ell$. Depending on the hadronic decay or  leptonic decay  of SM $W$, we call it di-lepton channel or tri-lepton channel respectively. The Feynman diagram of the signal channels are shown in Fig.~\ref{fig:signal}. 

In this section, based on our discussion in previous section, we first summarize couplings relevant to our collider study in Sec.~\ref{sec:analysis_results}. Then, we specify the choice of parameters used for actual analysis, together with related bounds. We then discuss production and dominant decay channels of heavy gauge bosons, i.e. $W_L$ and $W_R$. In particular, we show that $W_R \to N \tilde \ell$ is indeed the dominant decay channel for most of the parameter space of interest, providing abundance production of $N$ and $\tilde \ell$. We end our discuss by providing formulae for decay widths of $N$ and $\tilde \ell$.

\subsection{Relevant Couplings}
\label{subsec:relevant_couplings}

There are three types of couplings that we need to consider: (1) couplings between $W_L$/$W_R$ and SM fermions (2) couplings of $W_L$/$W_R$ to $N$-$\tilde \ell$ pair, and (3) couplings among $N$ ($\tilde \ell$) -- SM $H$, longitudinal $W/Z$ -- SM lepton $\ell$ ($\nu$) via Yukawa coupling.

(1) The first type of coupling can be obtained by using Eq.~(\ref{eq:Lfermion}) and EWSB induced mixing Eq.~(\ref{eq:W_mass_basis}):
\bea
\delta \mathcal L_{(1)}= \frac{g_W^2}{g_{\star W}} c_\star W^+_{L\mu}\bar\psi_L \gamma^\mu \psi'_L + \frac{g_W^2}{g_{\star W}} s_\star W^+_{R \mu}\bar\psi_L \gamma^\mu \psi'_L + \textrm {h.c}
\label{eq:Wproduction}
\eea
These couplings are responsible for the production of $W_L$ and $W_R$ via light quarks fusion inside proton. Notice that they suppressed by the factor $\frac{g_W}{g_{\star W}}$ and mixing angle compared to 4D LR models. However, as we will show in Sec.~\ref{sec:analysis_results}, these couplings, even with such suppressions, still render large enough signal production to be discoverable in near future. %Also, interestingly, if no mixing between $W^{(1)}_L$ and $W^{(1)}_R$, $W_R$ can not be produced via light quarks. Therefore, this mixing is crucial to give our signal.

(2) The second type of coupling can be understood from Eq.~(\ref{eq:Lfermion}) and mixing induced by EWSB Eq.~(\ref{eq:W_mass_basis}): 
%\sh{In the second term, $\ell$ should be $\tilde{\ell}$ ? Also, for $\{ N \leftrightarrow \nu ; \ell \leftrightarrow \tilde\ell\}$ part, do you mean composite ``second'' lepton $SU(2)_R$ doublet, not elementary $SU(2)_L$ doublet ? I mean, if it is elementary doublet, there must be mixing/suppression factor, which is indeed shown in above equation ?}
%
%
\bea\label{eq:Wdecay}
\delta \mathcal L_{(2)}= g_{\star W} s_\star W^+_{L\mu}\bar N \gamma^\mu \tilde \ell + g_{\star W} c_\star W^+_{R \mu}\bar N  \gamma^\mu \tilde \ell +\textrm {h.c.}
%\{ N \leftrightarrow \nu ; \ell \leftrightarrow \tilde\ell\}
\eea
These couplings lead to the decays of $W_L$ and $W_R$ to $N$ and $\tilde \ell$. %Without mixing, $W^{(1)}_L$ can not decay to $N$ and $\tilde \ell$. Mixing provides such decay channels for both $W_L$ and $W_R$.

(3) The third type of couplings are similarly obtained from Eq.~(\ref{eq:Lfermion}), Eq.~(\ref{eq:Lhiggs}) and mixing induced by EWSB Eq.~(\ref{eq:N_nu_mass_basis})(\ref{eq:charged_lepton_mass_basis}):
\bea \label{eq:Ndecay}
\delta \mathcal L_{(3)}&=& g_{ W} V_{\ell N} W^+_{\mu}\bar N_L \gamma^\mu \ell_L+\{ N \leftrightarrow \nu ; \ell \leftrightarrow \tilde\ell\}  \nonumber \\
&+ &g_{ Z} V_{\ell N} Z_{\mu}\bar N_L\gamma^\mu \nu_L + y H \bar N_R \nu_L +\{ N \leftrightarrow \tilde \ell ; \nu \leftrightarrow \ell\} + \textrm {h.c.}
%g_{ W} V_{\nu N} W^+_{\mu}\bar \nu_L \gamma^\mu \tilde \ell_L+ g_{ Z} V_{\nu N} Z_{\mu}  \bar\ell_L \gamma^\mu \tilde\ell_L +y H \bar \ell_L \tilde \ell_R
\eea
These couplings lead to the decays of $N$ and $\tilde{\ell}$ to $H/W/Z$ and $\ell/\nu$.

\subsection{Parameter Choice}
\label{subsec:Parameter_Choice}

The composite sector generally contains many parameters, such as $g_\star$'s and $\tilde g_\star$'s. In our study, as our benchmark points, we assume all $\phi$'s are the same, i.e. the ratio $g/g_\star$ are the same for all SM gauge groups. This choice is mainly for the sake of simplicity, and other choices with small variations will not lead to much difference in the final results.  Besides, we fix $g_{\star W}=g_{\star R}$, or equivalently we assume there exists $Z_2$ symmetry connecting $SU(2)_{\rm L}$ and $SU(2)_{\rm R}$. This is well motivated by the consistency with EW precision tests, e.g. to suppress the corrections to the coupling $Z \to b \bar b $. With these choices, we are left with basically only one free gauge coupling in composite sector 
%\sh{$\leftarrow$ one dimensionless parameter in gauge sector ? there are other parameters as discussed below...}; 
$g_{\star W}$. The composite gauge coupling $g_{\star W}$ has a lower bound $\sim3$, which comes from the requirement that the Landau pole does not appear below the GUT scale. %In our study, $g_{\star W}$ appears in the denominator of $W_L$ and $W_R$ production, which means a smaller $g_{\star W}$ will enhance our signal. Therefore 
We choose $g_{\star W}=3$ as a benchmark points.

The mass parameter $m_\star$ for heavy gauge bosons is constrained by EW precision tests. With extended symmetry group $SU(2)_{\rm L} \times SU(2)_{\rm R} \times U(1)_{\rm X}$, the bound is given by $\gtrsim 3$ TeV in most parts of parameter space. Partly motivated by the discoverability at the LHC, we choose $m_\star=2$ TeV for our study. Such a low mass might be achieved in some 
%
%parts of
%
corners of the parameter space or by invoking additional effects in EW precision tests (see for example \cite{Contino:2015mha}). Also, we choose $\cos \phi_{Q^3_{L}}=0.21$, which may be on the edge of constraints from the EW precision test. This, again, can potentially be allowed by introducing additional structure in the model. 

Next, $\left| V_{\ell N} \right|^2$ is constrained by various experiments and the results are summarized in \cite{delAguila:2008pw} . Considering consistency with these experimental bounds, we choose the $\left| V_{\ell N} \right|^2=0.001$ for all three generations.

In order for $W_L$ and $W_R$ to be able to decay to the pair $N$-$\tilde \ell$, $m_{\tilde L_R}$ needs to be smaller than half of $m_\star$. In principle, this mass is also constrained correlated with constraints of $\left| V_{\ell N} \right|^2$. With the choice we make $\left| V_{\ell N} \right|^2=0.001$, however, there is no effective bound on $m_{\tilde L_R}$. Nevertheless, given that heavy gauge bosons, $N$, and $\tilde{\ell}$ all ``live'' in the same composite sector, too big hierarchy between $m_{\tilde L_R}$ and $m_\star$ will lead to unwanted tuning. Taking into account all these considerations, we choose $m_{\tilde L_R} = 750$ GeV in our study.

\subsection{$W_L/W_R$ production and decay}
\label{subsec:WL_WR_production_decay}

As mentioned already, $W_L$ and $W_R$ are produced via couplings in Eq.~(\ref{eq:Wproduction}). Decay width for dominant decay channels are shown below, which are computed using couplings Eq.~(\ref{eq:Wdecay}).
%.. \sh{We also need to give coupling of $W_{L/R}$ to $H/Z/W$ ? Also, to $t/b$ ? Peizhi ?}.
 Since the analytic expression for decay widths of mass eigenstates $W_L$ and $W_R$ are quite complicated, we instead provide expressions for gauge fields in gauge basis, namely  $W_L^{(1)}$ and $W_R^{(1)}$.
 %\sh{$\to$ ? You mean really ``before'' EWSB ? or just states in gauge basis ?}
 This will be sufficient for the purpose of our discussion. All the decay widths present in this paper are given with the assumption $m_\star > 2m_{\tilde L_R}\gg \textrm{mass of SM particles}$, thus masses of SM particles are reasonably neglected.  Decay widths for $W_L^{(1)}$ are given by
\bea \label{eq:WLwdith}
&&\Gamma(W_L^{(1)} \to W H/W Z) = g^2_{\star W} \frac{m_{\star}}{192 \pi \cos\phi_W}\nonumber \\
&&\Gamma(W_L^{(1)} \to t b)= g^2_{\star W}  \cos^2 \phi_{Q^3_L}   \frac{m_{\star}}{16 \pi \cos \phi_W} \\
&&\Gamma(W_L^{(1)} \to \psi \psi')= N_c \frac{g_W^4}{g^2_{\star W}} \frac{m_{\star}}{48 \pi \cos\phi_W}  \nonumber
\eea
where $\psi$ and $\psi'$ denote SM fermions, and %\sh{there is no $N_\psi$ above.. you mean $N_c$ ? Then, which one is correct ?} 
$N_c$ shows the degree of freedom of corresponding fermion $\psi$: 3 for quarks and 1 for leptons. Next, decay widths for $W_R^{(1)}$ are given by
\bea \label{eq:WRwidth}
&&\Gamma(W_R^{(1)} \to N_i \tilde \ell_i) = g^2_{\star W} \left(1+2\frac{m^2_{\tilde L_R}}{m^2_{\star}}\right) \sqrt{1-4\frac{m^2_{\tilde L_R}}{m^2_{\star}}} \frac{m_{\star}}{24\pi} \nonumber \\
~\nonumber\\
&&\Gamma(W_R^{(1)} \to W Z/W H) =  g^2_{\star W}  \frac{m_{\star}}{192\pi}
\eea
where subscript $i$ is generation index. %	 $e,~\mu,~\tau$.

\begin{figure}
\center
\includegraphics[width=0.42\linewidth]{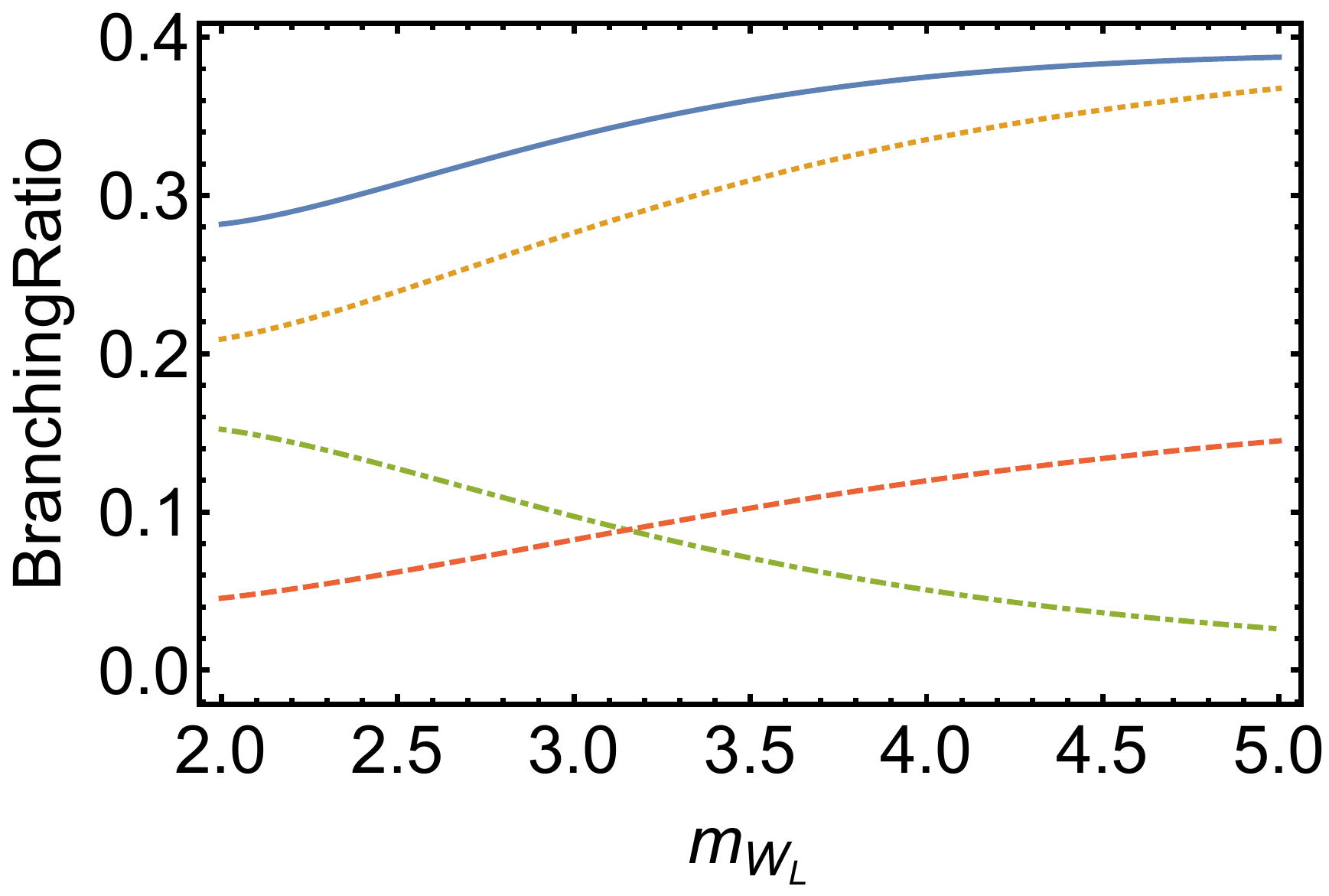}
\includegraphics[width=0.54\linewidth]{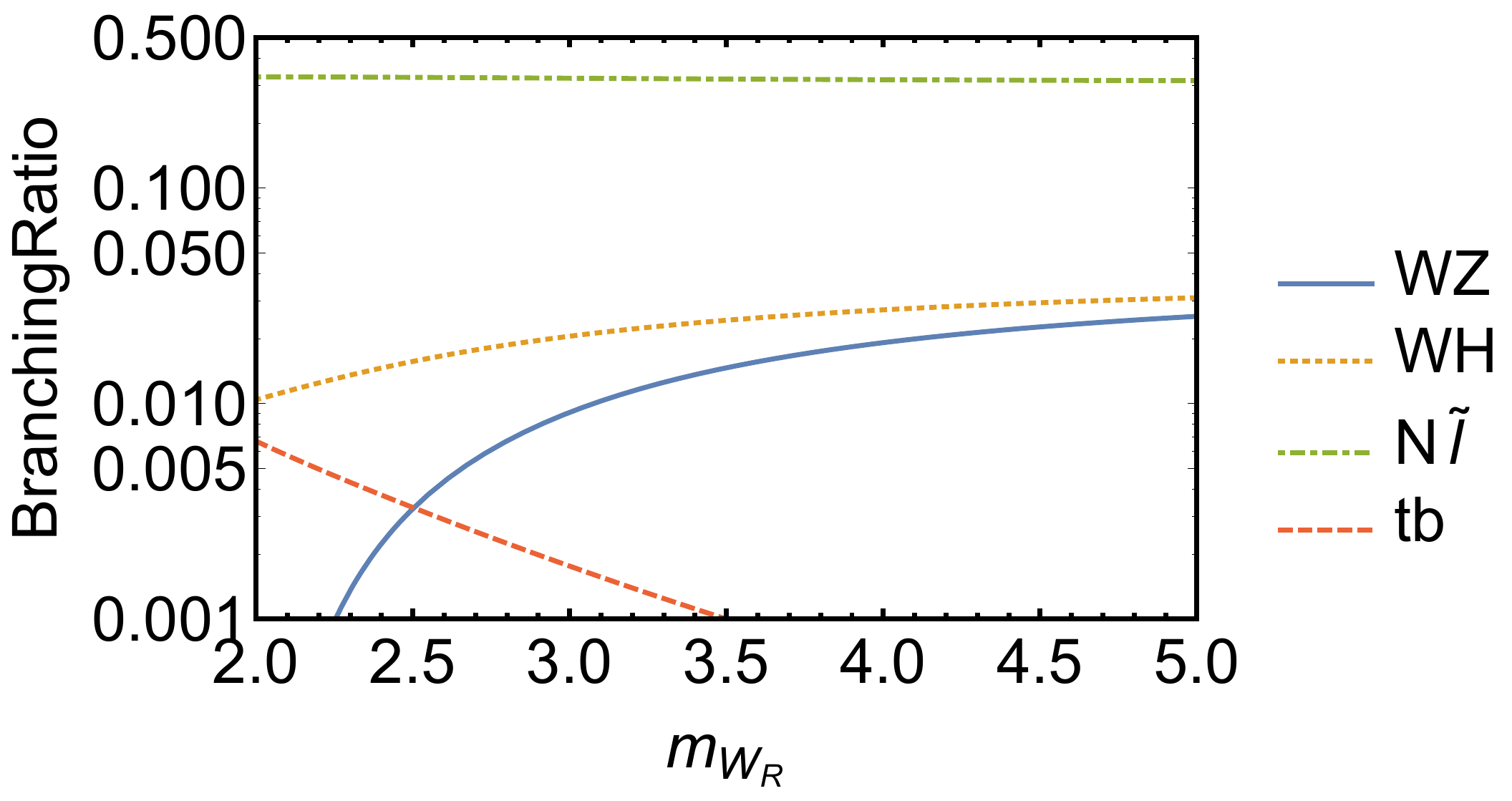}
\caption{The plot on the left (right) panel shows branching ratios of $W_L$ ($W_R$) as a function of its mass.}
\label{fig:WLRBR}
\end{figure}

From Eq.~(\ref{eq:WLwdith}) and Eq.~(\ref{eq:WRwidth}), we see that $W_R^{(1)}$ does not decay to quarks and $W_L^{(1)}$ does not decay to $N$-$\tilde \ell$ pair. All this is what we anticipated already. For the illustrative purpose, in Fig.~\ref{fig:WLRBR}, 
%\sh{Peizhi, please check the following sentence and see if it conveys the intended meaning.} 
we show the results ignoring $W^{(0)}$ component in the mixing, which would lead to  an error of the size $\frac{  g^{ 2}_{\star W} v^2   } {4 m_\star^2 }<0.1$. From there, we see that $W_R$ indeed decays dominantly to $N$-$\tilde \ell$ pair, providing production mechanism for them. This can be contrasted to the case of 4D LR models, where the dominant decay is into jets.
 For our collider study in Sec.~\ref{sec:analysis_results}, however, we used full model including Higgs induced mixing and mass splitting.

\subsection{$N$ and $\tilde \ell$ production and decay}
\label{subsec:N_ell_production_decay}

As mentioned in the last section, $N$ and $\tilde \ell$ are produced from on-shell decay of $W_L$ and $W_R$ via couplings in Eq.~(\ref{eq:Wdecay}). Decays of them are proceeded via the couplings Eq.~(\ref{eq:Ndecay}), resulting in decay widths:
\bea
&&\Gamma(N \to W \ell) = g^2_W \left| V_{\ell N} \right|^2\frac{m_{\tilde L _R}}{48\pi} \nonumber\\
&&\Gamma(N \to H/Z \nu) =g^2_W \left| V_{\ell N} \right|^2 \frac{m_{\tilde L _R}}{96\pi} \nonumber \\
&&\Gamma(\tilde \ell \to W \nu)= g^2_W \left| V_{\ell N} \right|^2 \frac{m_{\tilde L _R}}{48\pi} \\
&&\Gamma(\tilde \ell \to H/Z \ell)= g^2_W \left| V_{\ell N} \right|^2 \frac{m_{\tilde L _R}}{96\pi}. \nonumber
\eea
In principle, there will be three body decays via virtual $W_R$. However, we have checked that, for the choice of parameters we made, such three body decays are suppressed compared to 2 body decays.

%Signal channels mentioned before contain only charged gauge bosons $W_L$ and $W_R$. However, there are also neutral gauge bosons $Z_1$ and $Z'$ in this model. We can obtain all couplings related to production and decays of $Z_1$ and $Z'$  gauge bosons from Sec.~\ref{sec:twosite}. $Z_1$($Z'$) couple to light quarks with coupling $\frac{g^2_Z}{g_{\star Z}}$times mixing induced by EWSB. This gives production of $Z_1$ and $Z'$ via quarks inside proton. Since $Z_1$( $Z'$) have almost the same mass as  $W_L$($W_R$), production rate of $Z_1$and $Z'$ is not suppressed compared to $W_L$ and $W_R$. On the contrary, 4D LR inverse seesaw models have suppressed production of $Z'$ compared to $W_R$ due to $Z'$ being heavier than $W_R$. Moreover,  in this model, $Z_1$ and $Z'$ will have significant branching ratio to pairs of $N$, since $N$ is composite. In principle, there will be signal channels of $N$ via decays of $Z_1$ and $Z'$. However, we do not show detailed analysis of these channels here, because we will analyze the same final states via decays of neutral composite gauge bosons in the follow-up paper[cite], with slightly different assumptions of the model.

%\sh{Peizhi's last paragraph should go here.. It is already 3:30am.. and I am very tried.. Let me do it in the morning.. I am sorry..}

So far, we have focused on production and decay of charged gauge bosons, $W_L$ and $W_R$, and resulting production of singlet neutrino $N$. In addition to these, however, the model also contains neutral gauge bosons $Z_1$ and $Z'$ (see Sec.~\ref{sec:twosite}). The relevant couplings for these neutral gauge bosons can be obtained in a similar way as those for charged ones. In particular, just like Eq.~(\ref{eq:Wproduction}) for charged gauge bosons, $Z_1$ ($Z'$) couplings to light quarks is basically $\frac{g^2_Z}{g_{\star Z}}$ times a factor for EWSB induced mixing, and it is via this couplings that neutral gauge bosons are produced at the LHC. In our framework (i.e. 5D/composite LR model), since $Z_1$ and $Z'$ arise as composite vector mesons of the strong dynamics in the same way as the charged ones do, they have the same/comparable mass as $W_L$ ($W_R$). This, then, naturally leads to the comparable production rates for $Z_1$ and $Z'$, i.e. they are  not suppressed compared to $W_L$ and $W_R$. This feature can be contrasted to the case of 4D LR models, where production of $Z'$ is suppressed compared to $W_R$ due to the fact that $Z'$, as an elementary particle, is heavier than $W_R$. Moving onto the decays of the neutral gauge bosons, for the same reason for the charged gauge bosons, $Z_1$ and $Z'$ also have significant branching ratio to a pair of $N$. In this way, we see that, production and decay of these neutral gauge bosons provide another way to abundantly produce a pair of singlet neutrinos $N$. This signal channel, however, has almost the same process topology as 4D LR. Instead, we are planning to study the production of the singlet neutrino via on-shell decay of neutral gauge boson in our follow-up paper, but in a slightly different set up with interesting features/differences that only 5D framework can furnish.

%We mainly consider two channels, Di-lepton and Tri-lepton channels as typical channels for natural warped seesaw. Di-lepton channel is $p p > W_L/W_R > N l^\pm > l^\pm l^\mp W^\pm  H > l^\pm l^\mp j j b \bar b$ and Tri-lepton channel is $p p > W_L/W_R > N l^\pm > l^\pm l^\mp W^\pm  H > l^\pm l^\mp l^\pm b \bar b + \textrm{MET}$.

% Here we assume $N$ is pure Dirac, which means no lepton number violation. This is a good approximation for collider point of view, because Majorana splitting is much smaller than KK scale, TeV, as discussed in Sec.[ref]

%%%%%%%%%%%%%%%%%%%%%%%%%%%%%%%%%%%%%%%%%%%%%%%%%%%%%%%%%%%%%%%%%%%%%%%%%%%%%%%%%%%%%%%%%
%%%%%%%%%%%%%%%%%%%%%%%%%%%%%%%%%%%%%%%%%%%%%%%%%%%%%%%%%%%%%%%%%%%%%%%%%%%%%%%%%%%%%%%%%

%%%%%%%%%%%%%%%%%%%%%%%%%%%%%%%%%%%%%%%%%%%%%%%%%%%%%%%%%%%%%%%%%%%%%
%%%%%%%%%%%%%%%%%%%% Analysis Results %%%%%%%%%%%%%%%%%%%%%%%%%%%%%%%
%%%%%%%%%%%%%%%%%%%%%%%%%%%%%%%%%%%%%%%%%%%%%%%%%%%%%%%%%%%%%%%%%%%%%

\section{Discovery Potential}
\label{sec:analysis_results}

In this section, we present our results for phenomenological studies of the LHC signals for the model described in Sec.~\ref{sec:twosite}. In particular, we study the pair production of the singlet neutrino ($N$) and its $SU(2)_{\rm R}$ partner ($\tilde{\ell}$) via the one-shell decay of $W_R$ and $W_L$, and their subsequent decays to SM particles. 
%\sh{I assume we will discuss importance of $W^{(1)}_L-W^{(1)}_R$ mixing for production channel etc in earlier sections, so I won't repeat them here.}
We consider two benchmark points depending on how $N$ and $\tilde{\ell}$ cascade decay to SM particles: \emph{Di-lepton}- and \emph{Tri-lepton}-channels. 

%\noindent \sh{What I wrote below might be repetition of earlier parts. In that case, we can either remove or shorten.}

\noindent For Di-lepton channel, the production and the cascade decays of $N$ and $\tilde{\ell}$ are as follows:
\bea
p p > W_L / W_R > N \; \tilde{\ell}^{\pm}, \;\;\; N > \ell^{\pm} \; \left( W^{\mp} > j j \right), \;\;\; \tilde{\ell}^{\pm} > \ell^{\pm} \; \left( H/Z > b \bar{b} \right).
\eea
%
%where $W^{(1)}_L / W^{(1)}_R$ indicates the Higgs-induced mixing between $W^{(1)}_L$ and $W^{(1)}_R$. 
Hence, the final states of the Di-lepton channel consist of $\ell \ell j j b \bar{b}$, where, for the lepton pair, only opposite sign combination can arise since we are ignoring small Majorana splitting for $N$. That is, this process is lepton-number conserving. %This is because even though the charge of the lepton from the decay of $W^{(1)}_L / W^{(1)}_R$ is fixed and the same as that of $W^{(1)}_L / W^{(1)}_R$, the lepton from the decay $N_R > W \ell$ can have either sign. 
In particular, this channel contains \emph{two} leptons, and hence the name for the channel.

\noindent For Tri-lepton channel, on the other hand, we take the leptonic decay for SM $W$ boson from $N$. In detail, we get:
\bea
p p > W_L / W_R > N \; \tilde{\ell}^{\pm}, \;\;\; N > \ell^{\pm} \; \left( W^{\mp} > \ell^{\mp} \nu \right), \;\;\; \tilde{\ell}^{\pm} > \ell^{\pm} \; \left( H/Z > b \bar{b} \right).
\eea
Hence, the final states of the Tri-lepton channel consist of $\ell \ell \ell \nu b \bar{b}$. This time, it contains \emph{three} leptons, explaining the name of the channel.

\noindent Notice that in both channels, we add contributions from both $H$ and $Z$ decaying into $b \bar{b}$. This is because resolutions of LHC detectors may not be good enough to distinguish those two cases, and at the same time, we will achieve a slight increase in the signal rate. 

\begin{figure}
\center
\includegraphics[width=0.4\linewidth]{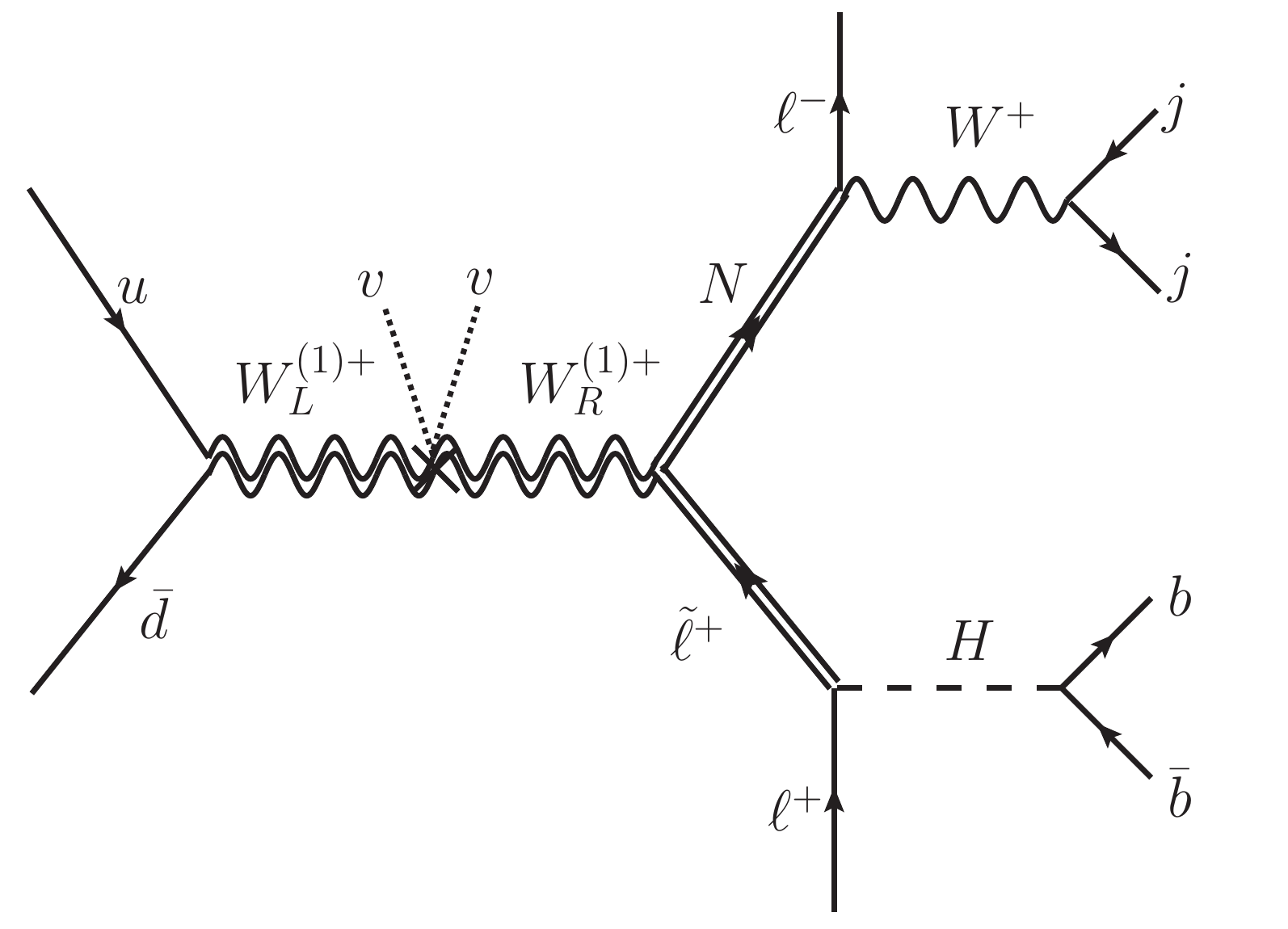}
\includegraphics[width=0.4\linewidth]{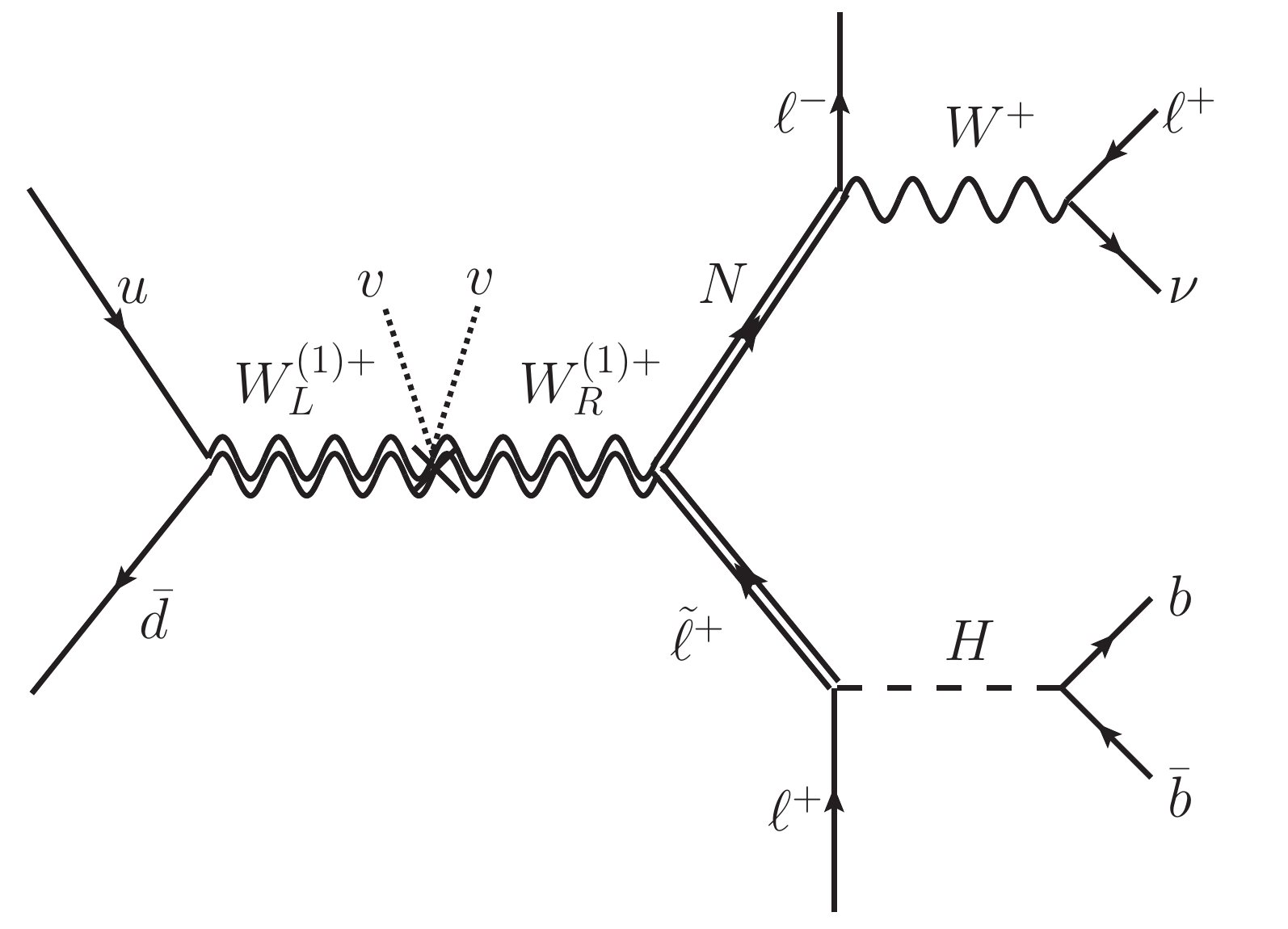}
\caption{The left panel shows Feynman diagram for the signal process of Di-lepton channel. The right panel shows Feynman diagram for the signal process of Tri-lepton channel. Double (single) lines denote composite (elementary) particles. Here composite gauge bosons are in gauge basis $W_L^{(1)}$ and $W_R^{(1)}$ in order to show the mixing induced by Higgs VEV explicitly. }
\label{fig:signal}
\end{figure} 

\noindent The Feynman diagrams for both signal processes are shown in Fig.~\ref{fig:signal}. The topology of our signal processes are characterized by several resonance peaks in various invariant mass variables. In particular, invariant masses of $W_R$ and $N / \tilde{\ell}$, which we take to be $M_{W_R} = 2$ TeV and $M_{N} = 750$ GeV in our study, will draw sharp distinctions between signal and SM backgrounds. For Tri-lepton channel, however, due to the presence of neutrino and the multiplicity of leptons (i.e. combinatorics issue), naively, one would think that resonance peaks are less pronounced. %although some of them will still provide very efficient cuts to reduce backgrounds.
However, as we show below, by reconstructing the longitudinal component of the neutrino's momentum and by figuring out the identification of each lepton, i.e. which lepton is to be paired with $b\bar{b}$, neutrino, and $\ell \nu$, respectively, we are able to construct all invariant mass peaks.

%%%%%%%%%%% Event Simulation %%%%%%%%%%%%%%%%%%%

\noindent Event simulations are performed by employing a sequence of simulation tools. We first created our two-site simplified model files using \textsc{FeynRules}~\cite{Alloul:2013bka} based on Heavy Vector Triplets models \cite{Pappadopulo:2014qza}. Then we used them as inputs model in a Monte Carlo event generator \textsc{MG5@aMC}~\cite{Alwall:2014hca} to generate parton level events. In this procedure, parton distribution functions parameterized by \textsc{NN23LO1}~\cite{Ball:2012cx} is used.  All the simulations are done at the leading order with a $\sqrt{s}=14$ TeV $pp$ collider. The generated parton level events are then streamlined to \textsc{Pythia 6.4}~\cite{Sjostrand:2006za} to take care of showering and hadronization/fragmentation. Since all our channels contain only regular jets, i.e. no boosted gauge bosons leading to fat jets, we directly pass on the output from \textsc{Pythia 6.4} to \textsc{Delphes 3}~\cite{deFavereau:2013fsa}. \textsc{Delphes 3}, interfaced with \textsc{FastJet}~\cite{Cacciari:2005hq, Cacciari:2011ma}, provides a way to incorporate the detector effects and jet formation. The jets are constructed with the anti-$k_t$ algorithm~\cite{Cacciari:2011ma} with a radius parameter $R=0.4$.

\noindent In Sec.~\ref{subsec:dilepton_channel}, we present our results for Di-lepton channel. Results for Tri-lepton channel follow in Sec.~\ref{subsec:trilepton_channel}. We also briefly discuss phenomenological distinctions between our 5D left-right symmetry model and that of 4D. In particular, we will point out several salient features of our case by which two frameworks can be distinguished once discovery is made.

\subsection{Dilepton $+$ dijet $+$ $H/Z$ channel}
\label{subsec:dilepton_channel}

%-- bottom-line final state

%-- SM background considered

%-- details of jet-smearing (crucial here)

%-- cut flow/table for signal and background

%-- final significance and signal size 

We begin by considering the production of $N-\tilde{\ell}$ pair and their decays at the LHC. In our current study, we consider $(N, \tilde{\ell})$ as a $SU(2)_{\rm R}$ doublet and as a consequence the production of this doublet pair should be proceeded via decay of $W^{(1)}_R$ gauge boson. However, since SM quarks are not charged under $SU(2)_{\rm R}$ gauge group, $W^{(1)}_R$ can only be produced via its mixing with $W^{(1)}_L$. Namely, once $W^{(1)}_L$ is produced via quark fusion inside the proton through its $SU(2)_{\rm L}$-coupling, EWSB-induced mixing between $W^{(1)}_L$ and $W^{(1)}_R$ leads to the production of $W^{(1)}_R$. $W^{(1)}_R$ then subsequently decays into $N-\tilde{\ell}$ pair. As shown in Sec.~\ref{sec:twosite}, the size of $W_R^{(1)} - W_L^{(1)}$ mixing angle is $\tan 2 \theta_{23} \approx \frac{-\frac{1}{2} g_{\star}^{W}g_{\star}^{R} v^2}{\left( M^2_{W_L^{(1)}}-M^2_{W_R^{(1)}} \right)}$ (see Eq.~(\ref{eq:mixing_WL_WR_2})), and when the mass splitting, $M^2_{W_L^{(1)}}-M^2_{W_R^{(1)}}$, is small enough we acquire significant mixing, leading to enhanced production for signal. This can be realized when the masses of $W^{(1)}_L$ and $W^{(1)}_R$ are approximately degenerate and the mass scale itself is low enough. Motivated by the consistency with the electroweak precision measurements (EWPM), our 5D warped extra-dimensional seesaw model or its two-site simplified model has built-in left-right symmetries, allowing desired mass degeneracy. In addition, the consistency with EWPM permits the mass of $M_{W_L^{(1)}} / M_{W_R^{(1)}}$ as low as $\mathcal{O}(2)$ TeV. Such a low mass for $W^{(1)}_L / W^{(1)}_R$ further allows, in addition to large mixing, resonance enhancement for the signal production cross section at the LHC. Moving onto the decay of $W^{(1)}_R$, as elaborated in Sec.~\ref{subsec:WL_WR_production_decay}, it will dominantly decay into $(N, \tilde{\ell})$ pair. Therefore, making use of all these features, we can secure enough statistics for signal production at 14 TeV LHC. In Di-lepton channel, $N$ decays to $W^{\pm} \ell^{\mp}$ and SM $W$ boson, in turn, decays hadronically producing two jets. On the other hand, $\tilde{\ell^{\pm}}$ decays to $\ell^{\pm} H/Z$, which is then followed by decay of $H/Z$ to $b \bar{b}$. As is evident from these cascade decays of $N$ and $\tilde{\ell^{\pm}}$, (i) signal process does not contain any neutrinos and hence no missing energy and (ii) there are several invariant mass variables which are all fully reconstructible. %, again due to absence of missing particle. 
%Those invariant mass variables include, $M_W$ from $jj$, $M_{H/Z}$ from $b \bar{b}$, $M_{N_R}$ for both $N_R$ and $\tilde{\ell^{\pm}}$ from either $jj\ell$ or $b\bar{b}\ell$, and finally $M_{\rm All}$ from all visible, reconstructed objects. If successfully reconstructed, the distribution of $M_{\rm All}$ variable with peaked near $M_{W_R}$.
Those invariant mass variables include, $M_{jj}$, $M_{b \bar{b}}$, $M_{jj\ell}$, $M_{b\bar{b}\ell}$, and $M_{\rm All}$, where $M_{\rm All}$ is the invariant mass of \emph{all} reconstructed/visible particles. If successfully reconstructed, for signal, the distributions of these variables will be peaked at $M_W$, $M_{H/Z}$, $M_{N}$, $M_{N}$, and $M_{W_R}$, respectively.

\noindent There are several SM backgrounds we need to consider and we describe them one by one now. \\

\noindent {\bf{(1) $\bf{t\bar{t}jj}$:}} The relevant process is $pp > t \bar{t} jj > \ell^- \ell^+ \nu \bar{\nu} b \bar{b} jj$, where $t > b \; ( W^+ > \ell^+ \nu)$, and similarly for $\bar{t}$, is considered. Being a purely QCD process, this is the background with largest cross section. Background reduction will be achieved by means of a combination of various invariant mass cuts. Particularly useful ones will be $M_{\rm All}$ and $M_{b\bar{b}\ell} / M_{jj\ell}$ cuts. In principle, missing transverse momentum $\slashed{E}_T$, the opposite of the vectorial $p_T$ sum of reconstructed objects in the event, can provide useful reduction, although we found other cuts are more efficient.
 \\

\noindent {\bf{(2) $\bf{t\bar{t} H/Z}$:}} The relevant process is $pp > t \bar{t} H/Z > \ell^- \ell^+ \nu \bar{\nu} b b \bar{b} \bar{b}$, where $t > b \; ( W^+ > \ell^+ \nu)$, and similarly for $\bar{t}$, and $H/Z > b \bar{b}$ are considered. If two $b$'s in the signal process are b-tagged as a part of selection criteria, then in order for this background to pass the selection criteria, two of four $b$'s must be un-tagged as regular two jets, leading to a large reduction of the background. Moreover, $M_{\rm All}$, $M_{b\bar{b}\ell} / M_{jj\ell}$ and $M_{jj}$ cuts will be useful. \\

\noindent {\bf{(3) $\bf{j j \ell \ell H/Z}$:}} The relevant process is $pp > j j \ell^- \ell^+ H/Z, \; H/Z > b \bar{b}$, where the lepton pair comes mostly from decay of on-shell $Z$ (and off-shell photon). Therefore, in this process, the distribution of the di-lepton invariant mass, $M_{\ell\ell}$, will be sharply peaked at the mass of the $Z$ boson, $M_Z$. However, since two leptons in the signal process do not reconstruct $M_Z$, the condition $M_{\ell\ell} \neq M_Z$ will remove most of this background. In addition, $M_{\rm All}$, $M_{b\bar{b}\ell} / M_{jj\ell}$ and $M_{jj}$ cuts will be useful. \\

\noindent {\bf{(4) irred (irreducible background):}} The relevant process is $pp > \ell^- \ell^+ W^{\pm} H/Z, \; W^{\pm} > jj, \; H/Z > b \bar{b}$. Similarly to $\bf{j j \ell \ell H/Z}$ background, the lepton pair will mostly arise from the on-shell decay of $Z$ and the cut $M_{\ell\ell} \neq M_Z$ will significantly reduce this events. Even though $jj$ ($b\bar{b}$) will successfully reconstruct $M_W (M_{H/Z})$, $M_{\rm All}$ and $M_{b\bar{b}\ell} / M_{jj\ell}$ cuts will still provide additional significant reduction of this background events. \\
%
%
%%%%%%%%%%%%%%% PLOTS %%%%%%%%%%%%%%%%%%%%%%%%%%%%%%%%%%%%%%%%%%%%%%%%%%%%%%%%

\begin{figure}%[h]
    \centering
    \includegraphics[width = 7.5 cm]{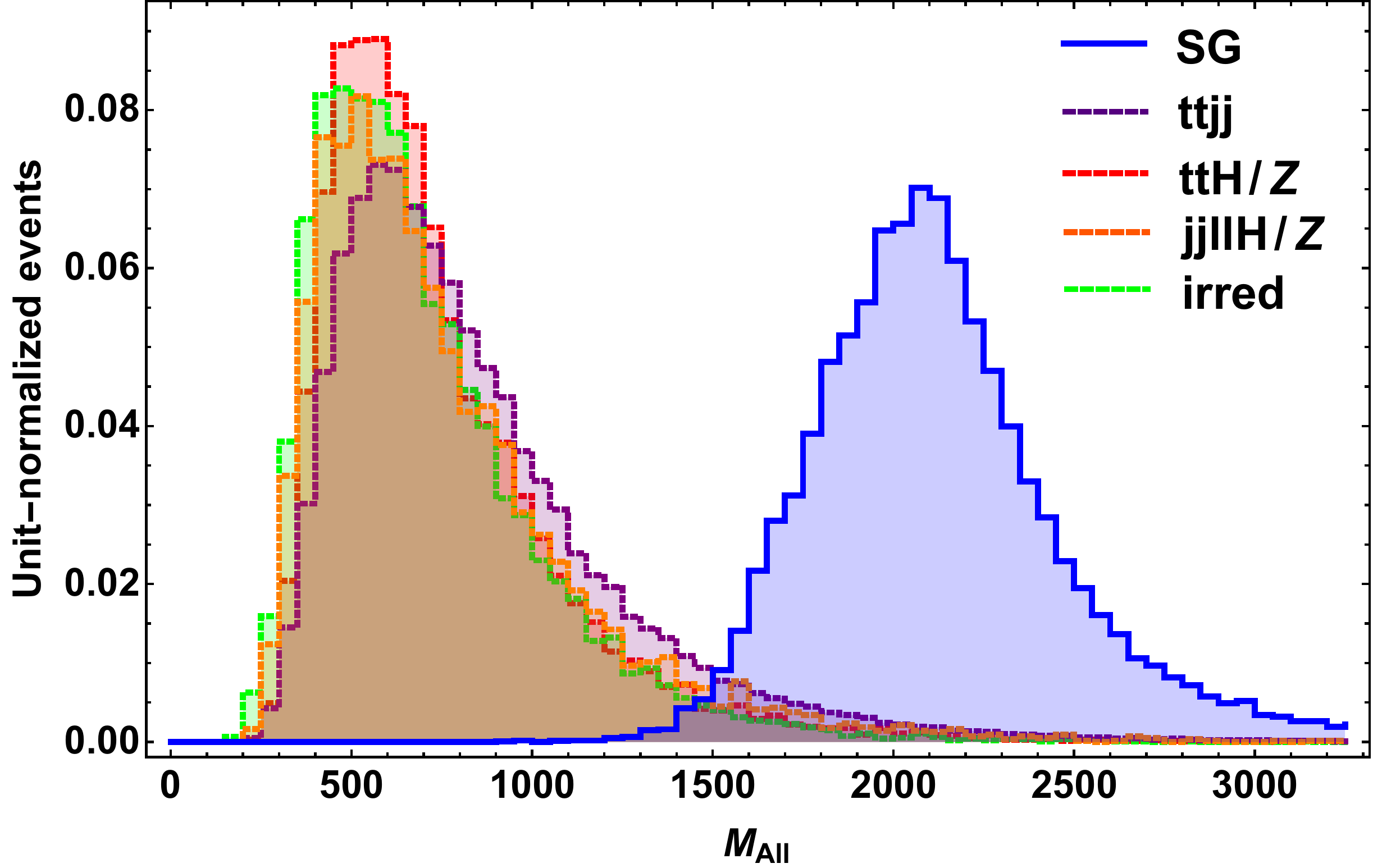}
    \includegraphics[width = 7.5 cm]{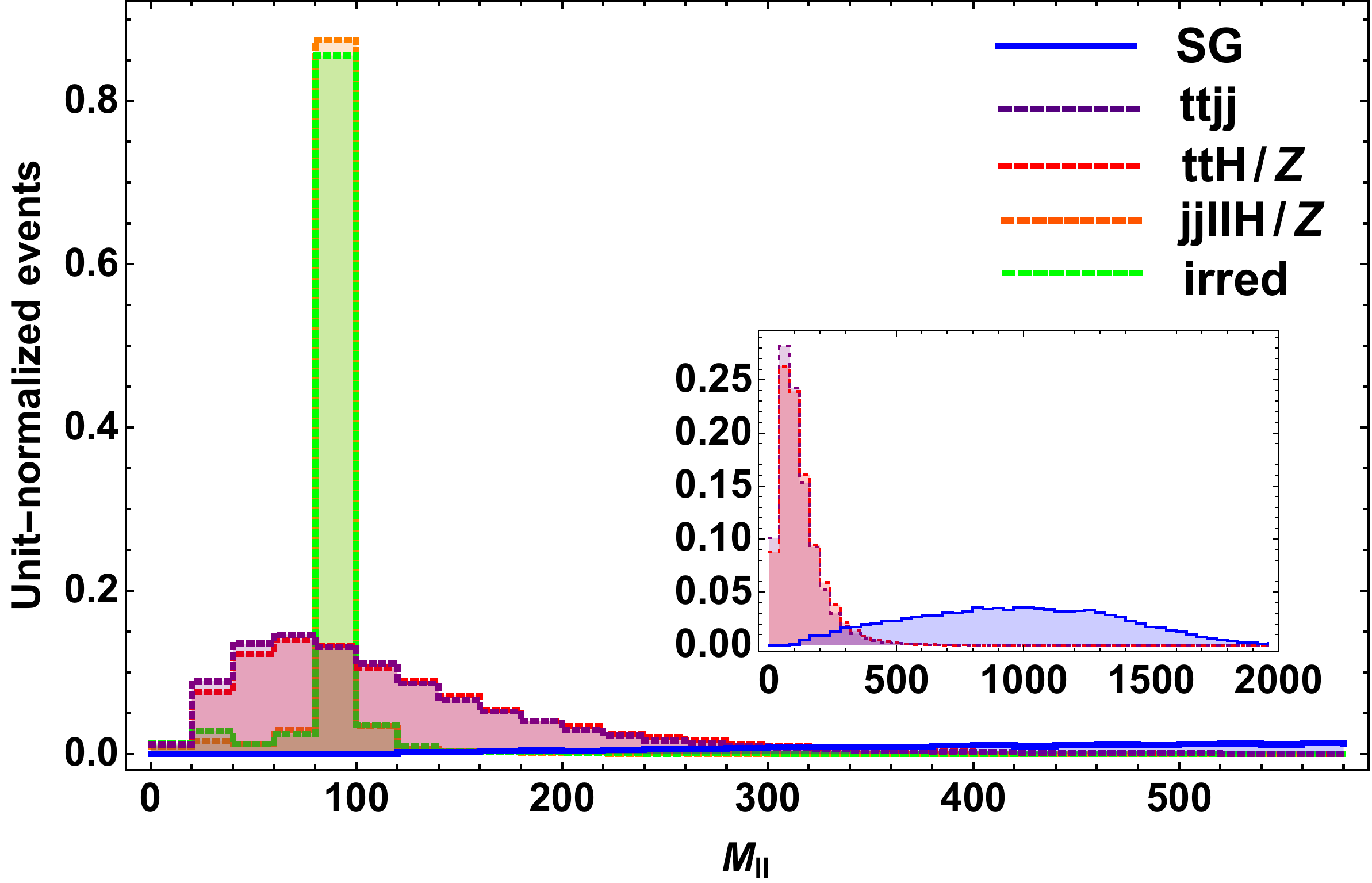}
    \includegraphics[width = 7.5 cm]{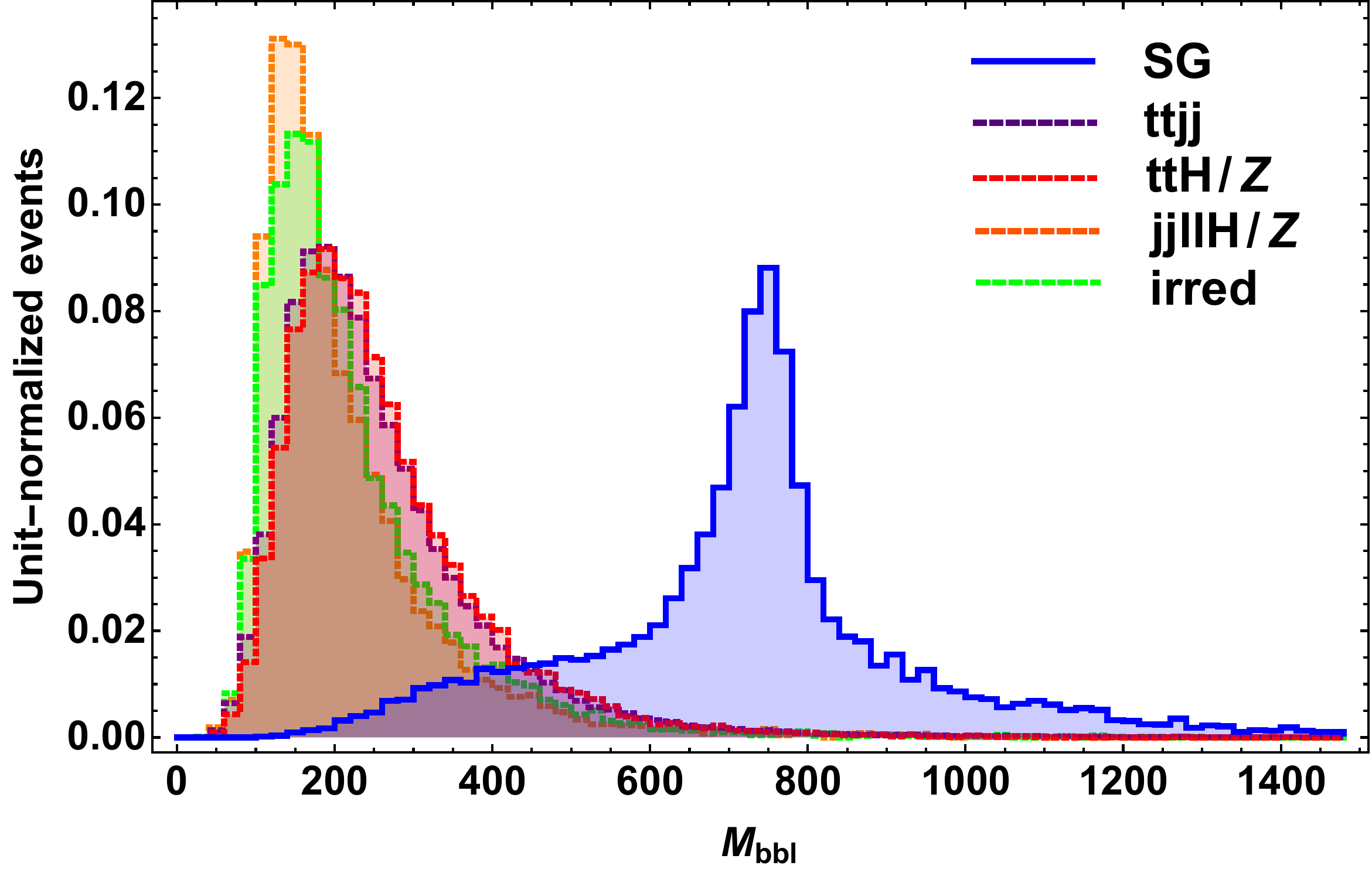}
    \includegraphics[width = 7.5 cm]{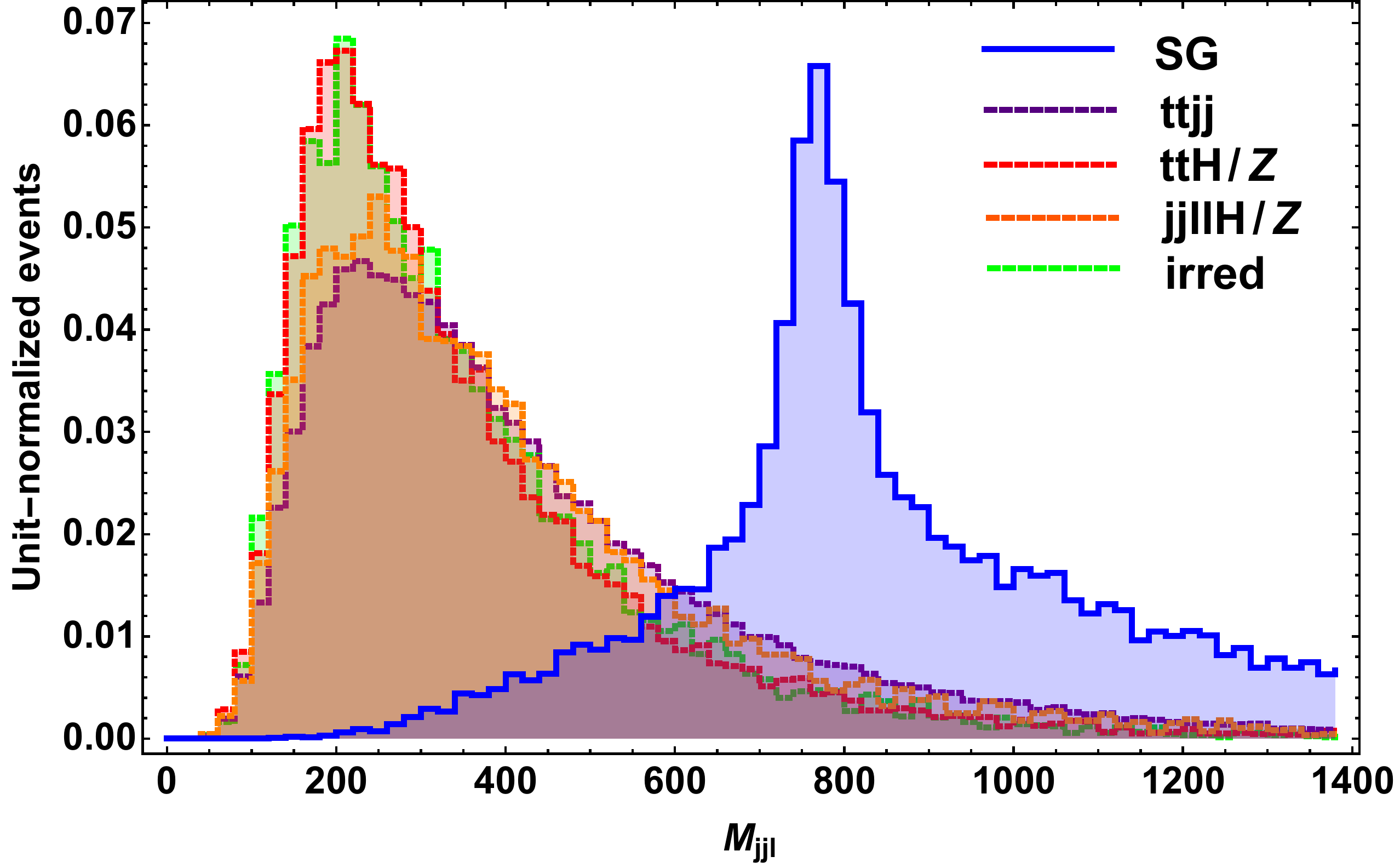}
    \includegraphics[width = 5 cm]{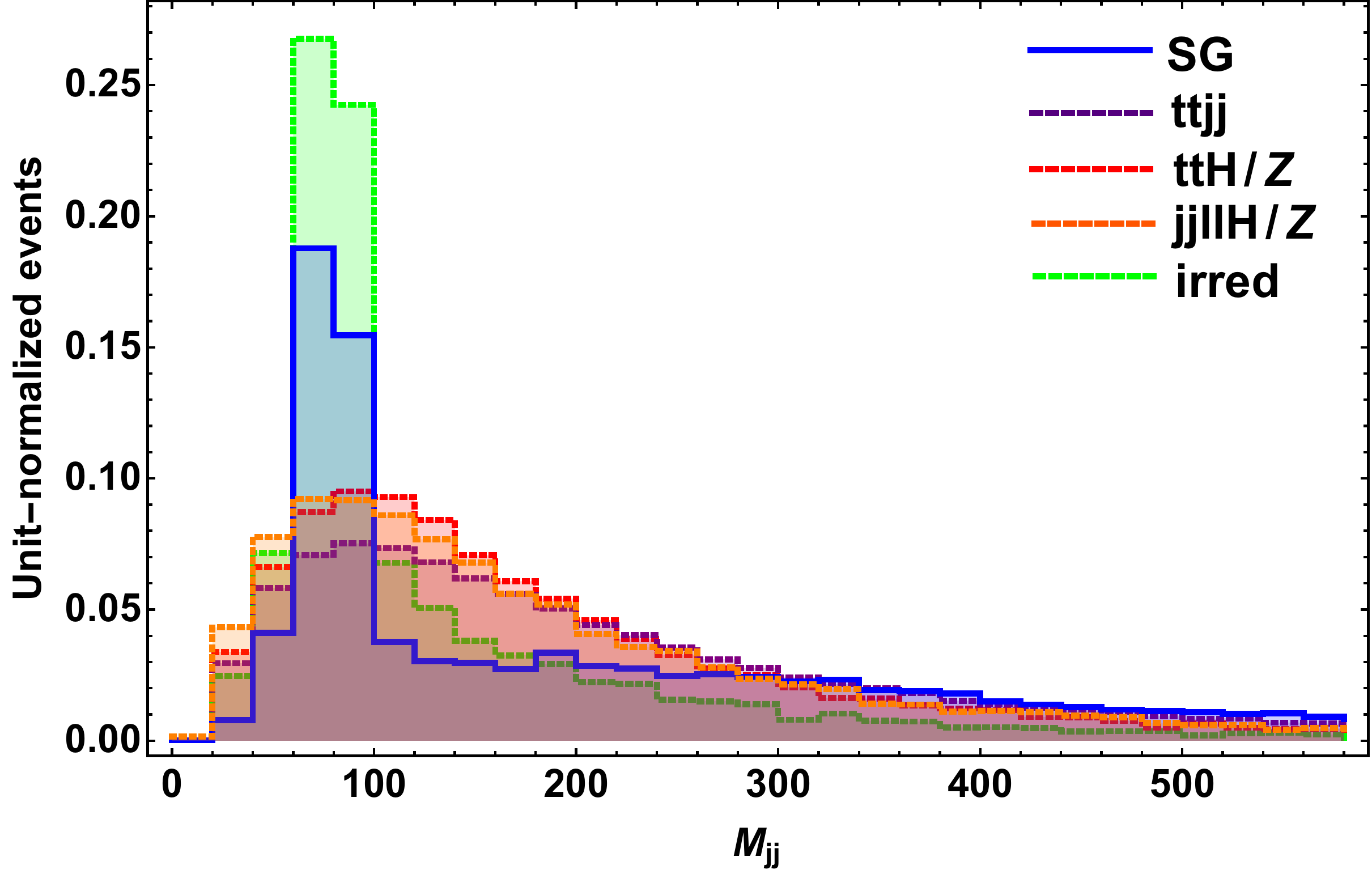}
    \includegraphics[width = 5 cm]{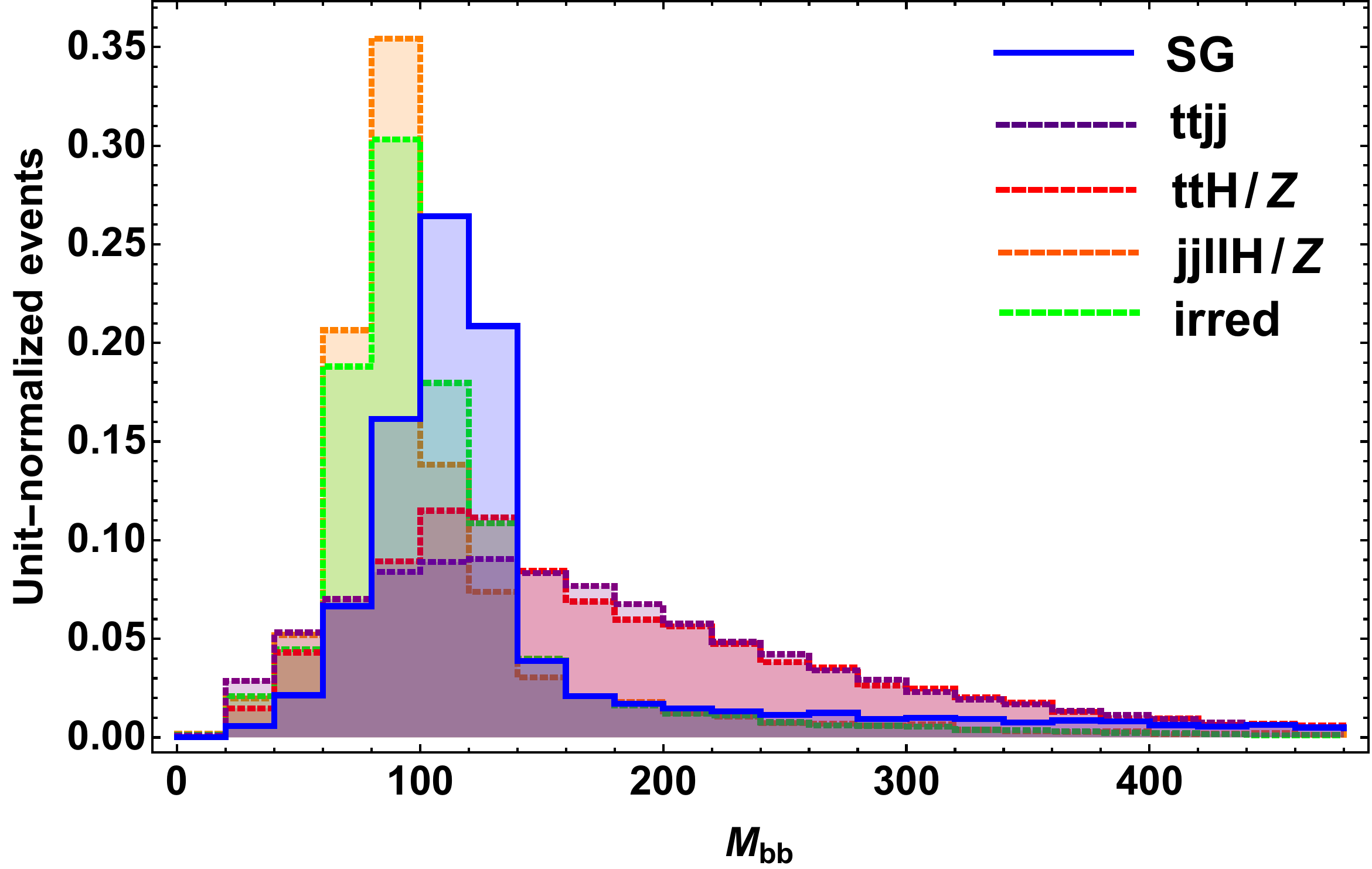}
    \includegraphics[width = 5 cm]{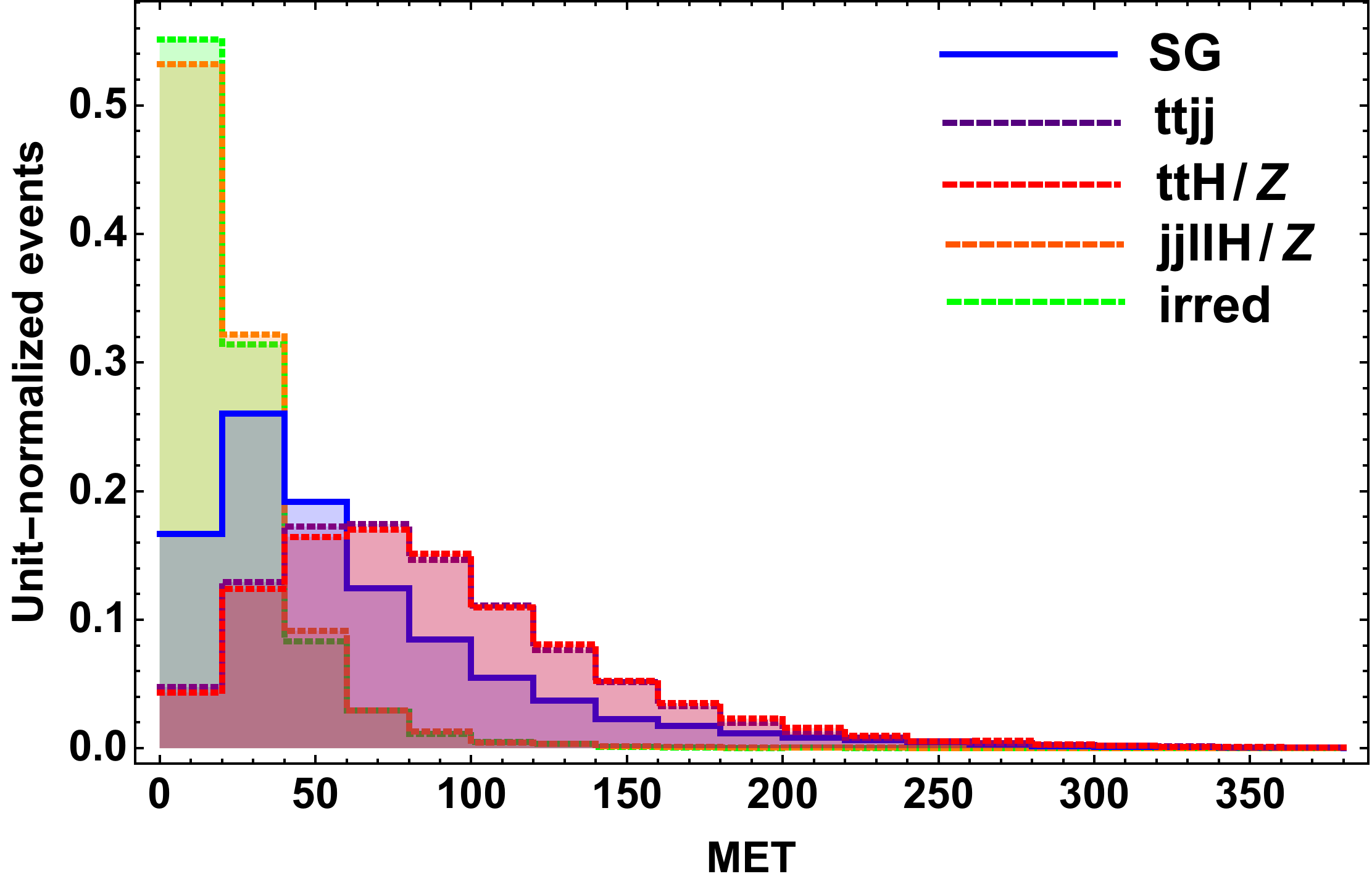}
    \caption{Di-lepton Channel: Distributions of variables:
$M_{\rm All}$ (top row, left), $M_{\ell\ell}$ (top row, right), $M_{b\bar{b}\ell}$ (mid row, left), $M_{jj\ell}$ (mid row, left), $M_{jj}$ (bottom row, left), $M_{b\bar{b}}$ (bottom row, mid) and MET $(\slashed{E}_T)$ (bottom row, right) for signal (solid blue) and backgrounds (dotted, $t\bar{t}jj$-purple, $t\bar{t} H/Z$-red, ${j j \ell \ell H/Z}$-orange, irred-green)
}
\label{fig:Di-lepton_plots}
\end{figure}
\noindent Defining $N_\ell$, $N_b$ and $N_j$ as the number of isolated leptons, b-tagged jets and non-b-tagged jets, respectively, %we restrict ourselves to the events satisfying
we select events using the following selection criteria:
\bea
& N_\ell & > 1 \;\;\; \text{with} \; \vert \eta_\ell \vert < 2.5 \nonumber \\
& N_b & > 1 \;\;\; \text{with} \; \vert \eta_b \vert < 3 \label{eq:selection_criteria}\\
& N_j & > 1 \;\;\; \text{with} \; \vert \eta_j \vert < 3. \nonumber
\eea
In addition, we impose a set of basic cuts $p_{Tj} / p_{Tb} > 20$ GeV and $p_{T\ell} > 10$ GeV at parton level event simulation, partly to avoid possible IR-divergence issues for background simulations. We reimpose such cuts on objects (hardest two jets, two $b$-jets, and two leptons) that pass selection criteria of Eq.~(\ref{eq:selection_criteria}). We use $p_T$ to evaluate hardness of the reconstructed objects and take the hardest two. %\sh{Should we mention $\Delta R > 0.4$ cuts ?} 
In Fig.~\ref{fig:Di-lepton_plots}, we show distributions of various variables for signal and background events that pass selection criteria and basic cuts. In particular, we see that $M_{\rm All}$ (top row, left), the invariant mass of \emph{all} reconstructed objects, i.e. hardest two $j$'s $+$ two $b$'s $+$ two $\ell$'s, for signal is peaked at 2 TeV, the mass of $W_R$ we take, and is well-separated from all backgrounds, providing a strong cut to reduce backgrounds. Similar sharp distinctions are drawn for $M_{jj\ell}$ (mid row, right) and $M_{b\bar{b}\ell}$ (mid row, left), but with slightly larger overlap with backgrounds. These two variables reconstruct the mass of $N$ and $\tilde{\ell}$, respectively. It may be worth describing the way we reconstruct these variables. The subtlety might be that since there are two leptons in the final states, it would be crucial to figure out which lepton is to be paired with $b$-pair, and similarly for $j$-pair. We found, for example, that naively plotting the invariant mass of $b$-pair with both leptons (similarly $j$-pair with both leptons) does not reveal sharp peak at $M_{N}$ and resulting distribution is broadly extended with large overlap with background distributions. In order to achieve sharper distinction, we make use of the fact that the masses of $N_R$ and $\tilde{\ell}_R$ are equal due to $SU(2)_{\rm R}$ invariance. Namely, we identify the lepton that goes with $b$-pair ($\ell_b$) and the one that goes with $j$-pair ($\ell_j$) by minimizing
\bea
\vert M_{b\bar{b}\ell_b} - M_{jj\ell_j} \vert.
\label{eq:l_b_vs_l_j}
\eea
As can be seen from Fig.~\ref{fig:Di-lepton_plots}, this criterion successfully reconstructs $M_{N}$ for majority of events, albeit imperfect. In this way, both $M_{b\bar{b}\ell}$ and $M_{jj\ell}$ provide another set of very useful cuts. Next very useful variable is $M_{\ell \ell}$ (top row, right). As anticipated above while we discuss each backgrounds, $M_{\ell \ell}$ distributions for {$\bf{j j \ell \ell H/Z}$} and {\bf{irred}} backgrounds are sharply localized at $M_Z$. In addition, other backgrounds also tend to be distributed over smaller $M_{\ell \ell}$ values compared to signal (see the inset plot of $M_{\ell \ell}$ distribution of Fig.~\ref{fig:Di-lepton_plots}). The bottom row of Fig.~\ref{fig:Di-lepton_plots} shows $M_{jj}$, $M_{b\bar{b}}$ and $\slashed{E}_T$ distributions. We see that $M_{jj}$ ($M_{b\bar{b}}$) distribution for signal events develops a peak at $M_W$ ($M_{H/Z}$) as expected. This is not true, on the other hand, for two major backgrounds: $\bf{t\bar{t}jj}$ and $\bf{t\bar{t}H/Z}$. Therefore, these variables will supplement above described variables to attain additional suppression of background events. %The missing transverse momentum variable, however, does not make obvious contrast between signal and backgrounds. Nonetheless, this is not necessarily in contradiction with the expectation that the backgrounds $\bf{t\bar{t}jj}$ and $\bf{t\bar{t}H/Z}$ have larger $\slashed{E}_T$ than signal. Indeed, $\slashed{E}_T$ distribution shows a slightly larger $\slashed{E}_T$ values for those two backgrounds than signal. 
Finally, the missing transverse momentum variable also helps a bit. This is expected based on the insight that the backgrounds $\bf{t\bar{t}jj}$ and $\bf{t\bar{t}H/Z}$ have larger $\slashed{E}_T$ than signal. We provide the cut flows for signal and the major SM backgrounds in Table~\ref{tab:Di-lepton_channel}. We find that the Di-lepton channel may provide a sensitivity to uncover warped seesaw nature by $\sim 3.5 \sigma$ with an integrated luminosity of $\mathcal{L} = 300 \; {\rm fb}^{-1}$ and even by $\sim 11 \sigma$ with $\mathcal{L} = 3000 \; {\rm fb}^{-1}$.

%%%%%%%%%%%%%%%%%%%%% Cut Flow Table %%%%%%%%%%%%%%%%%%%%%%%%%%%%%%%%%%%%%%%

\begin{table}[t]
\centering
\begin{tabular}{|c|c|c|c|c|c|}
\hline 
Cuts & {\bf{Signal}} & $\bf{t\bar{t}jj}$ & $\bf{t\bar{t}H/Z}$ & {$\bf{j j \ell \ell H/Z}$} & {\bf{irred}}  \\
\hline \hline
No cuts & 0.76  & $18.2 \times 10^3$ & 18.1 & 46.8 & 0.32 \\
$N_{\ell}>1$, $N_j>1$, $N_b>1$ with basic cuts & 0.12 & $2.0 \times 10^3$ & 3.11 & 3.97 & 0.030  \\
$M_{\ell \ell} \in [400,\,\infty]$ GeV & 0.11 & 25.63 & 0.045 & 0.0094 & 0 \\
$M_{\rm All} \in [1600,\,\infty]$ GeV & 0.11 & 6.50 & 0.01 & 0.0028 & 0 \\
$M_{b\bar{b}} \in [0,\,200]$ GeV & 0.09 & 2.04 & 0.0034 & 0.0014 & 0  \\	
$M_{b\bar{b}\ell} \in [550,\,\infty]$ GeV & 0.07 & 0.055 & 0.00091 & 0.00047 & 0  \\	
$\slashed{E}_T \in [0,\, 100]$ GeV & 0.058 & 0.018 & 0.00072 & 0.00047 & 0  \\	
\hline
$S/B$ & 3.02 & -- & -- & -- & -- \\
$S/\sqrt{S+B}$ ($\mathcal{L}=300$ fb$^{-1}$) & 3.62 & -- & -- & -- & -- \\
$S/\sqrt{S+B}$ ($\mathcal{L}=3000$ fb$^{-1}$) & 11.4 & -- & -- & -- & -- \\
\hline
\end{tabular}
\caption{Cut flows for signal and major background events in terms their cross sections. The cross sections are in fb. The numbers in the first row (``No cuts'') are cross sections obtained with basic cuts at the generation level to avoid divergence (for both signal and backgrounds). In the second row, the same basic cuts are reimposed to both signal and background events along with multiplicity requirements for b-jet, non-b-jet and leptons. Once the cross section decreases such that the net number of events at $\mathcal{L}=3000$ fb$^{-1}$ is less than 1, we report it as ``0''.  \label{tab:Di-lepton_channel} }
\end{table}

\subsection{Tri-lepton $+$ $H/Z$ channel}
\label{subsec:trilepton_channel}

%-- bottom-line final state 

%-- SM background considered

%-- details of smearing (less important here, since mostly leptons)...

%-- cut flow/table for signal and background

%-- final significance and signal size 

In this section, we present the results for Tri-lepton channel. Similarly to the Di-lepton channel discussed in previous section, $N-\tilde{\ell}$ pair is produced via the decay of $W^{(1)}_{R}$ using large mixing between $W^{(1)}_{R}$ and $W^{(1)}_{L}$. In Tri-lepton channel, $N$ decays to $W^{\pm} \ell^{\mp}$ and SM $W$ boson, in turn, decays leptonically producing $\ell \nu$. Like in Di-lepton channel, $\tilde{\ell}^{\pm}$ decays to $\ell^{\pm} H/Z$, with subsequent decay of $H/Z$ to $b \bar{b}$. As is evident from these cascade decays of $N$ and $\tilde{\ell}^{\pm}$, (i) signal process now does contain neutrino, leading to missing energy and (ii) there are three leptons in final states. The existence of neutrino (or missing particle in general) and the large multiplicity of leptons can be a potential obstacle in reconstruction of resonance peaks. However, we will show below that such difficulty can be, at least partly, overcome by reconstruction of longitudinal momentum of neutrino and by cleverly figuring out lepton identifications. Once these are done, various invariant mass variables can be successfully reconstructed and used to reduce backgrounds. 
Those invariant mass variables include $M_{\ell\nu}$, $M_{b \bar{b}}$, $M_{\ell\ell\nu}$, $M_{b\bar{b}\ell}$, and $M_{W_R}^{\rm recon}$, where $M_{W_R}^{\rm recon}$ is the invariant mass constructed from all reconstructed visible particles \emph{and} reconstructed neutrino four momentum. When properly  reconstructed, signal distribution of these variables will be peaked at $M_W$, $M_{H/Z}$, $M_{N}$, $M_{N}$, and $M_{W_R}$, respectively. Additional invariant mass variables exist: $M_{\ell\ell}$, $M_{\ell\ell\ell}$, and $M_{\rm All}$, where $M_{\rm All}$ is the invariant mass of \emph{all} reconstructed/visible particles without neutrino. These variables do not correspond to any of resonance peaks appeared in the signal process. However, they will still provide very strong distinctions between the signal and backgrounds.

\noindent There are several SM backgrounds we need to consider and we describe them one by one now. \\

\noindent {\bf{(1) $\bf{t\bar{t}W}$:}} The relevant process is $pp > t \bar{t} W^{\pm} > \ell^- \ell^+ \ell^{\pm} \nu \bar{\nu} \nu (\bar{\nu}) b \bar{b}$, where $t > b \; ( W^+ > \ell^+ \nu)$, and similarly for $\bar{t}$, is considered. All SM $W$'s decay leptonically: $W^{\pm} > \ell^{\pm} \nu (\bar{\nu})$. \\

\noindent {\bf{(2) irred (irreducible background):}} The relevant process is $pp > \ell^- \ell^+ W^{\pm} H/Z, \; W^{\pm} > l^{\pm} \nu (\bar{\nu}), \; H/Z > b \bar{b}$. \\

\noindent {\bf{(3) $\bf{\ell^-\ell^+W jj}$:}} The relevant process is $pp > \ell^-\ell^+ W^{\pm} jj, \; W^{\pm} > l^{\pm} \nu (\bar{\nu})$. Since we will select events with two $b$'s are tagged, only very small fraction of events with two regular jets mis-tagged as b-tagged jets will contribute to the backgrounds. Mistage rate is typically $\lesssim 1 \%$ \cite{Aad:2015ydr} and $uds$-jet mistag rate can even be as small as $0.3 \%$ \cite{Tomalin:2007zz}. The cross section of the process is $\sigma \sim 180$ fb and the surviving events with two mistagging is $\sim \mathcal{O}(0.01)$ fb. This corresponds to roughly $\sim \mathcal{O}(3)$ events at an integrated luminosity of $\mathcal{L} = 300 \; {\rm fb}^{-1}$. It will be very unlikely that any of these events will in the signal region given the number of invariant mass cuts that it should pass. Hence we will not explicitly consider this background for our analysis.  \\

%
%
%%%%%%%%%%%%%%% PLOTS %%%%%%%%%%%%%%%%%%%%%%%%%%%%%%%%%%%%%%%%%%%%%%%%%%%%%%%%

\begin{figure}%[h]
    \centering
    \includegraphics[width = 7.5 cm]{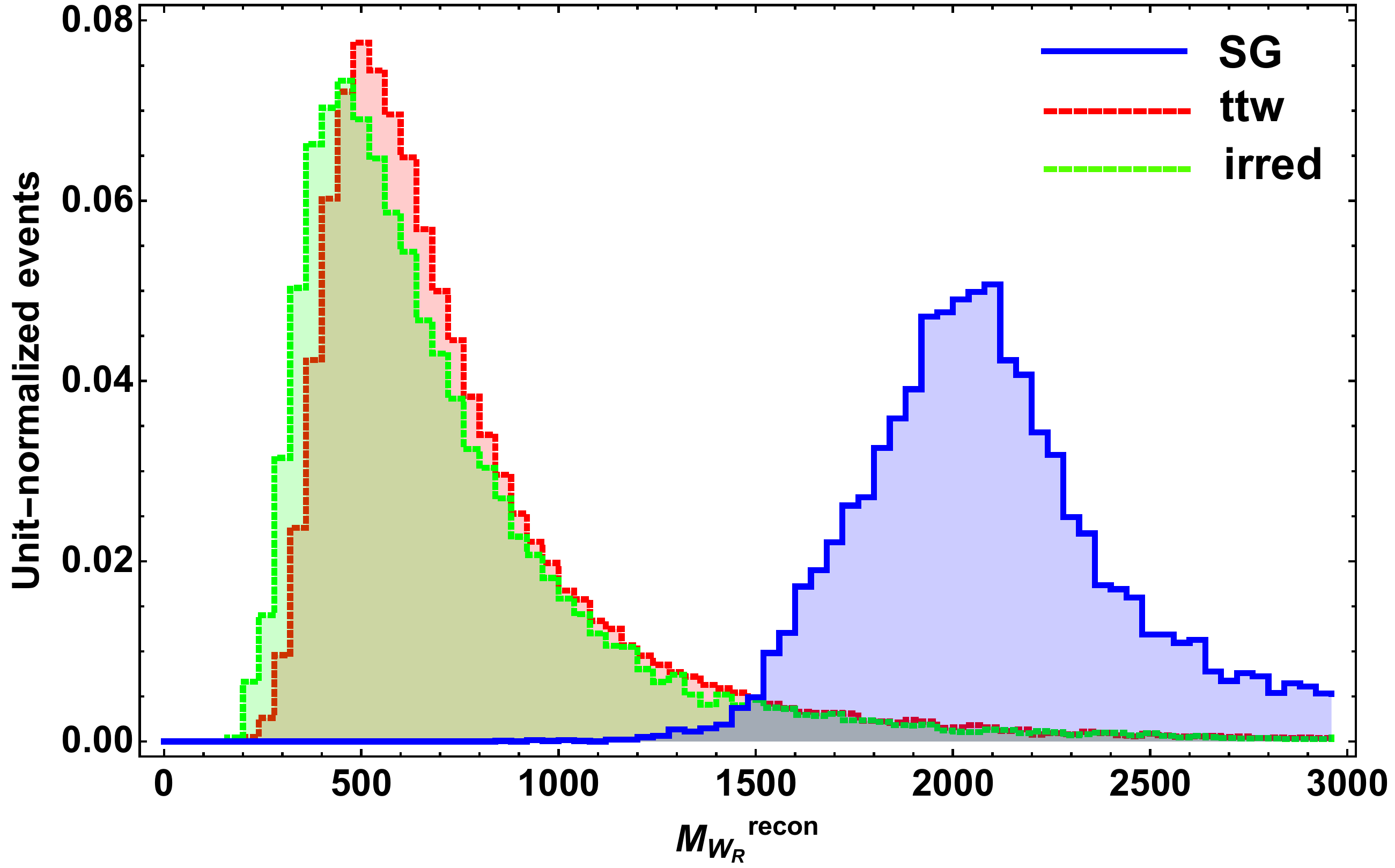}
    \includegraphics[width = 7.5 cm]{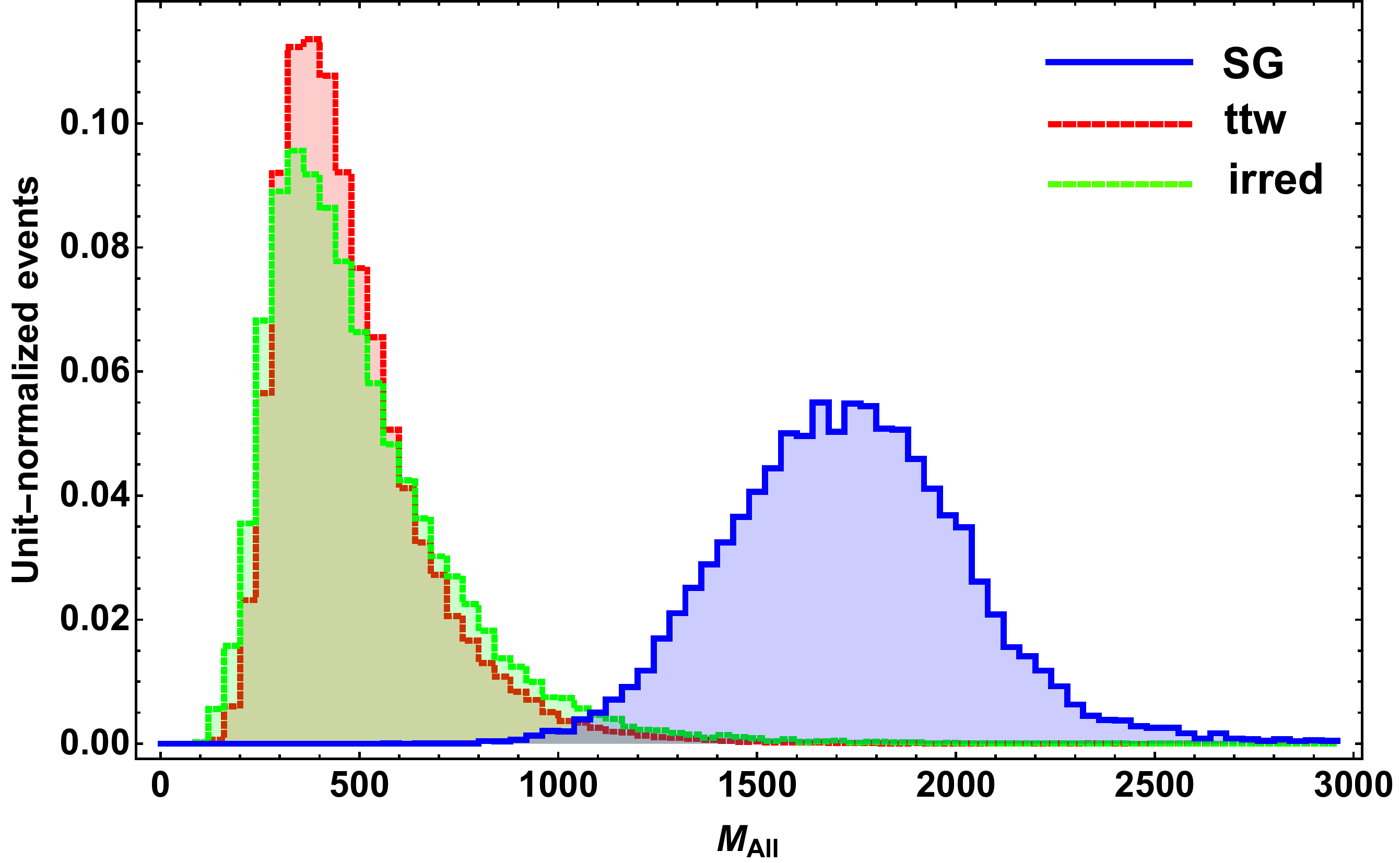}
    \includegraphics[width = 7.5 cm]{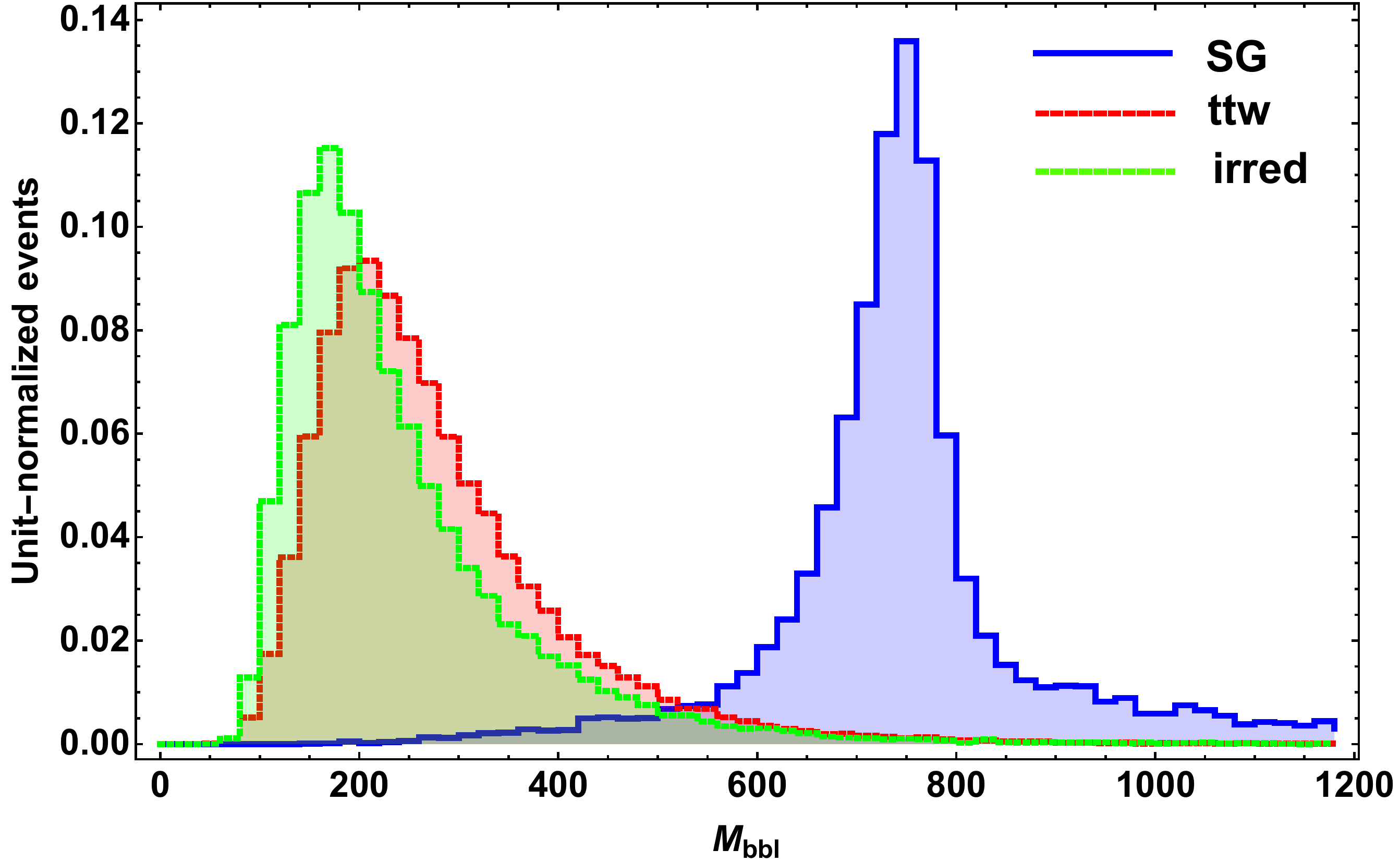}
    \includegraphics[width = 7.5 cm]{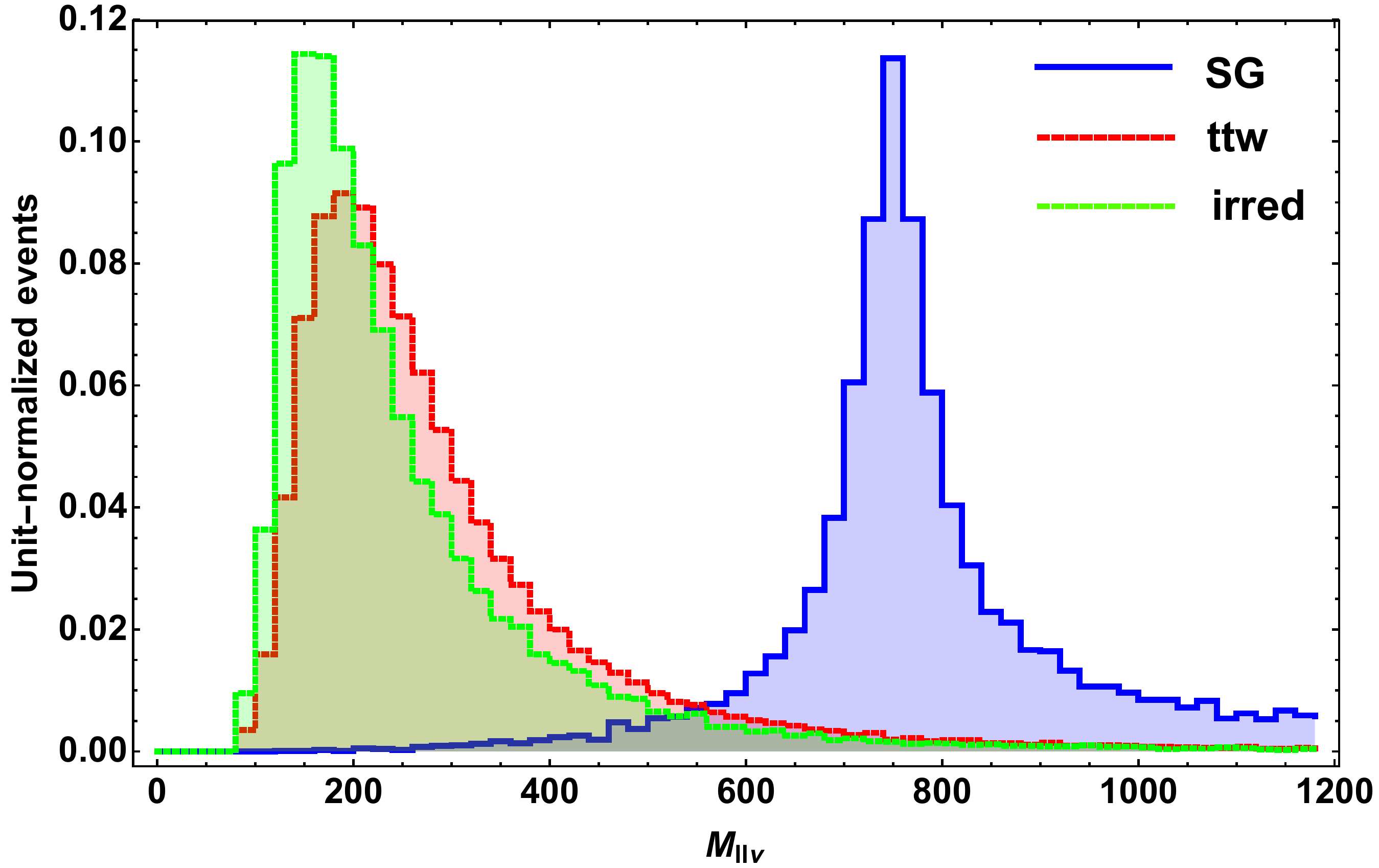}
    \includegraphics[width = 7.5 cm]{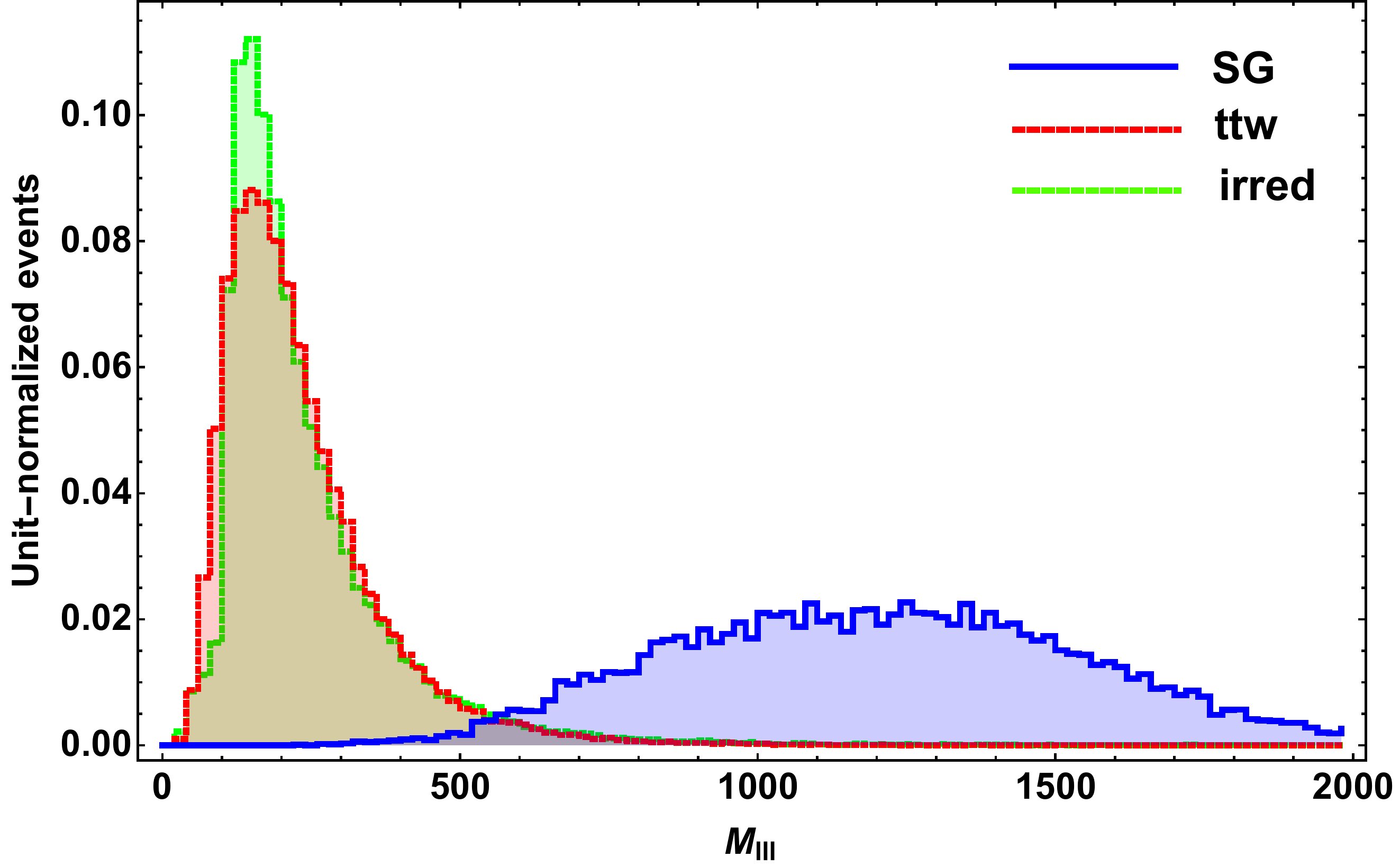}
    \includegraphics[width = 7.5 cm]{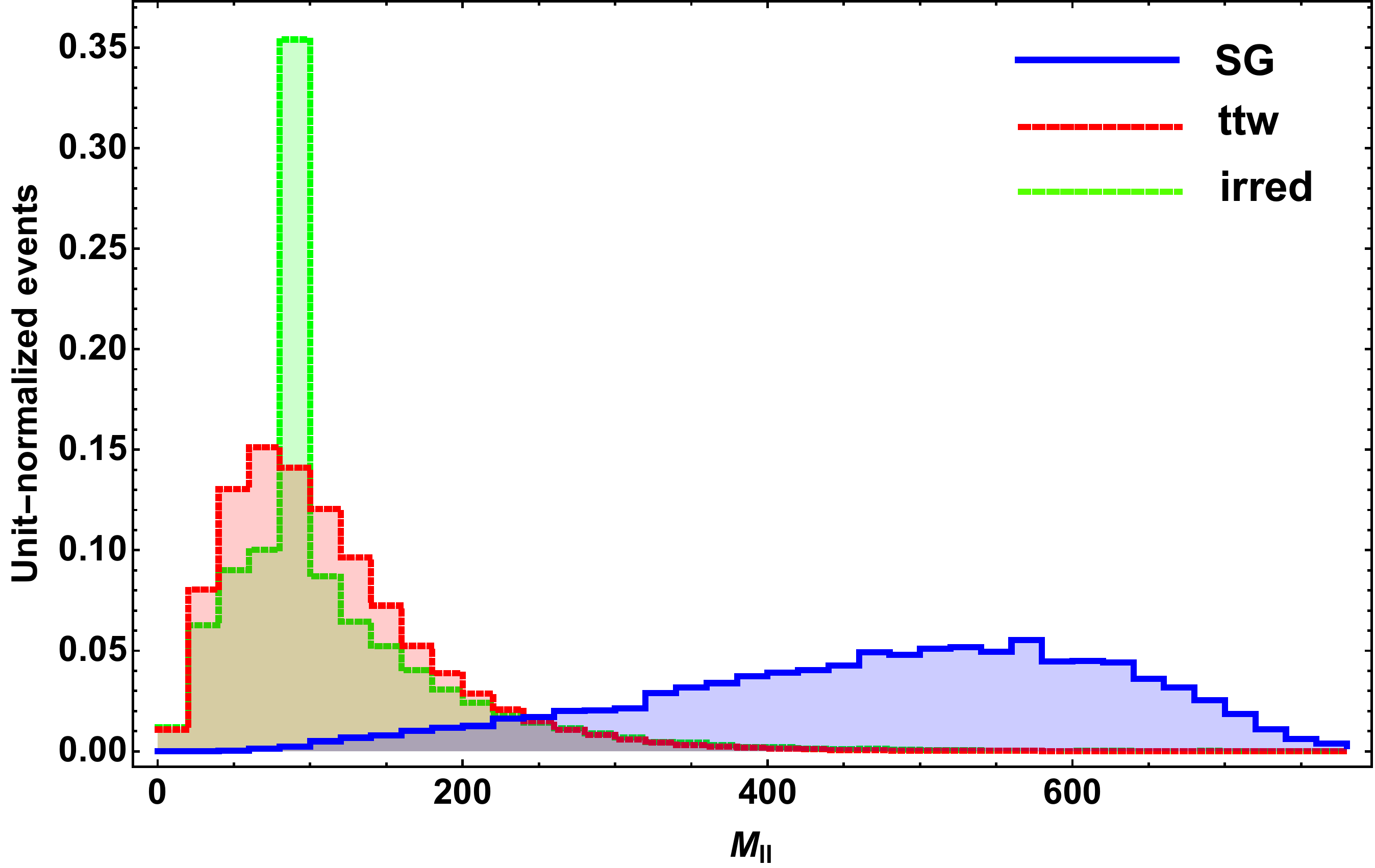}
    \caption{Tri-lepton Channel: Distributions of variables:
$M_{W_R}^{\rm recon}$ (top row, left), $M_{\rm All}$ (top row, right), $M_{b\bar{b}\ell}$ (mid row, left), $M_{\ell\ell\nu}$ (mid row, right), $M_{\ell\ell\ell}$ (bottom row, left), and $M_{\ell\ell}$ (bottom row, right) for signal (solid blue) and backgrounds (dotted, $t\bar{t}W$-red and irred-green)}
\label{fig:Tri-lepton_plots}
\end{figure}
\noindent Defining $N_\ell$ and $N_b$ as the number of isolated leptons and b-tagged jets, respectively, %we restrict ourselves to the events satisfying
we select events using the following selection criteria:
\bea
& N_\ell & > 2 \;\;\; \text{with} \; \vert \eta_\ell \vert < 2.5 \nonumber \\
& N_b & > 1 \;\;\; \text{with} \; \vert \eta_b \vert < 3 \label{eq:selection_criteria_tri-lepton}
\eea
In addition, we impose a set of basic cuts $p_{Tb} > 20$ GeV and $p_{T\ell} > 10$ GeV at parton level event simulation, partly to avoid possible IR-divergence issues for background simulations. We reimpose such cuts on objects (hardest two $b$-jets, and three leptons) that pass selection criteria of Eq.~(\ref{eq:selection_criteria_tri-lepton}). We use $p_T$ to evaluate hardness of the reconstructed objects. %\sh{Should we mention $\Delta R > 0.4$ cuts ?} 

\noindent Next, we discuss the way we reconstruct longitudinal component of neutrino's four momentum. Together, we also discuss how we figure out lepton identifications. Namely, we want to know, out of three leptons selected as described above, which one is produced together with $b\bar{b}$ from the decay of $\tilde{\ell}^{\pm}$ (we call it $\ell_b$) and which one is produced directly from the decay of $N$ (we call it $\ell_W$), and finally which one is the decay product of SM $W$ (we call it $\ell_{\nu}$).\footnote{The subscript is designed to indicate a set of particles that the lepton accompanies.} First of all, for a given choice of lepton (a candidate for $\ell_\nu$), the z-component of the neutrino's momentum can be obtained by requiring
\bea
M_W^2 = (p^\nu_\mu + p^\ell_\mu)^2
\label{eq:pvz_reconstruction}
\eea
where $M_W$ is the mass of the SM $W$ boson. For $p^{\nu}_{\mu}$, we use the fact that neutrino is massless, $(p^\nu_\mu)^2 = 0$. Then, the above equation is a quadratic equation for the z-component of $p^\nu_\mu$, and if solutions exist, there are two solutions, unless determinant vanishes by numerical coincidence. In this case, we pick up $p^\nu_z$ that minimizes the sum of z-component of all particles' momenta, i.e. sum of $p_z$ of two $b$'s, three $\ell$'s and $\nu$. This is based on the insight that $W_R$ is mostly produced at rest. In case when the determinant of the quadratic equation is less than 0, so that no solution exists, we set 
\bea
p^\nu_z = - \sum_{\rm all \; visible} p_z.
\label{eq:pvz_no_solution}
\eea
This choice again is motivated by the intuition that  $W_R$ is mostly produced at rest. Once z-component of neutrino's momentum (or equivalently full $p^\nu_\mu$) is reconstructed this way for a given choice of $\ell$ (again a candidate for $\ell_\nu$), we then determine $\ell_b$ and $\ell_W$ by minimizing
\bea
\vert M_{b\bar{b}\ell_b} - M_{\ell_\nu \ell_W \nu} \vert.
\label{eq:l_b_vs_l_W}
\eea
This criteria is motivated as before by $SU(2)_{\rm R}$ invariance and resulting mass degeneracy. In this way, for each choice of $\ell_\nu$, we determine full $p^\nu_\mu$ and identify, for the remaining two leptons, which lepton is $\ell_b$ and which lepton is $\ell_W$. We repeat this procedure for all three possible choices of $\ell_\nu$. Final decision is made for the combination $\{\ell_\nu, \ell_W, \ell_b\}$ that renders minimum value for Eq.~(\ref{eq:l_b_vs_l_W}). In Fig.~\ref{fig:Tri-lepton_plots}, we show distributions of various invariant mass variables for signal and background events that pass selection criteria and basic cuts. These invariant mass variables are constructed using $\{\ell_\nu, \ell_W, \ell_b\}$-identification and full $p^\nu_\mu$ reconstructed as described above. In particular, in addition to $M_{\rm All}$, $M_{\ell\ell}$, $M_{b\bar{b}}$ and $M_{\ell\ell\ell}$, which are all possible without knowing detailed information about lepton identification and full momentum for neutrino, now we can also explicitly compute $M_{W_R^{\rm recon}}$, $M_{b\bar{b}\ell_b}$, and $M_{\ell_\nu \ell_W \nu}$. These latter variables would not be possible without figuring out lepton identification, i.e. $\{\ell_\nu, \ell_W, \ell_b\}$, and full momentum for neutrino. To be more precise, we can actually calculate $M_{b\bar{b}\ell_b}$ and $M_{\ell_\nu \ell_W \nu}$ by simply considering all possible combination of two leptons, and we found that such computed distributions do not reveal any sharp peak and rather show very broad distributions, failing to provide strong cuts to reduce background. In Fig.~\ref{fig:Tri-lepton_plots}, we show distributions of $M_{b\bar{b}\ell_b}$ (mid row, left) and $M_{\ell_\nu \ell_W \nu}$ (mid row, right). Both distributions are sharply peaked at 750 GeV, a input value for $M_{N}$. In the case of $M_{W_R^{\rm recon}}$, we really need to know \emph{full} $p^\nu_\mu$ to be able to compute it. The left panel of the top row in Fig.~\ref{fig:Tri-lepton_plots} shows $M_{W_R^{\rm recon}}$ distribution and it is indeed peaked at/near 2 TeV, a input value for $W_R^{(1)}$. This is to be compared to the $M_{\rm All}$ distribution shown in the right panel of the top row in Fig.~\ref{fig:Tri-lepton_plots}. Again, $M_{\rm All}$ is the invariant mass for all reconstructed \emph{visible} particles, i.e. two $b$'s and three $\ell$'s, but without neutrino. Although $M_{\rm All}$ distribution also develops a peak with good separation from background distributions (dotted red ($\bf{ttw}$) and dotted green ({\bf{irred}})), the position of the peak is shifted toward the smaller value, reflecting the existence of neutrino. We found that both $M_{W_R^{\rm recon}}$ and $M_{\rm All}$, separately, provide very efficient cuts. Overall, we see that above described prescription for reconstructing $\{\ell_\nu, \ell_W, \ell_b\}$-identification and full $p^\nu_\mu$ is very effective and successful. We also note that $M_{\ell\ell}$ distribution for {\bf{irred}} is sharply peaked at $M_Z$ showing that two leptons come from on-shell decay of $Z$ boson. Finally, $M_{\ell\ell\ell}$-distribution for backgrounds are clustered for smaller values and well-separated from that of signal. %
We provide the cut flows for signal and the major SM backgrounds in Table~\ref{tab:Tri-lepton_channel}. We find that the Tri-lepton channel may provide a sensitivity to discover $N$, $\tilde{\ell}$ and $W_R$ by $\sim 4 \sigma$ with an integrated luminosity of $\mathcal{L} = 300 \; {\rm fb}^{-1}$ and even by $\sim 13 \sigma$ with $\mathcal{L} = 3000 \; {\rm fb}^{-1}$.

%%%%%%%%%%%%%%%%%%%%% Cut Flow Table %%%%%%%%%%%%%%%%%%%%%%%%%%%%%%%%%%%%%%%

\begin{table}[t]
\centering
\begin{tabular}{|c|c|c|c|}
\hline 
Cuts & {\bf{Signal}} & $\bf{t\bar{t}W}$ & {\bf{irred}}  \\
\hline \hline
No cuts & 0.42  & $2.50$ & 0.12 \\
$N_{\ell}>2$, $N_b>1$ with basic cuts & 0.060 & 0.30 & 0.011  \\
$M_{W_R^{\rm recon}} \in [1400,\,\infty]$ GeV & 0.60 & 0.022 & 0.00074 \\
$M_{\rm All} \in [1000,\,\infty]$ GeV & 0.059 & 0.0040 & 0 \\
$M_{\ell\ell\ell} \in [500,\,\infty]$ GeV & 0.059 & 0.0030 & 0  \\	
\hline
$S/B$ & 19.67 & -- & --  \\
$S/\sqrt{S+B}$ ($\mathcal{L}=300$ fb$^{-1}$) & 4.10 & -- & --  \\
$S/\sqrt{S+B}$ ($\mathcal{L}=3000$ fb$^{-1}$) & 13.00 & -- & -- \\
\hline
\end{tabular}
\caption{Cut flows for signal and major background events in terms their cross sections. The cross sections are in fb. The numbers in the first row (``No cuts'') are cross sections obtained with basic cuts at the generation level to avoid divergence (for both signal and backgrounds). In the second row, the same basic cuts are reimposed to both signal and background events along with multiplicity requirements for b-jet and leptons. Once the cross section decreases such that the net number of events at $\mathcal{L}=3000$ fb$^{-1}$ is less than 1, we report it as ``0''.  \label{tab:Tri-lepton_channel} }
\end{table}

%\sh{I still have to compose these sentences..It is already 3:30 am and my brain stopped working.. Let me do it in the morning. I am sorry..}

We close our discussion by pointing out several phenomenological features that can draw distinction between 4D LR and 5D/composite LR models.\footnote{For distinguishing between various 4D seesaw models, see, for example, \cite{Chen:2013fna}}

\begin{itemize}
\item[$\blacktriangleright$] First of all, the production of $W_R^{ \pm }$ in 4D LR models is via the \emph{unsuppressed} coupling to quarks, whereas in the case of 5D LR, it is via suppressed/smaller couplings, leading to smaller production rate. 

\item[$\blacktriangleright$] For 5D/composite LR, the production of $N$ via the decay of $W_R^{ \pm }$ accompanies its $SU(2)_{\rm R}$ partner $\tilde{\ell}$. This, in turn, renders additional Higgs/$Z$. Therefore, in 5D/composite LR models, there are two extra resonance bumps, those of $\tilde{\ell}$ and Higgs/$Z$. Both structures were crucial in reducing background. Perhaps more importantly, once discovery is made, these extra resonance peaks will be critical in discriminating 4D vs. 5D LR nature.

\item[$\blacktriangleright$] The distribution of the di-lepton invariant mass will have (i) different shape and (ii) different dependence of endpoints on $M_{ W_R^{ \pm } }$ and $M_N$. To be more specific, for usual 4D LR, the signal process is two-step cascade decay, leading to smooth distribution, except perhaps at {\em endpoint}, where, depending on spin correlations, there could be a sharp/``vertical'' drop \cite{Miller:2005zp}. For 5D/composite LR, on the other hand, having heavy $\tilde{\ell}$, in addition to $N$, in the decay of $M_{W_R^{ \pm }}$, the shape of the distribution will be that of antler with a cusp, i.e., a derivative discontinuity, in roughly {\em middle} of distribution \cite{Han:2009ss}. The end point for 4D LR is located at $\sim \sqrt{M_{W_R}^2 - M_N^2}$, that of 5D LR being different from this.
\end{itemize}

%-- focus here on discovery, that too in spite of smaller coupling to $W_R^{ \pm }$ vs.~usual/4D LR models...

%-- post-discovery question: how to distinguish the two\footnote{For distinguishing between various 4D seesaw models, see, for example, \cite{Chen:2013fna}}...

%-- already mentioned additional Higgs/$Z$ in composite seesaw vs.~usual/4D LR models ...

%-- invariant mass of dileptons will have different shape/dependence of endpoints on $M_{ W_R^{ \pm } }$ and mass of singlet neutrino: 2-step cascade in usual/4D LR case (smooth distribution, except perhaps at {\em endpoint}, where
%-- depending on spin correlations -- there could be a sharp/``vertical" drop) \cite{Miller:2005zp} vs.~antler [with a cusp, i.e., a derivative discontinuity, in (roughly) {\em middle} of distribution] \cite{Han:2009ss} in composite seesaw...

%%%%%%%%%%%%%%%%%%%%%%%%%%%%%%%%%%%%%%%%%%%%%%%%%%%%%%%%%%%%%%%%%%%%%%%%%%%%%%%%%%%%%%%%%%%
%%%%%%%%%%%%%%%%%%%%%%%%%%%%%%%%%%%%%%%%%%%%%%%%%%%%%%%%%%%%%%%%%%%%%%%%%%%%%%%%%%%%%%%%%%%

\section{Conclusions and Outlook}
\label{conclude}

Searches have been done (and are ongoing) at the LHC for {\em TeV}-mass SM singlet neutrinos involved in the generation of super-small SM neutrino mass via various 4D models of {\em seesaw}.
However, we have tried to present a case here that many these require a small parameter in order to obtain the right size of the SM neutrino mass, thus in some cases reducing the original attraction of the seesaw. % which was to obtain tiny SM neutrino masses with{\em out} such tuning. 
In fact, we feel that there might not be any strong motivation for singlets in these models to be at $\sim$ TeV other than getting a signal at the LHC from them. In earlier work, some of us had demonstrated that a completely natural realization of TeV-scale seesaw occurs instead in a warped {\em extra}-dimensional framework, which is dual (as per the AdS/CFT correspondence) to the SM Higgs being a {\em composite} particle arising from some new strong dynamics.

In this paper (and a follow-up), we initiated the study of the LHC phenomenology of this  framework of a natural TeV-scale seesaw.
In particular, here, we showed that signals similar to the 4D models arise in this warped/composite framework as well. At the same time, the details of the phenomenology are different in an interesting manner. % (see below for more details of the above claims).
Hence, one can suitably adapt {\em existing} searches for singlet neutrinos in 4D models to the natural 5D one. %(i.e., more natural) one. %, i.e., fortunately the previous effort spent on such signals  (again, from {\em tuned} models) would still be relevant.
%
%%%%%%%%%%%%%%%%%%%%%%%%

The easiest way to see how these features arise is using a (effective) two-site picture of this framework.
Namely, we have two sectors of the theory: elementary and composite. The SM Higgs is contained in the composite sector, whose characteristic mass scale is $\sim$ TeV so as to address the Planck-weak hierarchy problem; whereas, the rest of SM particles are admixtures of those in the two sectors, i.e., {\em partially} composite/elementary.
Specifically, the degree of compositeness of the non-Higgs SM particles reflects the size of their mass, i.e., the top quark is significantly composite, while the light quarks are negligibly so.
Moreover, lepton-number is preserved by the composite sector, but broken at the UV cut-off in the elementary sector. %itself (which is a Planckian scale) 
So, if we include an elementary SM singlet RH neutrino ($N_R$), then it will naturally have a super-large, even Planck-scale, Majorana mass.
However, by itself, this lepton-number violation is {\em not} quite sufficient to induce Majorana mass for {\em SM} neutrino, since we {\em also} require EWSB/Higgs VEV for this purpose.
Thus, this information about lepton-number violation has to be transmitted from the elementary to the composite sector, where the SM Higgs resides.
In this way, one can ``sew'' together the two necessary ingredients in order to generate the SM neutrino mass.

A simple and natural way for sharing lepton-number violation between the two sectors is for the above elementary $N_R$ to also mix with {\em composite} sector TeV-mass singlets. % (i.e., just like happens for SM-charged fermions).
%
%One can show that 
%
These singlet states are purely Dirac to begin with, but as a result of the above coupling to elementary  $N_R$, they acquire a tiny Majorana mass component.
It can be shown that  it is the exchange of these (now {\em pseudo}-Dirac) singlet states generates -- without any tuning -- the right size of the SM neutrino mass.
Thus, the TeV-mass singlets play a crucial role in this entire process: %(in particular, acting as ``messengers'' of lepton-number violation between elementary and composite sectors)
their observation at the LHC would provide a vital test of this mechanism of the SM neutrino mass generation.
Just to emphasize, the {\em TeV}-mass for these composite singlets is natural, being directly related to the electroweak scale (cf.~usual 4D models, where some {\em extra} assumptions are typically needed in order to get such a mass for the singlet neutrinos).

The obvious next question is how to produce these TeV-mass {\em composite} neutrinos $N_R$ at the LHC, given that they are SM {\em singlets}.
The analogous 4D models provide a recipe: typically this is achieved in these models in the context of extending the SM EW symmetry to the left-right (LR) structure, i.e., $SU(2)_{\rm L} \times SU(2)_{\rm R} \times U(1)_{ {\rm B} - {\rm L} }$, with $SU(2)_{\rm R} \times U(1)_{ {\rm B} - {\rm L} }$ broken down to SM hypercharge at the TeV scale. % (in this process, charged $W_R$ acquires a $\sim$ TeV mass).
The point is that $N_R$ -- while being SM singlet -- is a doublet of $SU(2)_R$, thus can be produced via 
decay of charged $W_R$.  $W_R$ is, in turn, produced via $q \bar{q}$ annihilation with the associated $W_R$ couplings of SM EW strength.

Indeed, a similar LR symmetric pattern is motivated in the warped/composite Higgs framework, albeit for a different reason (i.e., than parity restoration in usual 4D models). The purpose of the extra symmetry is to protect $\rho$ parameter %(i.e., the SM prediction of ratio of $W/Z$ masses) 
from receiving large corrections. % (which would be in conflict with the data).
So, we assume this extension only in the {\em composite} sector as simply a {\em global} symmetry. There is then {\em no} elementary charged $W_R$ {\em gauge} boson (unlike for the SM $W_L$), but we do have {\em composite} charged $W_R$'s.
However, in this way, it seems naively that we do not have a way to produce $W_R$, since the SM quarks inside proton are mostly elementary, leading to a negligible direct coupling to composite-sector $W_R$.

Remarkably, we found that elementary-composite $W_L$ mixing, followed by composite $W_L$-$W_R$ mixing via Higgs VEV, induces the required coupling of composite charged $W_R$'s to quarks. It is the degeneracy %at leading order 
among spin-1 composites which ensures that the second mixing effect is rather large for a few TeV composite $W$'s.
%
% factor not small
%
The end result is that coupling of light quarks to $W_R$ in these models is suppressed compared to the typical SM EW coupling, but it still sizable. %not negligible (cf.~the naive expectation that it is tiny, based on elementary-composite nature of the involved particles), i.e., 
Consequently, although production rates for $W_R$ are smaller than in 4D LR case, as we showed here, it is still enough for discovery.
We would like to emphasize here that this subtle effect %(i.e., generation of sizable coupling between light, mostly elementary quarks and composite spin-1 particles, in this case $W_R$) 
has been discussed earlier in the context of LHC signals for these spin-1 states in general, i.e., independent of neutrino mass considerations. However, this feature was not really exploited before, in the sense that decay modes of $W_R$ studied in that context (for example, $W/Z/$Higgs) were also accessible via $W_L$, i.e., production of $W_R$ was not really ``needed'' (cf.~here $N_R$ only couples to $W_R$).

Note that, in the $W_R$ decay, the composite $N_R$ is accompanied by {\em composite} charged lepton, since the associated coupling is, for example, larger than coupling to one composite and one elementary states (cf.~in 4D models, it would be simply the SM charged lepton).
Composite charged lepton decays into SM charged lepton, {\em plus} Higgs/longitudinal $Z$, while $N_R$ decays (just like in 4D models) into SM charged lepton and $W$, latter decaying either leptonically or hadronically. % into hadrons.
Thus, the final state is either (di-lepton + $W$-jet $+$ Higgs/$Z$) or (tri-lepton $+$ MET $+$ Higgs/$Z$).
Note that the dileptons in first channel are of {\em opposite} sign, given the pseudo-Dirac nature of these singlets (cf.~same-sign dileptons from {\em Majorana} singlets in some 4D LR models).
%
%\sh{I've changed ``discovery'' to ``significant excess'', since ~3 $\sigma$ is not that close to discovery... ? What do you think ? Anyway, see the changed sentence below.}

We performed a detailed analyses of both these channels for singlet neutrino production via decay of composite $W_R$, % (including above-mentioned new features vs.~4D LR models), 
finding that, for both channels, significant evidence can be observed for $\sim 2$ TeV $W_R$ and composite $N_R$/composite charged lepton of mass 750 GeV, with an integrated luminosity of $300$ ${\rm fb}^{-1}$, and even discovery with slightly more integrated luminosity. %, respectively (for certain choices of other parameters).
It is clear that the extra boson in final state permits distinguishing this framework from 4D LR models. In addition, this feature is crucial for reducing the SM background, especially given smaller rate than in 4D 
LR models {\em and} the absence of the ``smoking-gun'', i.e., same-sign dileptons; indeed, it is noteworthy that in spite of these seeming challenges, we are able to extract a reasonable signal.

%%%%%%%%%%%%%%%%%%%%%%%%

Finally, we would like to provide a ``preview'' of part II, where we will consider signals of singlet neutrinos from production and decay of particles absent in 4D LR models.
In particular, one idea is to relax the degeneracy of spin-1 composites that was assumed here. In the light of the above discussion, this direction actually results in suppressing the charged $W_R$ signal, but we will show that a ``new'' type of signal appears from a neutral heavy boson, i.e., which is {\em not} accompanied by a charged channel (unlike in the 4D LR case, where charged spin-1 channel is actually dominant, $W_R$ being lighter than the corresponding extra neutral gauge boson).
We will also study production of composite $SU(2)_{\rm L}$ {\em doublet} leptons inherent to this framework (cf.~absent in the 4D LR models); singlet neutrinos can be produced in their decays via a Yukawa coupling, i.e., {\em in}dependent of the couplings of $N_R$ to $W_R$, thus of the representation of $N_R$ under the extended EW symmetry [cf.~signals studied earlier do reply on singlet being charged under $SU(2)_{\rm R} \times U(1)_{ {\rm B} - {\rm L} }$].
Overall, our work leads to a new perspective on the nature and relevance of LHC signals of TeV-scale singlet neutrinos.

%%%%%%%%%%%%%%%%%%%%%%%%%%%%%%%%%%%%%%%%%%%%%%%%%%%%%%%%%%%%%%%%%%%%%%%%%%%%%%%%%%%%%%%%%%%%
%%%%%%%%%%%%%%%%%%%%%%%%%%%%%%%%%%%%%%%%%%%%%%%%%%%%%%%%%%%%%%%%%%%%%%%%%%%%%%%%%%%%%%%%%%%%
%%%%%%%%%%%%%%%%%%%%%%%%%%%%%%%%%%%%%%%%%%%%%%%%%%%%%%%%%%%%%%%%%%%%%%%%%%%%%%%%%%%%%%%%%%%%

\section*{Acknowledgements}

We would like to thank Chien-Yi Chen, Roberto Contino, Bhupal Dev, Shrihari Gopalakrishna, Doojin Kim and...for discussions and David Curtin for help with simulations. 
This work was supported in part by NSF Grant No.~PHY-1315155 and the Maryland Center for Fundamental Physics. SH was also supported in part by a fellowship from The Kwanjeong Educational Foundation.

\end{document}